\documentclass[showpacs,aps,floatfix,prx,reprint,superscriptaddress]{revtex4-1}
\usepackage{xfrac}
\usepackage{amssymb}
\usepackage{braket}
\usepackage{graphicx}
\usepackage{verbatim}
\usepackage{etoolbox}
\usepackage{amsfonts} 
\usepackage{amsmath} 
\usepackage[utf8]{inputenc} 
\usepackage{mathtools}
\usepackage{soul,xcolor,colortbl}
\usepackage{hyperref} \hypersetup{colorlinks=true,citecolor=blue,linkcolor=blue,urlcolor=blue} 

\usepackage{bm} 
\usepackage{array}
\newcolumntype{C}[1]{>{\centering\arraybackslash}p{#1}}\usepackage{soul}
\definecolor{Gray}{gray}{0.85}

\usepackage{color} 
\usepackage{morefloats}

\definecolor{Gray}{gray}{0.9}
\definecolor{LightCyan}{rgb}{0.88,1,1}
\definecolor{green}{rgb}{0.5451,0.2706,0.0745}

\def\EXP{{\mathrm{exp}}}
\def\acoustic{{\mathrm{a}}}

\def\WmK{{\footnotesize (W/(m$\cdot$K))}}
\def\K{{\footnotesize (K)}}
\def\GPa{{\footnotesize (GPa)}}
\def\EOS{{\small EOS}}
\def\VRH{{\small VRH}}
\def\AGL{{\small AGL}}
\def\AEL{{\small AEL}}
\def\AFLOW{{\small AFLOW}}
\def\RESTAPI{{\small REST-API}}
\def\GIBBS{{\small GIBBS}}
\def\RMSrD{{\small RMSrD}}

\def\sDebye{{\substack{\scalebox{0.6}{D}}}}
\def\sD{{\substack{\scalebox{0.6}{D}}}}
\def\sDFT{{\substack{\scalebox{0.6}{DFT}}}}
\def\sAGL{{\substack{\scalebox{0.6}{AGL}}}}
\def\sAEL{{\substack{\scalebox{0.6}{AEL}}}}
\def\sMP{{\substack{\scalebox{0.6}{MP}}}}
\def\sBM{{\substack{\scalebox{0.6}{BM}}}}
\def\sBCN{{\substack{\scalebox{0.6}{BCN}}}}
\def\sVIN{{\substack{\scalebox{0.6}{Vinet}}}}
\def\sVRH{{\substack{\scalebox{0.6}{VRH}}}}
\def\sVoigt{{\substack{\scalebox{0.6}{Voigt}}}}
\def\sReuss{{\substack{\scalebox{0.6}{Reuss}}}}
\def\sStatic{{\substack{\scalebox{0.6}{Static}}}}

\def\svib{{\substack{\scalebox{0.6}{vib}}}}
\def\sopt{{\substack{\scalebox{0.6}{opt}}}}
\def\sL{{\substack{\scalebox{0.6}{L}}}}
\def\sT{{\substack{\scalebox{0.6}{T}}}}
\def\sB{{\substack{\scalebox{0.6}{B}}}}
\def\sS{{\substack{\scalebox{0.6}{S}}}}
\def\sV{{\substack{\scalebox{0.6}{V}}}}
\def\sGGA{{\substack{\scalebox{0.6}{GGA}}}}
\def\sLDA{{\substack{\scalebox{0.6}{LDA}}}}

\makeatletter  \renewcommand\frontmatter@abstractwidth{\dimexpr\textwidth\relax} \makeatother  

\begin{document}
\title{Combining the AFLOW GIBBS and Elastic Libraries\\ for efficiently and robustly screening thermo-mechanical properties of solids}

\author{Cormac Toher}
\affiliation{Department of Mechanical Engineering and Materials Science, Duke University, Durham, North Carolina 27708, USA}
\author{Corey Oses}
\affiliation{Department of Mechanical Engineering and Materials Science, Duke University, Durham, North Carolina 27708, USA}
\author{Jose J. Plata}
\affiliation{Department of Mechanical Engineering and Materials Science, Duke University, Durham, North Carolina 27708, USA}
\author{David Hicks}
\affiliation{Department of Mechanical Engineering and Materials Science, Duke University, Durham, North Carolina 27708, USA}
\author{Frisco Rose}
\affiliation{Department of Mechanical Engineering and Materials Science, Duke University, Durham, North Carolina 27708, USA}
\author{Ohad Levy}
\affiliation{Department of Mechanical Engineering and Materials Science, Duke University, Durham, North Carolina 27708, USA}
\affiliation{Department of Physics, NRCN, {Beer-Sheva}, 84190, Israel}
\author{Maarten de Jong}
\affiliation{Department of Materials Science and Engineering, University of California, Berkeley, 210 Hearst Memorial Mining Building, Berkeley, USA}
\affiliation{Space Exploration Technologies, 1 Rocket Road, Hawthorne, CA 90250}
\author{Mark Asta}
\affiliation{Department of Materials Science and Engineering, University of California, Berkeley, 210 Hearst Memorial Mining Building, Berkeley, USA}
\author{Marco Fornari}
\affiliation{Department of Physics, Central Michigan University, Mount Pleasant, MI 48858, USA }
\author{Marco Buongiorno Nardelli}
\affiliation{Department of Physics and Department of Chemistry, University of North Texas, Denton TX }
\author{Stefano Curtarolo}
\email[]{stefano@duke.edu}
\affiliation{Materials Science, Electrical Engineering, Physics and Chemistry, Duke University, Durham NC, 27708}

\date{\today}

\begin{abstract}
  \noindent
  Thorough characterization of the thermo-mechanical properties of materials requires difficult and time-consuming experiments. 
  This severely limits the availability of data {and is} one of the main obstacles for the development of effective accelerated materials design strategies.
  The rapid screening of new potential {materials} requires highly integrated, sophisticated and robust computational approaches. 
  We tackled the challenge by {developing an automated, integrated workflow with robust error-correction} 
  within the \AFLOW\ framework {which combines} the newly developed ``Automatic Elasticity Library'' with the previously implemented \GIBBS\ method.
  The first extracts the mechanical properties from automatic self-consistent stress-strain calculations, while the latter employs those mechanical properties to evaluate the thermodynamics within the Debye model. 
  {This} new thermo-elastic {workflow} is benchmarked against a set of 74 experimentally characterized systems to pinpoint a 
  robust computational methodology for the evaluation of bulk and shear moduli, Poisson ratios, Debye temperatures, Gr{\"u}neisen parameters, and thermal conductivities of a wide variety of materials. 
  The effect of different choices of equations of state {and exchange-correlation functionals} is examined and the optimum combination of properties for the
  Leibfried-Schl{\"o}mann prediction of thermal conductivity is identified, leading to improved agreement with experimental results than the \GIBBS-only approach. 
  {The framework has been applied to the \AFLOW.org data repositories to compute the thermo-elastic properties 
    of over 3500 unique materials. The results are now available online by using an expanded version of the REST-API described in the appendix.}
\end{abstract}
\maketitle

\section{Introduction}

Calculating the thermal and elastic properties of materials is
important for predicting the thermodynamic and mechanical stability of structural
phases \cite{Greaves_Poisson_NMat_2011, Poirier_Earth_Interior_2000,
 Mouhat_Elastic_PRB_2014, curtarolo:art106} and assessing their importance for a variety of applications.
Elastic and mechanical properties such as the shear and bulk moduli are important for predicting the hardness of materials
\cite{Chen_hardness_Intermetallics_2011}, and thus their resistance to
wear and distortion.
Thermal properties, such as specific heat capacity and lattice thermal conductivity, are important for applications including thermal barrier coatings, 
thermoelectrics \cite{zebarjadi_perspectives_2012, curtarolo:art84, Garrity_thermoelectrics_PRB_2016}, and heat sinks \cite{Watari_MRS_2001, Yeh_2002}. 

{\bf Elasticity.} There are two main methods for calculating the elastic constants,
based on the response of either the stress tensor or the total energy to a set of
applied strains \cite{Mehl_TB_Elastic_1996, Mehl_Elastic_1995, Golesorkhtabar_ElaStic_CPC_2013, curtarolo:art100, Silveira_Elastic_CPC_2008, Silveira_Elastic_CPC_2008, Silva_Elastic_PEPI_2007}.
In this study, we obtain the elastic constants from the calculated stress tensors for a set of independent deformations of the crystal lattice.
This method is implemented within the \AFLOW\ framework for
computational materials design
\cite{curtarolo:art65,curtarolo:art49,curtarolo:art87}, where it is referred to as the 
\underline{A}utomatic \underline{E}lasticity \underline{L}ibrary (\AEL). 
{A similar} implementation within the Materials
Project \cite{curtarolo:art100} {allows} extensive
screening studies by combining data from these two large
repositories of computational materials data.

{\bf Thermal properties.} The determination of the thermal conductivity of materials from first principles requires either calculation of anharmonic
\underline{i}nteratomic \underline{f}orce \underline{c}onstants (IFCs) for use in the 
\underline{B}oltzmann \underline{T}ransport \underline{E}quation (BTE) \cite{Broido2007, Wu_PRB_2012, ward_ab_2009, ward_intrinsic_2010,
  Zhang_JACS_2012, Li_PRB_2012, Lindsay_PRL_2013, Lindsay_PRB_2013}, {or molecular dynamics} simulations in combination with
the Green-Kubo formula \cite{Green_JCP_1954,Kubo_JPSJ_1957}, both of
which are highly demanding computationally even within multiscale approaches \cite{curtarolo:art12}.
These methods are  unsuitable for rapid generation and screening of large databases of materials properties in order to identify trends
and simple descriptors \cite{curtarolo:art81}. 
Previously, we have implemented the ``\GIBBS'' quasi-harmonic Debye model
\cite{Blanco_CPC_GIBBS_2004, Blanco_jmolstrthch_1996} within both the
\underline{A}utomatic \underline{{\small G}}{\small IBBS} \underline{L}ibrary (\AGL) \cite{curtarolo:art96} of the
\AFLOW\ \cite{curtarolo:art65, curtarolo:art75, curtarolo:art92, curtarolo:art104,curtarolo:art110} and
Materials Project \cite{materialsproject.org,APL_Mater_Jain2013,CMS_Ong2012b} frameworks.
This approach does not require large supercell calculations since it
relies merely on first-principles calculations of the energy as a function of unit cell volume. It is thus
much more tractable computationally and eminently suited to investigating the thermal properties of 
entire classes of materials in a highly-automated {fashion
to identify} promising candidates for more in-depth experimental and computational analysis.

The data set of computed thermal and elastic properties
produced for this study is available in the \AFLOW\ 
\cite{curtarolo:art75} online data repository, either using the \AFLOW\
\underline{RE}presentational \underline{S}tate \underline{T}ransfer \underline{A}pplication \underline{P}rogramming \underline{I}nterface
(REST-API) \cite{curtarolo:art92} or via the {\sf aflow.org} web portal \cite{curtarolo:art75,curtarolo:art58}.

\section{The AEL-AGL Methodology}

The \AEL-\AGL\ methodology combines elastic constants calculations, in
the Automatic Elasticity Library (\AEL), with the calculation of
thermal properties within the Automatic \GIBBS\ Library (\AGL\
\cite{curtarolo:art96}) - ``\GIBBS'' \cite{Blanco_CPC_GIBBS_2004} implementation of the Debye model.
This integrated software library includes automatic {error correction} to facilitate high-throughput
computation of thermal and elastic materials properties within the
\AFLOW\ framework \cite{curtarolo:art65, curtarolo:art75, curtarolo:art92, curtarolo:art104,curtarolo:art53,curtarolo:art57,curtarolo:art63,curtarolo:art67,curtarolo:art54}.
The principal ingredients of the calculation are described in the following Sections.

\subsection{Elastic properties}
\label{aelmethod}

The elastic constants are evaluated from the stress-strain relations  
\begin{equation}
  \left( \begin{array}{l} s_{11} \\ s_{22} \\ s_{33} \\ s_{23} \\ s_{13} \\ s_{12} \end{array} \right) =
  \left( \begin{array}{l l l l l l} c_{11}\ c_{12}\ c_{13}\ c_{14}\ c_{15}\ c_{16} \\
           c_{12}\ c_{22}\ c_{23}\ c_{24}\ c_{25}\ c_{26} \\
           c_{13}\ c_{23}\ c_{33}\ c_{34}\ c_{35}\ c_{36} \\
           c_{14}\ c_{24}\ c_{34}\ c_{44}\ c_{45}\ c_{46} \\
           c_{15}\ c_{25}\ c_{35}\ c_{45}\ c_{55}\ c_{56} \\
           c_{16}\ c_{26}\ c_{36}\ c_{46}\ c_{56}\ c_{66} \end{array} \right)
       \left( \begin{array}{c} \epsilon_{11} \\ \epsilon_{22} \\ \epsilon_{33} \\ 2\epsilon_{23} \\ 2\epsilon_{13} \\ 2\epsilon_{12} \end{array} \right)
\end{equation}
with stress tensor elements $s_{ij}$ calculated
for a set of independent normal and shear strains $\epsilon_{ij}$. The elements of the
elastic stiffness tensor $c_{ij}$, written in the 6x6 Voigt notation using the mapping \cite{Poirier_Earth_Interior_2000}: 
$11 \mapsto 1$, $22 \mapsto 2$, $33 \mapsto 3$, $23 \mapsto 4$, $13 \mapsto 5$, $12 \mapsto 6$;
are derived from polynomial fits for each independent strain, where the polynomial degree
is automatically set to be less than the number of strains applied in each independent {direction to} avoid overfitting.
The elastic constants are then used to compute the bulk and shear
moduli, using either the Voigt approximation
\begin{equation}
  \label{bulkmodvoigt}
  B_{\sVoigt} = \frac{1}{9} \left[ (c_{11} + c_{22} + c_{33}) + 2 (c_{12} + c_{23} + c_{13}) \right]
\end{equation}
for the bulk modulus, and
\begin{multline}
  \label{shearmodvoigt}
  G_{\sVoigt} = \frac{1}{15} \left[ (c_{11} + c_{22} + c_{33}) -  (c_{12} + c_{23} + c_{13}) \right] + \\
  + \frac{1}{5} (c_{44} + c_{55} + c_{66})
\end{multline}
for the shear modulus; or the Reuss approximation, which uses the elements of the compliance tensor $s_{ij}$ (the inverse of the stiffness tensor),
where the bulk modulus is given by
\begin{equation}
  \label{bulkmodreuss}
  \frac{1}{B_{\sReuss}} =  (s_{11} + s_{22} + s_{33}) + 2 (s_{12} + s_{23} + s_{13})
\end{equation}
and the shear modulus is
\begin{multline}
  \label{shearmodreuss}
  \frac{15}{G_{\sReuss}} = 4(s_{11} + s_{22} + s_{33}) - 4 (s_{12} + s_{23} + s_{13}) + \\
  + 3 (s_{44} + s_{55} + s_{66}).
\end{multline}
For polycrystalline materials, the Voigt approximation {corresponds to assuming that the strain is uniform and that the stress is supported by the individual grains in parallel, giving} the upper bound on the elastic moduli{;} while the Reuss approximation {assumes that the stress is uniform and that the strain is the sum of the strains of the individual grains in series, giving} the lower bound {on the elastic moduli \cite{Poirier_Earth_Interior_2000}}.
The two approximations can be combined in the \underline{V}oigt-\underline{R}euss-\underline{H}ill (\VRH) \cite{Hill_elastic_average_1952} averages for the bulk modulus
\begin{equation}
  \label{bulkmodvrh}
  B_{\sVRH} = \frac{B_{\sVoigt} + B_{\sReuss}}{2};
\end{equation}
and the shear modulus
\begin{equation}
  \label{shearmodvrh}
  G_{\sVRH} = \frac{G_{\sVoigt} + G_{\sReuss}}{2}.
\end{equation}
The Poisson ratio $\sigma$ is then obtained by:
\begin{equation}
  \label{Poissonratio}
  \sigma = \frac{3 B_{\sVRH} - 2 G_{\sVRH}}{6 B_{\sVRH} + 2 G_{\sVRH}}
\end{equation}

These elastic moduli can also be used to compute the speed of sound for the transverse and longitudinal waves, as well as the 
average speed of sound in the material \cite{Poirier_Earth_Interior_2000}.
The speed of sound for the longitudinal waves is
\begin{equation}
  \label{longitudinalsoundspeed}
  v_\sL = \left(\frac{B + \frac{4}{3}G}{\rho}\right)^{\frac{1}{2}}\!\!\!,
\end{equation}
and for the transverse waves
\begin{equation}
  \label{transversesoundspeed}
  v_\sT = \left(\frac{G}{\rho}\right)^{\frac{1}{2}}\!\!\!,
\end{equation}
where $\rho$ is the mass density of the material. The average speed of
sound is then evaluated by
\begin{equation}
  \label{speedsound}
  {\overline v} = \left[\frac{1}{3} \left( \frac{2}{v_\sT^3} + \frac{1}{v_\sL^3} \right) \right]^{-\frac{1}{3}}\!\!\!.
\end{equation}

\subsection{The \AGL\ quasi-harmonic Debye-Gr{\"u}neisen model}

The Debye temperature of a solid can be written as \cite{Poirier_Earth_Interior_2000}
\begin{equation}
  \label{debyetempv}
  \theta_\sDebye = \frac{\hbar}{k_\sB}\left[\frac{6 \pi^2 n}{V}\right]^{1/3} \!\! {\overline v},
\end{equation}
where $n$ is the number of atoms in the cell, $V$ is its volume, and
${\overline v}$ is the average speed of sound of Eq.\ (\ref{speedsound}).
It can be shown by combining Eqs. (\ref{Poissonratio}), (\ref{longitudinalsoundspeed}),  (\ref{transversesoundspeed}) and (\ref{speedsound}) that ${\overline v}$ is equivalent to  \cite{Poirier_Earth_Interior_2000}
\begin{equation}
  \label{speedsoundB}
  {\overline v}  =  \sqrt{\frac{B_\sS}{\rho}} f(\sigma).
\end{equation}
where $B_\sS$ is the adiabatic bulk modulus, $\rho$ is the density, and $f(\sigma)$ is a function of the Poisson ratio $\sigma$:
\begin{equation}
  \label{fpoisson}
  f(\sigma) = \left\{ 3 \left[ 2 \left( \frac{2}{3} \!\cdot\! \frac{1 + \sigma}{1 - 2 \sigma} \right)^{3/2} \!\!\!\!\!\!\!+ \left( \frac{1}{3} \!\cdot\! \frac{1 + \sigma}{1 - \sigma} \right)^{3/2} \right]^{-1} \right\}^{\frac{1}{3}}\!\!\!\!,
\end{equation}
In an earlier version of \AGL\ \cite{curtarolo:art96}, the Poisson ratio in Eq.\ (\ref{fpoisson}) was assumed to have the {constant
value $\sigma = 0.25$ which} is the ratio for a Cauchy solid. This was found to be a reasonable approximation, producing
good correlations with experiment.
The \AEL\ approach, Eq.\ (\ref{Poissonratio}), directly evaluates $\sigma$ assuming only that it is independent of temperature and pressure.
Substituting Eq.\ (\ref{speedsoundB}) into Eq.\ (\ref{debyetempv}), the
Debye temperature is obtained as
\begin{equation}
  \label{debyetemp}
  \theta_\sDebye = \frac{\hbar}{k_\sB}[6 \pi^2 V^{1/2} n]^{1/3} f(\sigma) \sqrt{\frac{B_\sS}{M}},
\end{equation}
where $M$ is the mass of the unit cell.
The bulk modulus $B_\sS$ is obtained from a set of DFT calculations for different volume cells, either by fitting the resulting $E_\sDFT(V)$
data to a phenomenological equation of state or by taking the numerical second derivative of
a polynomial fit
\begin{eqnarray}
  \label{bulkmod}
  B_\sS (V) &\approx& B_{\mathrm{static}} (\vec{x}) \approx B_{\mathrm{static}}(\vec{x}_\sopt(V)) =\\ \nonumber
            &=&V \left( \frac{\partial^2 E(\vec{x}_\sopt (V))}{\partial V^2} \right) = V \left( \frac{\partial^2 E(V)}{\partial V^2} \right).
\end{eqnarray}
Inserting Eq.\ (\ref{bulkmod}) into Eq.\ (\ref{debyetemp}) gives the Debye temperature as a function of volume $\theta_\sDebye(V)$, for each value of
pressure, $p$, and temperature, $T$. 

The equilibrium volume at any particular $(p, T)$ point is obtained by minimizing the Gibbs free energy with
respect to volume. First, the vibrational Helmholtz free energy, $F_\svib(\vec{x}; T)$, is calculated in the quasi-harmonic approximation
\begin{equation}
  F_\svib(\vec{x}; T) \!=\!\! \int_0^{\infty} \!\!\left[\frac{\hbar \omega}{2} \!+\! k_\sB T\ \mathrm{log}\!\left(1\!-\!{\mathrm e}^{- \hbar \omega / k_\sB T}\right)\!\right]\!g(\vec{x}; \omega) d\omega,
\end{equation}
where $g(\vec{x}; \omega)$ is the phonon density of states and $\vec{x}$ describes the geometrical configuration of the system. In the Debye-Gr{\"u}neisen model, $F_\svib$ can be expressed
in terms of the Debye temperature $\theta_\sDebye$
\begin{equation}
  \label{helmholtzdebye}
  F_\svib(\theta_\sDebye; T) \!=\! n k_\sB T \!\left[ \frac{9}{8} \frac{\theta_\sDebye}{T} \!+\! 3\ \mathrm{log}\!\left(1 \!-\! {\mathrm e}^{- \theta_\sDebye / T}\!\right) \!\!-\!\! D\left(\frac{\theta_\sDebye}{T}\right)\!\!\right],
\end{equation}
where $D(\theta_\sDebye / T)$ is the Debye integral
\begin{equation}
  D \left(\theta_\sDebye/T \right) = 3 \left( \frac{T}{\theta_\sDebye} \right)^3 \int_0^{\theta_\sDebye/T} \frac{x^3}{e^x - 1} dx.
\end{equation}
The Gibbs free energy is calculated as
\begin{equation}
  \label{gibbsdebye}
  {\sf G}(V; p, T) = E_\sDFT(V) + F_\svib (\theta_\sDebye(V); T)  + pV,
\end{equation}
and fitted by a polynomial of $V$. The equilibrium volume, $V_{\mathrm{eq}}$, is that which minimizes ${\sf G}(V; p, T)$. 

Once $V_{\mathrm{eq}}$ has been determined, $\theta_\sDebye$ can be determined, and then other thermal properties including the Gr{\"u}neisen parameter and thermal
conductivity can be calculated as described in the following Sections.

\subsection{Equations of State}
\label{eqnsofstate}

Within \AGL\, the bulk modulus can be determined either numerically from the second derivative of the polynomial fit of $E_\sDFT(V)$,
Eq.\ (\ref{bulkmod}), or by fitting the $(p,V)$ data to a
phenomenological equation of state (\EOS). Three different analytic \EOS\ have been implemented within
\AGL: the Birch-Murnaghan \EOS\ \cite{Birch_Elastic_JAP_1938, Poirier_Earth_Interior_2000, Blanco_CPC_GIBBS_2004}; the Vinet \EOS\ \cite{Vinet_EoS_JPCM_1989, Blanco_CPC_GIBBS_2004};
and the Baonza-C{\'a}ceres-N{\'u}{\~n}ez spinodal \EOS\ \cite{Baonza_EoS_PRB_1995, Blanco_CPC_GIBBS_2004}.

The Birch-Murnaghan \EOS\ is
\begin{equation}
  \label{birch}
  \frac{p}{3 f (1 + 2 f)^\frac{5}{2}} = \sum_{i=0}^2 a_i f^i ,
\end{equation}
where $p$ is the pressure, $a_i$ are polynomial coefficients, and $f$ is the ``compression''  given by
\begin{equation}
  \label{birchf}
  f = \frac{1}{2} \left[\left(\frac{V}{V_0} \right)^{-\frac{2}{3}}- 1 \right].
\end{equation}
The zero pressure bulk modulus is equal to the coefficient $a_0$.

The Vinet \EOS\ is \cite{Vinet_EoS_JPCM_1989, Blanco_CPC_GIBBS_2004}
\begin{equation}
  \label{vinet}
  \log \left[ \frac{p x^2}{3 (1 - x)} \right] = \log B_0 + a (1 - x),
\end{equation}
where $a$ and $\log B_0$ are fitting parameters and
\begin{equation}
  \label{vinetx}
  x = \left(\frac{V}{V_0} \right)^{\frac{1}{3}}\!\!\!, \ 
  a = 3 (B_0' - 1) / 2.
\end{equation}
The isothermal bulk modulus $B_\sT$ is given by \cite{Vinet_EoS_JPCM_1989, Blanco_CPC_GIBBS_2004}
\begin{equation}
  \label{vinetBT}
  B_\sT = - x^{-2} B_0 e^{a(1-x)} f(x),
\end{equation}
where
\begin{equation*}
  \label{vinetfx}
  f(x)  = x - 2 - ax (1 - x).
\end{equation*}

The Baonza-C{\'a}ceres-N{\'u}{\~n}ez spinodal equation of state has the form \cite{Baonza_EoS_PRB_1995, Blanco_CPC_GIBBS_2004}
\begin{equation}
  \label{bcn}
 V = V_{\mathrm{sp}} \exp \left[ - \left(\frac{K^*}{1 - \beta} \right) (p - p_{\mathrm{sp}})^{1 - \beta} \right], 
\end{equation}
where $K^*$, $p_{\mathrm{sp}}$ and $\beta$ are the fitting parameters, and $V_{\mathrm{sp}} $ is given by
\begin{equation*}
 V_{\mathrm{sp}}  = V_0 \exp \left[ \frac{\beta}{\left(1 - \beta \right) B_0'} \right],
\end{equation*}
where $B_0 = [K^*]^{-1} (-p_{\mathrm{sp}})^{\beta}$ and $B_0' = (-p_{\mathrm{sp}})^{-1}\beta B_0$.
The isothermal bulk modulus $B_\sT$ is then given by \cite{Baonza_EoS_PRB_1995, Blanco_CPC_GIBBS_2004}
\begin{equation}
  \label{bcnBT}
  B_\sT = \frac{(p - p_{\mathrm{sp}})^{\beta}}{K^*}.
\end{equation}

{Note that \AGL\ uses $B_\sT$ instead of $B_\sS$ in Eq. \ref{debyetemp} when one of these phenomenological \EOS\ 
is selected. $B_\sS$ can then be calculated as 
\begin{equation}
 \label{BsBT}
 B_\sS = B_\sT(1 + \alpha \gamma T),
\end{equation}
where $\gamma$ is the Gr{\"u}neisen parameter (described in Section \ref{sect:gruneisen} below), and $\alpha$ is the thermal expansion
\begin{equation}
 \label{thermal_expansion}
\alpha = \frac{\gamma C_\sV}{B_\sT V},
\end{equation}
where $C_\sV$ is the heat capacity at constant volume, given by
\begin{equation}
 \label{heat_capacity}
C_\sV = 3 n k_\sB \left[4 D\left(\frac{\theta_\sD}{T}\right) - \frac{3 \theta_\sD / T}{\exp(\theta_\sD / T) - 1} \right].
\end{equation}
}

\subsection{The Gr{\"u}neisen Parameter}
\label{sect:gruneisen}

The Gr{\"u}neisen parameter describes the variation of the thermal properties of a material with the unit cell size, and contains
information about higher order phonon scattering which is important
for calculating the lattice thermal conductivity
\cite{Leibfried_formula_1954, slack, Morelli_Slack_2006, Madsen_PRB_2014, curtarolo:art96},
and thermal expansion \cite{Poirier_Earth_Interior_2000, Blanco_CPC_GIBBS_2004, curtarolo:art114}.
It is defined as the phonon frequencies dependence on the unit cell volume
\begin{equation}
  \label{gamma_micro}
  \gamma_i = - \frac{V}{\omega_i} \frac{\partial \omega_i}{\partial V}.
\end{equation}
Debye’s theory assumes that the volume dependence of all mode
frequencies is the same as that of the cut-off Debye frequency, so the Gr{\"u}neisen parameter can be expressed in terms of $\theta_\sDebye$
\begin{equation}
  \label{gruneisen_theta}
  \gamma = - \frac{\partial \ \mathrm{log} (\theta_\sDebye(V))}{\partial \ \mathrm{log} V}.
\end{equation}

This macroscopic definition of the Debye temperature is a weighted
average of Eq.\ (\ref{gamma_micro}) with the heat capacities for each branch of the phonon spectrum
\begin{equation}
  \gamma = \frac{\sum_i \gamma_i C_{V, i}} {\sum_i C_{V,i}}.
\end{equation}

{
Within \AGL\ \cite{curtarolo:art96}, the Gr{\"u}neisen parameter can
be calculated in several different ways, including direct evaluation of Eq. \ref{gruneisen_theta},
by using the more stable Mie-Gr{\"u}neisen equation \cite{Poirier_Earth_Interior_2000},
\begin{equation}
  \label{miegruneisen}
  p - p_{T=0} = \gamma \frac{U_\svib}{V},
\end{equation}
where $U_\svib$ is the vibrational internal energy \cite{Blanco_CPC_GIBBS_2004}
\begin{equation}
  \label{Uvib}
  U_\svib = n k_\sB T\left[ \frac{9}{8} \frac{\theta_\sDebye}{T} + 3D \left( \frac{\theta_\sDebye}{T} \right)\right].
\end{equation}
The ``Slater gamma'' expression  \cite{Poirier_Earth_Interior_2000}
\begin{equation}
\label{slatergamma}
  \gamma = - \frac{1}{6} + \frac{1}{2} \frac{\partial B_\sS}{\partial p}
\end{equation}
is the default method in the automated workflow used
for the \AFLOW\ database.
}

\subsection{Thermal conductivity}

In the \AGL\ framework, the thermal conductivity is calculated using the
Leibfried-Schl{\"o}mann equation \cite{Leibfried_formula_1954, slack, Morelli_Slack_2006} 
\begin{eqnarray}
  \label{thermal_conductivity}
  \kappa_{\mathrm l} (\theta_\acoustic) &=& \frac{0.849 \times 3 \sqrt[3]{4}}{20 \pi^3(1 - 0.514\gamma_\acoustic^{-1} + 0.228\gamma_\acoustic^{-2})} \times \\ \nonumber
                                        & &\times \left( \frac{k_\sB \theta_\acoustic}{\hbar} \right)^2 \frac{k_\sB m V^{\frac{1}{3}}}{\hbar \gamma_\acoustic^2}.
\end{eqnarray}
where $V$ is the volume of the unit cell and $m$ is the average atomic mass.
It should be noted that the Debye temperature and Gr{\"u}neisen parameter in this formula, $\theta_\acoustic$ and $\gamma_\acoustic$, are slightly
different {from} the traditional Debye temperature, $\theta_\sDebye$, calculated in Eq.\ (\ref{debyetemp}) and Gr{\"u}neisen parameter, $\gamma$, obtained from
Eq.\ (\ref{slatergamma}). Instead, $\theta_\acoustic$ and $\gamma_\acoustic$  are obtained by only considering the acoustic modes, based on the assumption that the optical
phonon modes in crystals do not contribute to heat transport \cite{slack}. This $\theta_\acoustic$ is referred to as the ``acoustic'' Debye temperature
\cite{slack, Morelli_Slack_2006}. It can be derived directly from the phonon DOS by integrating only over the acoustic modes \cite{slack,
  Wee_Fornari_TiNiSn_JEM_2012}. Alternatively, it can be calculated from the traditional Debye temperature $\theta_\sDebye$ \cite{slack, Morelli_Slack_2006}
\begin{equation}
  \label{acousticdebyetemp}
  \theta_\acoustic = \theta_\sDebye n^{-\frac{1}{3}}.
\end{equation}

{There is no simple way to extract the ``acoustic''  Gr{\"u}neisen parameter from the traditional Gr{\"u}neisen parameter.} 
Instead, it must be calculated from Eq.\ (\ref{gamma_micro}) for each phonon branch separately and summed over the acoustic branches \cite{curtarolo:art114, curtarolo:art119}.
This requires using the quasi-harmonic phonon approximation which involves calculating the full phonon spectrum for different
volumes \cite{Wee_Fornari_TiNiSn_JEM_2012, curtarolo:art114, curtarolo:art119}, and is therefore too computationally demanding to be used for
high-throughput screening, particularly for large, low symmetry systems. Therefore, we use the approximation
$\gamma_\acoustic = \gamma$ in the \AEL-\AGL\ approach to {calculate} the thermal conductivity. The dependence of the expression in Eq.
(\ref{thermal_conductivity}) on $\gamma$ is weak \cite{curtarolo:art96, Morelli_Slack_2006}, thus
the evaluation of $\kappa_l$ using the traditional  Gr{\"u}neisen parameter introduces just a small systematic error which is insignificant for
screening purposes \cite{curtarolo:art119}.

The thermal conductivity at temperatures other than $\theta_\acoustic$ is estimated by \cite{slack, Morelli_Slack_2006, Madsen_PRB_2014}:
\begin{equation}
  \label{kappa_temperature}
  \kappa_{\mathrm l} (T) = \kappa_{\mathrm l}(\theta_\acoustic) \frac{\theta_\acoustic}{T}.
\end{equation}

\subsection{DFT calculations and workflow details}

The {\small DFT} calculations to obtain $E(V)$ and the strain tensors were performed using
the {\small VASP} software \cite{kresse_vasp} with projector-augmented-wave
pseudopotentials \cite{PAW} and the PBE parameterization of the
generalized gradient approximation to the exchange-correlation
functional \cite{PBE}, using the {parameters described} in the \AFLOW\
Standard \cite{curtarolo:art104}. The energies were calculated at zero
temperature and pressure, with spin polarization and without zero-point motion or lattice
vibrations. The initial crystal structures were fully relaxed (cell
volume and shape and the basis atom coordinates inside the cell).

For the \AEL\ calculations, 4 strains were applied in each independent lattice direction
(two compressive and two expansive) with a maximum strain of 1\% in each direction,
for a total of 24 configurations \cite{curtarolo:art100}. For cubic systems,
the crystal symmetry was used to reduce the number of required strain configurations
to 8. For each configuration, two ionic positions \AFLOW\ Standard {\verb!RELAX!} \cite{curtarolo:art104}
calculations at fixed cell volume and shape were followed by a single \AFLOW\ Standard  {\verb!STATIC!}  \cite{curtarolo:art104}
calculation.
The elastic constants are then calculated by fitting the elements of stress tensor obtained for each independent strain.
The stress tensor from the zero-strain configuration
(i.e. the initial unstrained relaxed structure) can also be {included in the set of fitted strains}, although this was found to have negligible effect on the results.
Once these calculations are complete, it is verified that the eigenvalues of the stiffness tensor are all positive,
that the stiffness tensor obeys the appropriate symmetry rules for the lattice type \cite{Mouhat_Elastic_PRB_2014}, and
that the applied strain is still within the linear regime, using the method described by de Jong et al. \cite{curtarolo:art100}.
If any of these conditions fail, the calculation is repeated with
adjusted applied strain.

The \AGL\ calculation of $E(V)$ is fitted to the energy at 28 different
volumes of the unit cell obtained by increasing or decreasing the relaxed lattice parameters in fractional
increments of 0.01, with a single \AFLOW\ Standard
{\verb!STATIC!} \cite{curtarolo:art104} calculation at each volume.
The resulting $E(V)$ data is checked for convexity and to verify that the minimum energy is at the
initial volume (i.e. at the properly relaxed cell size). If any of these
conditions fail, the calculation is repeated with adjusted parameters,
e.g. increased k-point grid density.


\subsection{Correlation Analysis}

Pearson and Spearman correlations {are used to}
analyze the results for entire sets of materials. The {Pearson coefficient} $r$ is a measure of the linear
correlation between two variables, $X$ and $Y$. It is calculated by
\begin{equation}
  \label{Pearson}
  r = \frac{\sum_{i=1}^{n} \left(X_i - \overline{X} \right) \left(Y_i - \overline{Y} \right) }{ \sqrt{\sum_{i=1}^{n} \left(X_i - \overline{X} \right)^2} \sqrt{\sum_{i=1}^{n} \left(Y_i - \overline{Y} \right)^2}},
\end{equation}
where $\overline{X}$ and $\overline{Y}$ are the mean values of $X$ and $Y$.

The {Spearman coefficient} $\rho$ is a measure of the monotonicity of the relation between two variables.
The raw values of the two variables $X_i$ and $Y_i$ are sorted in ascending order, and are assigned rank values $x_i$ and $y_i$ which
are equal to their position in the sorted list. If there is more than one variable with the same value, the average of the position values
are assigned to {all duplicate entries}. The correlation coefficient is then given by
\begin{equation}
  \label{Spearman}
  \rho = \frac{\sum_{i=1}^{n} \left(x_i - \overline{x} \right) \left(y_i - \overline{y} \right) }{ \sqrt{\sum_{i=1}^{n} \left(x_i - \overline{x} \right)^2} \sqrt{\sum_{i=1}^{n} \left(y_i - \overline{y} \right)^2}}.
\end{equation}
It is useful for determining how well the ranking order of the values of one variable predict the ranking order of the values of the other variable.


The discrepancy between the \AEL-\AGL\ predictions and experiment is
evaluated in terms  normalized root-mean-square relative deviation
\begin{equation}
  \label{RMSD}
  {\small \mathrm{RMSrD}}  = \sqrt{\frac{ \sum_{i=1}^{n} \left( \frac{X_i - Y_i}{X_i} \right)^2 }{N - 1}} ,
\end{equation}
{In contrast} to the correlations described above, lower values of the \RMSrD\ indicate better agreement with experiment. This measure is particularly useful for
comparing predictions of the same property using different
methodologies that may have very similar correlations with, but different
deviations from, the experimental results.

\section{Results}

We used the \AEL-\AGL\ methodology to calculate the mechanical and thermal properties, including the bulk modulus,
shear modulus, Poisson ratio,  Debye temperature, Gr{\"u}neisen parameter and thermal conductivity for a set of 74 materials
with structures including diamond, zincblende, rocksalt, wurzite, rhombohedral and body-centred tetragonal.
The results have been compared to experimental values (where available), and the correlations between the calculated and
experimental values were deduced.
In cases where multiple experimental values are present in the literature, we used the most recently reported
value, unless otherwise specified.

In Section \ref{aelmethod}, three different approximations for the bulk and shear moduli are described: Voigt (Eqs. (\ref{bulkmodvoigt}), (\ref{shearmodvoigt})),
Reuss (Eqs. (\ref{bulkmodreuss}), (\ref{shearmodreuss})), and the Voigt-Reuss-Hill (\VRH) average (Eqs. (\ref{bulkmodvrh}), (\ref{shearmodvrh})).
These approximation{s give very similar values for the
bulk modulus} for the set of materials included in this work, particularly those with cubic symmetry. 
Therefore only {$B_\sVRH^\sAEL$} 
is explicitly cited in the following listed results
(the values obtained for all three  approximations are available in the \AFLOW\ database entries for
these materials). The values for the shear modulus in these three
approximations exhibit larger variations, and are therefore all listed and compared to experiment.
In several cases, the experimental values of the bulk and shear moduli have been calculated
from the measured  elastic constants using Eqs. (\ref{bulkmodvoigt}) through (\ref{shearmodvrh}), and an experimental Poisson ratio $\sigma^\EXP$
was calculated from these values using Eq.\ (\ref{Poissonratio}).

As described in Section \ref{eqnsofstate}, the bulk modulus in \AGL\ can be calculated from a polynomial fit of the $E(V)$ data as shown in Eq.\ (\ref{bulkmod}), 
or by fitting the $E(V)$ data to one of three empirical equations
of state: Birch-Murnaghan (Eq.\ (\ref{birch})), Vinet (Eq.\ (\ref{vinet})), and the Baonza-C{\'a}ceres-N{\'u}{\~n}ez
(Eq.\ (\ref{bcn})). We compare the results of these four methods, labeled $B_\sStatic^\sAGL$,  $B_\sStatic^\sBM$, $B_\sStatic^\sVIN$, and
$B_\sStatic^\sBCN$, respectively,  with the experimental values $B^\EXP$ and those obtained from the
elastic calculations $B_\sVRH^\sAEL$.
The Debye temperatures, Gr{\"u}neisen parameters and thermal conductivities depend on the calculated bulk modulus and are
therefore also cited below for each of the equations of state.
Also included are the Debye temperatures derived from the calculated
elastic constants and speed of sound as given by Eq.\ (\ref{speedsound}).
The Debye temperatures, $\theta_\sDebye^\sBM$
(Eq.\ (\ref{birch})), $\theta_\sDebye^\sVIN$ (Eq.\ (\ref{vinet})),
$\theta_\sDebye^\sBCN$, Eq.\ (\ref{bcn})), calculated using the Poisson ratio $\sigma^\sAEL$ obtained from
Eq.\ (\ref{Poissonratio}), are compared to $\theta_\sDebye^\sAGL$, obtained from the numerical fit
of $E(V)$ (Eq.\ (\ref{bulkmod})) using both $\sigma^\sAEL$ and the approximation $\sigma =
0.25$ used in Ref. \onlinecite{curtarolo:art96}, to
$\theta_\sDebye^\sAEL$, calculated with the speed of sound obtained
using Eq.\ (\ref{speedsound}),
and to the experimental values $\theta^\EXP$.
The values of the acoustic Debye temperature ($\theta_\acoustic$, Eq.\ (\ref{acousticdebyetemp}))
are shown, where available, in parentheses below the traditional Debye temperature value.

The experimental Gr{\"u}neisen parameter, $\gamma^\EXP$, is compared to $\gamma^\sAGL$ (Eq.\ (\ref{bulkmod})),  obtained using the numerical
polynomial fit of $E(V)$ and both values of the Poisson ratio
($\sigma^\sAEL$ and the approximation $\sigma = 0.25$ from
Ref. \onlinecite{curtarolo:art96}), and to $\gamma^\sBM$
(Eq.\ (\ref{birch})), $\gamma^\sVIN$ (Eq.\ (\ref{vinet})), and $\gamma^\sBCN$ (Eq.\ (\ref{bcn})), calculated
using $\sigma^\sAEL$ only. Similarly, the experimental lattice thermal
conductivity $\kappa^\EXP$ is compared to $\kappa^\sAGL$ (Eq.\ (\ref{bulkmod})),
obtained using the numerical polynomial fit and both the calculated
and approximated values of $\sigma$, and to $\kappa^\sBM$
(Eq.\ (\ref{birch})), $\kappa^\sVIN$ (Eq.\ (\ref{vinet})), and $\kappa^\sBCN$
(Eq.\ (\ref{bcn})), calculated using only $\sigma^\sAEL$.

The \AEL\ method has been been previously implemented in the Materials Project framework for calculating 
elastic constants \cite{curtarolo:art100}. {Data from} the Materials Project database are included
in the tables below for comparison {for the bulk modulus $B_\sVRH^\sMP$, shear modulus $G_\sVRH^\sMP$, and Poisson ratio $\sigma^\sMP$.}

\subsection{Zincblende and diamond structure materials}

The mechanical and thermal properties were calculated for a set of materials with the zincblende
(spacegroup: F$\overline{4}3$m,\ $\#$216; Pearson symbol: cF8; 
\AFLOW\ prototype: {\sf AB\_cF8\_216\_c\_a} \cite{curtarolo:art121}) and diamond
(Fd$\overline{3}$m,\ $\#$227; cF8; {\sf A\_cF8\_227\_a}
\cite{curtarolo:art121}) structures. This {is the same set of materials
as} in Table I of Ref. \onlinecite{curtarolo:art96}, which in {turn are from} Table II of 
Ref. \onlinecite{slack} and Table 2.2 of Ref. \onlinecite{Morelli_Slack_2006}.

\begin{figure*}[t!]
  \includegraphics[width=0.98\textwidth]{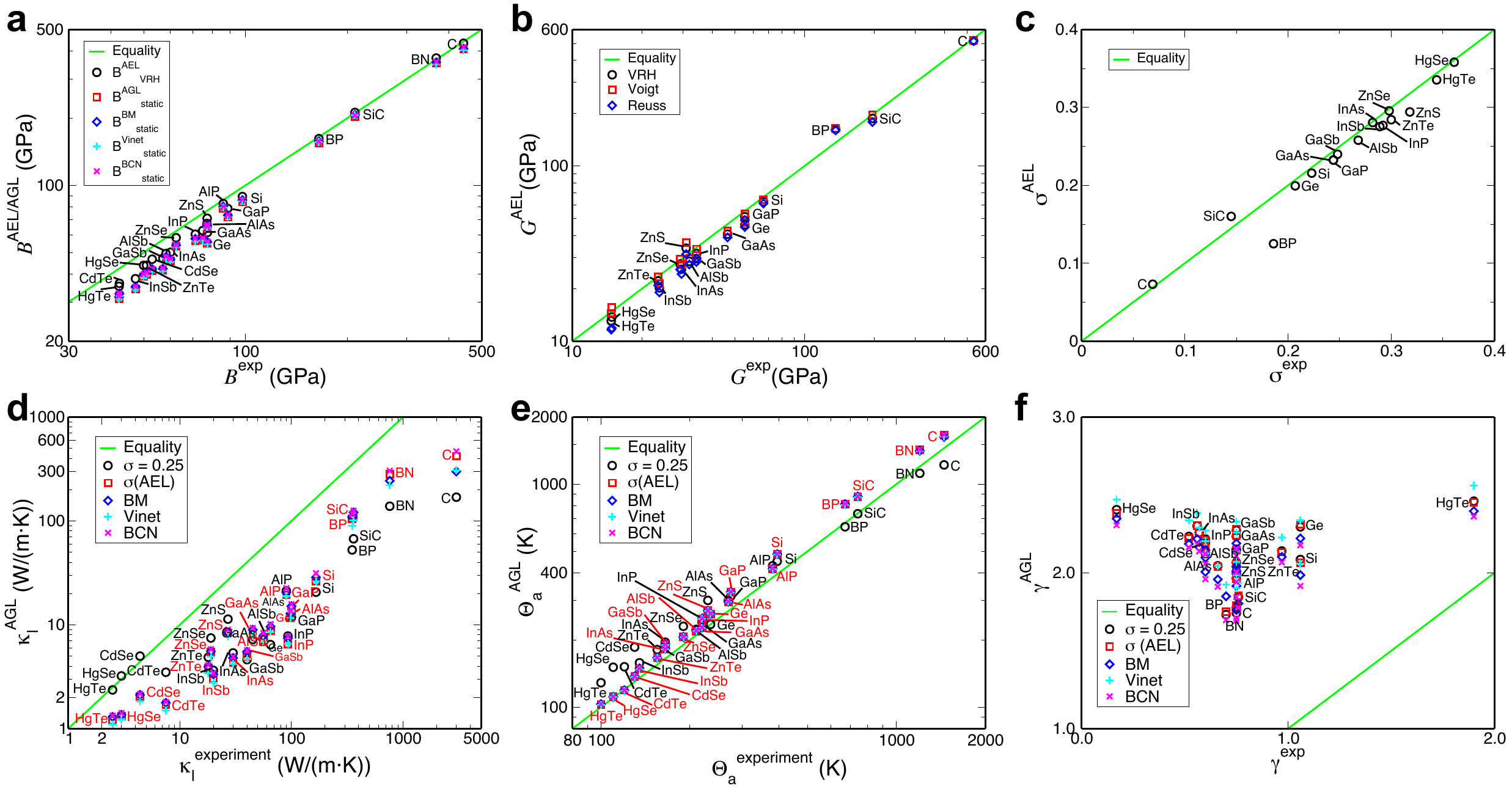}
  \vspace{-4mm}
  \caption{\small 
    {\bf (a)}  Bulk modulus,
    {\bf (b)} shear modulus,
    {\bf (c)} Poisson ratio, 
    {\bf (d)} lattice thermal conductivity at 300K,
    {\bf (e)} acoustic Debye temperature and
    {\bf (f)} Gr{\"u}neisen parameter of zincblende  (\AFLOW\ prototype: {\sf AB\_cF8\_216\_c\_a} \cite{curtarolo:art121}) and diamond ({\sf A\_cF8\_227\_a} \cite{curtarolo:art121}) structure
    semiconductors. }
  \label{fig:zincblende_thermal_elastic}
\end{figure*}

The elastic {properties bulk modulus}, shear modulus and Poisson {ratio calculated} using \AEL\ and \AGL\ are shown
in Table \ref{tab:zincblende_elastic} and Fig. \ref{fig:zincblende_thermal_elastic}, together
with experimental values from the literature where available. As can be seen
from the results in Table \ref{tab:zincblende_elastic} and Fig. \ref{fig:zincblende_thermal_elastic}(a), the $B_\sVRH^\sAEL$ values are
generally closest to experiment as shown by the \RMSrD\ value of $0.13$, producing an underestimate of the order of 10\%. The \AGL\ values from both the numerical
fit and the empirical equations of state are generally very similar to each other, while being slightly less than the $B_\sVRH^\sAEL$
values.

\begin{table*}[t!]
  \caption{\small Bulk modulus, shear modulus and Poisson ratio of
    zincblende (\AFLOW\ prototype: {\sf AB\_cF8\_216\_c\_a} \cite{curtarolo:art121}) and diamond ({\sf A\_cF8\_227\_a} \cite{curtarolo:art121}) structure semiconductors.
    ``N/A''= Not available for that source.
    Units: $B$ and $G$ in \GPa.
  }
  \label{tab:zincblende_elastic}
  {\footnotesize
    \begin{tabular}{c c c c c c c c c c c c c c c c}
      \hline
      Comp. & $B^\EXP$  & $B_\sVRH^\sAEL$ & $B_\sVRH^\sMP$ & $B_\sStatic^\sAGL$ & $B_\sStatic^\sBM$ & $B_\sStatic^\sVIN$ &  $B_\sStatic^\sBCN$ & $G^\EXP$ & $G_\sVoigt^\sAEL$ & $G_\sReuss^\sAEL$ &  $G_\sVRH^\sAEL$ & $G_\sVRH^\sMP$ & $\sigma^\EXP$ & $\sigma^\sAEL$ & $\sigma^\sMP$      \\
      \hline
      C & 442 \cite{Semiconductors_BasicData_Springer,
          Lam_BulkMod_PRB_1987, Grimsditch_ElasticDiamond_PRB_1975} & 434 & N/A & 408 & 409 & 403 & 417 & 534 \cite{Semiconductors_BasicData_Springer, Grimsditch_ElasticDiamond_PRB_1975}  & 520 & 516 & 518 & N/A & 0.069 \cite{Semiconductors_BasicData_Springer, Grimsditch_ElasticDiamond_PRB_1975} & 0.073 & N/A \\
      SiC & 248 \cite{Strossner_ElasticSiC_SSC_1987} & 212 & 211 & 203 & 207 & 206 & 206 & 196 \cite{Fate_ShearSiC_JACeramS_1974} & 195 & 178 & 187 & 187 & 0.145 \cite{Lam_BulkMod_PRB_1987, Fate_ShearSiC_JACeramS_1974} & 0.160 & 0.16 \\
            & 211 \cite{Semiconductors_BasicData_Springer, Lam_BulkMod_PRB_1987} & & & & & & & 170 \cite{Semiconductors_BasicData_Springer} & & & & & 0.183 \cite{Semiconductors_BasicData_Springer} & & \\
      Si & 97.8 \cite{Semiconductors_BasicData_Springer, Hall_ElasticSi_PR_1967} & 89.1 & 83.0 & 84.2 & 85.9 & 85.0 & 86.1 & 66.5 \cite{Semiconductors_BasicData_Springer, Hall_ElasticSi_PR_1967} & 64 & 61 & 62.5 & 61.2 & 0.223 \cite{Semiconductors_BasicData_Springer, Hall_ElasticSi_PR_1967} & 0.216 & 0.2  \\
            & 98 \cite{Lam_BulkMod_PRB_1987}  & & & & & \\
      Ge & 75.8 \cite{Semiconductors_BasicData_Springer, Bruner_ElasticGe_PRL_1961} & 61.5 & 59.0 & 54.9 & 55.7 & 54.5 & 56.1 & 55.3 \cite{Semiconductors_BasicData_Springer, Bruner_ElasticGe_PRL_1961} & 47.7 & 44.8 & 46.2 & 45.4 & 0.207 \cite{Semiconductors_BasicData_Springer, Bruner_ElasticGe_PRL_1961} & 0.199 & 0.19 \\
            & 77.2 \cite{Lam_BulkMod_PRB_1987}  & & & & & \\
      BN & 367.0 \cite{Lam_BulkMod_PRB_1987} & 372 & N/A & 353 & 356 & 348 & 359 & N/A & 387 & 374 & 380 & N/A & N/A & 0.119 & N/A \\
      BP & 165.0  \cite{Semiconductors_BasicData_Springer, Lam_BulkMod_PRB_1987} & 162 & 161 & 155 & 157 & 156 & 157 & 136 \cite{Semiconductors_BasicData_Springer, Wettling_ElasticBP_SSC_1984} & 164 & 160 & 162 & 162 & 0.186 \cite{Semiconductors_BasicData_Springer, Wettling_ElasticBP_SSC_1984} & 0.125 & 0.12 \\
            & 267 \cite{Semiconductors_BasicData_Springer, Suzuki_ElasticBP_JAP_1983}  & & & & & \\
            & 172 \cite{Semiconductors_BasicData_Springer, Wettling_ElasticBP_SSC_1984} & & & & & \\
      AlP & 86.0  \cite{Lam_BulkMod_PRB_1987} & 82.9 & 85.2 & 78.9 & 80.4 & 79.5 & 80.4 & N/A & 48.6 & 44.2 & 46.4 & 47.2 & N/A & 0.264 & 0.27 \\
      AlAs & 77.0  \cite{Lam_BulkMod_PRB_1987} & 67.4 & 69.8 & 63.8 & 65.1 & 64.0 & 65.3 & N/A & 41.1 & 37.5 & 39.3 & 39.1 & N/A & 0.256 & 0.26 \\
            & 74 \cite{Greene_ElasticAlAs_PRL_1994}  & & & & &  \\
      AlSb & 58.2  \cite{Lam_BulkMod_PRB_1987, Semiconductors_BasicData_Springer, Bolef_ElasticAlSb_JAP_1960, Weil_ElasticAlSb_JAP_1972} &  49.4 & 49.2 & 46.5 & 47.8 & 46.9 & 47.8 &  31.9 \cite{Semiconductors_BasicData_Springer, Bolef_ElasticAlSb_JAP_1960, Weil_ElasticAlSb_JAP_1972} & 29.7 & 27.4 & 28.5 & 29.6 & 0.268 \cite{Semiconductors_BasicData_Springer, Bolef_ElasticAlSb_JAP_1960, Weil_ElasticAlSb_JAP_1972}  & 0.258 & 0.25 \\
      GaP & 88.7  \cite{Lam_BulkMod_PRB_1987} & 78.8 & 76.2 & 71.9 & 73.4 & 72.2 & 73.8 & 55.3 \cite{Boyle_ElasticGaPSb_PRB_1975} & 53.5 & 49.1 & 51.3 & 51.8 & 0.244 \cite{Boyle_ElasticGaPSb_PRB_1975} & 0.232 & 0.22 \\
            & 89.8 \cite{Boyle_ElasticGaPSb_PRB_1975} & & & & & \\
      GaAs & 74.8  \cite{Lam_BulkMod_PRB_1987} & 62.7 & 60.7 & 56.8 & 57.7 & 56.6 & 58.1 & 46.6 \cite{Bateman_ElasticGaAs_JAP_1975} & 42.6 & 39.1 & 40.8 & 40.9 & 0.244 \cite{Bateman_ElasticGaAs_JAP_1975} & 0.233 & 0.23 \\
            & 75.5 \cite{Bateman_ElasticGaAs_JAP_1975} & & & & & \\
      GaSb & 57.0  \cite{Lam_BulkMod_PRB_1987} & 47.0 & 44.7 & 41.6 & 42.3 & 41.2 & 42.6 & 34.2  \cite{Boyle_ElasticGaPSb_PRB_1975} & 30.8 & 28.3 & 29.6 & 30.0 &  0.248  \cite{Boyle_ElasticGaPSb_PRB_1975} & 0.240 & 0.23 \\
            & 56.3 \cite{Boyle_ElasticGaPSb_PRB_1975} & & & & & \\
      InP  & 71.1  \cite{Lam_BulkMod_PRB_1987, Nichols_ElasticInP_SSC_1980} & 60.4 & N/A & 56.4 & 57.6 & 56.3 & 57.8 & 34.3 \cite{Nichols_ElasticInP_SSC_1980}  & 33.6 & 29.7 & 31.6 & N/A & 0.292 \cite{Nichols_ElasticInP_SSC_1980} & 0.277 & N/A \\
      InAs & 60.0  \cite{Lam_BulkMod_PRB_1987} & 50.1 & 49.2 & 45.7 & 46.6 & 45.4 & 46.9 & 29.5 \cite{Semiconductors_BasicData_Springer, Gerlich_ElasticAlSb_JAP_1963} & 27.3 & 24.2 & 25.7 & 25.1 & 0.282 \cite{Semiconductors_BasicData_Springer, Gerlich_ElasticAlSb_JAP_1963} & 0.281 & 0.28 \\
            & 57.9 \cite{Semiconductors_BasicData_Springer, Gerlich_ElasticAlSb_JAP_1963} & & & & & \\
      InSb & 47.3  \cite{Lam_BulkMod_PRB_1987, DeVaux_ElasticInSb_PR_1956} & 38.1 & N/A & 34.3 & 35.0 & 34.1 & 35.2 & 22.1 \cite{DeVaux_ElasticInSb_PR_1956} & 21.3 & 19.0 & 20.1 & N/A & 0.298 \cite{DeVaux_ElasticInSb_PR_1956} & 0.275 & N/A \\
            & 48.3  \cite{Semiconductors_BasicData_Springer, Slutsky_ElasticInSb_PR_1959} & & & & & & & 23.7 \cite{Semiconductors_BasicData_Springer, Slutsky_ElasticInSb_PR_1959} & & & & & 0.289  \cite{Semiconductors_BasicData_Springer, Slutsky_ElasticInSb_PR_1959} & \\
            & 46.5 \cite{Vanderborgh_ElasticInSb_PRB_1990} & & & & & \\
      ZnS & 77.1  \cite{Lam_BulkMod_PRB_1987} & 71.2 & 68.3 & 65.8 & 66.1 & 65.2 & 66.6 & 30.9 \cite{Semiconductors_BasicData_Springer} & 36.5 & 31.4 & 33.9 & 33.2 & 0.318 \cite{Semiconductors_BasicData_Springer} & 0.294 & 0.29 \\
            & 74.5 \cite{Semiconductors_BasicData_Springer} & & & & & \\
      ZnSe & 62.4  \cite{Lam_BulkMod_PRB_1987, Lee_ElasticZnSeTe_JAP_1970} & 58.2 & 58.3 & 53.3 & 53.8 & 52.8 & 54.1 & 29.1 \cite{Lee_ElasticZnSeTe_JAP_1970} & 29.5 & 25.6 & 27.5 & 27.5 & 0.298 \cite{Lee_ElasticZnSeTe_JAP_1970} & 0.296 & 0.3\\
      ZnTe & 51.0  \cite{Lam_BulkMod_PRB_1987, Lee_ElasticZnSeTe_JAP_1970} & 43.8 & 46.0 & 39.9 & 40.5 & 39.4 & 40.7 & 23.4 \cite{Lee_ElasticZnSeTe_JAP_1970} & 23.3 & 20.8 & 22.1 & 22.4 & 0.30 \cite{Lee_ElasticZnSeTe_JAP_1970} & 0.284 & 0.29 \\
      CdSe & 53.0  \cite{Lam_BulkMod_PRB_1987} & 46.7 & 44.8 & 41.5 & 42.1 & 41.1 & 42.3 & N/A & 16.2 & 13.1 & 14.7 & 15.3 & N/A & 0.358 & 0.35 \\
      CdTe & 42.4  \cite{Lam_BulkMod_PRB_1987} & 36.4 & 35.3 & 32.2 & 32.7 & 31.9 & 32.8 & N/A & 14.2 & 11.9 & 13.0 & 13.6 & N/A & 0.340 & 0.33 \\
      HgSe & 50.0  \cite{Lam_BulkMod_PRB_1987} & 43.8 & 41.2 & 39.0 & 39.7 & 38.5 & 39.9 & 14.8 \cite{Lehoczky_ElasticHgSe_PR_1969} & 15.6 & 11.9 & 13.7 & 13.3 & 0.361  \cite{Lehoczky_ElasticHgSe_PR_1969} & 0.358 & 0.35 \\
            & 48.5 \cite{Lehoczky_ElasticHgSe_PR_1969} & & & & & \\
      HgTe & 42.3  \cite{Lam_BulkMod_PRB_1987, Semiconductors_BasicData_Springer, Cottam_ElasticHgTe_JPCS_1975} & 35.3 & N/A & 31.0 & 31.6 & 30.8 & 31.9  & 14.7 \cite{Semiconductors_BasicData_Springer, Cottam_ElasticHgTe_JPCS_1975} & 14.4 & 11.6 & 13.0 & N/A & 0.344 \cite{Semiconductors_BasicData_Springer, Cottam_ElasticHgTe_JPCS_1975} & 0.335 & N/A \\
      \hline
    \end{tabular}
  }
\end{table*}

For the shear modulus, the experimental values $G^\EXP$  are compared to the \AEL\ values $G_\sVoigt^\sAEL$,
$G_\sReuss^\sAEL$ and $G_\sVRH^\sAEL$. As can be seen from the values in
Table \ref{tab:zincblende_elastic} and Fig. \ref{fig:zincblende_thermal_elastic}(b), the agreement with the experimental values is generally
good with a very low \RMSrD\ of 0.111 for $G_\sVRH^\sAEL$, with the Voigt approximation tending to overestimate and the Reuss approximation tending to underestimate, as would be
expected. The experimental values of the Poisson ratio $\sigma^\EXP$ and the \AEL\ values $\sigma^\sAEL$ (Eq.\ (\ref{Poissonratio})) are
also shown in Table \ref{tab:zincblende_elastic} and Fig. \ref{fig:zincblende_thermal_elastic}(c), and the values are generally in good
agreement. The Pearson (i.e. linear, Eq.\ (\ref{Pearson})) and Spearman (i.e. rank order, Eq.\ (\ref{Spearman})) correlations between all of
the \AEL-\AGL\ elastic property values and experiment are shown in Table \ref{tab:zincblende_correlation}, and are generally
very high for all of these properties, ranging from 0.977 and 0.982 respectively for $\sigma^\EXP$ vs. $\sigma^\sAEL$, up to 0.999
and 0.992 for $B^\EXP$ vs. $B_\sVRH^\sAEL$. These very high correlation values demonstrate the validity of using the \AEL-\AGL\
methodology to predict the elastic and mechanical properties of
materials.

The Materials Project values of $B_\sVRH^\sMP$, $G_\sVRH^\sMP$ and $\sigma^\sMP$ for diamond and zincblende structure materials are also shown in 
Table \ref{tab:zincblende_elastic}, where available. The Pearson correlations values for the experimental results with the available values of 
$B_\sVRH^\sMP$, $G_\sVRH^\sMP$ and $\sigma^\sMP$ were calculated to be 0.995, 0.987 and 0.952, respectively, while the respective Spearman correlations
were 0.963, 0.977 and 0.977, and the \RMSrD\ values were 0.149, 0.116 and 0.126. For comparison, the corresponding Pearson correlations for the same 
subset of materials for $B_\sVRH^\sAEL$, $G_\sVRH^\sAEL$ and $\sigma^\sAEL$ are 0.997, 0.987, and 0.957 respectively,  while the respective Spearman correlations
were 0.982, 0.977 and 0.977, and the \RMSrD\ values were 0.129, 0.114 and 0.108. These correlation values are very similar, and the general close agreement
{for $B_\sVRH^\sAEL$, $G_\sVRH^\sAEL$ and $\sigma^\sAEL$ with $B_\sVRH^\sMP$, $G_\sVRH^\sMP$ and $\sigma^\sMP$} 
demonstrate that the small differences in the parameters used for the DFT calculations make little difference to the results, 
indicating that the parameter set used here is robust for high-throughput calculations.

The thermal {properties Debye} temperature, Gr{\"u}neisen parameter and thermal conductivity calculated using \AGL\ for this set of materials are
compared to the experimental values  taken from the literature in Table \ref{tab:zincblende_thermal} and are also plotted in Fig. \ref{fig:zincblende_thermal_elastic}.
For the Debye temperature, the experimental values $\theta^\EXP$ are compared {to
$\theta_\sDebye^\sAGL$, $\theta_\sDebye^\sBM$, $\theta_\sDebye^\sVIN$ and $\theta_\sDebye^\sBCN$} in Fig. \ref{fig:zincblende_thermal_elastic}(e), while {the values} for
the empirical equations of state are provided in the supplementary information.
Note that the $\theta^\EXP$ values taken from Ref. \onlinecite{slack} and
Ref. \onlinecite{Morelli_Slack_2006} are for $\theta_\acoustic$, and generally are in good agreement with the $\theta_\acoustic^\sAGL$ values. The
values obtained using the numerical $E(V)$ fit and the three different equations of state are also in good agreement with each other, whereas
the values of $\theta_\sDebye^\sAGL$ calculated using different $\sigma$ values differ significantly, indicating that for this property the value
of $\sigma$ used is far more important than the equation of state used.  The correlation
between $\theta^\EXP$ and the various \AGL\ values is also very high,
of the order of 0.999, and the \RMSrD\ is low, of the order of 0.13.

\begin{table*}[t!]
  \caption{\small Thermal properties lattice thermal conductivity at
    300K, Debye temperature and Gr{\"u}neisen parameter of 
    zincblende (\AFLOW\ prototype: {\sf AB\_cF8\_216\_c\_a} \cite{curtarolo:art121}) and diamond ({\sf A\_cF8\_227\_a} \cite{curtarolo:art121})
    structure semiconductors, comparing the effect of using the
    calculated value of the Poisson ratio to the previous approximation of
    $\sigma = 0.25$. The values  listed for $\theta^{\mathrm{exp}}$ are
    $\theta_\acoustic$, except 141K for HgTe which is $\theta_{\mathrm D}$ \cite{Snyder_jmatchem_2011}.
    Units: $\kappa$ in \WmK, $\theta$ in \K.
  }
  \label{tab:zincblende_thermal}
  {\footnotesize
    \begin{tabular}{c c c c c c c c c c c}
      \hline
      Comp. & $\kappa^\EXP$  & $\kappa^\sAGL $ & $\kappa^\sAGL$ & $\theta^\EXP$  & $\theta_\sDebye^\sAGL$ & $\theta_\sDebye^\sAGL$ & $\theta_\sDebye^\sAEL$ & $\gamma^\EXP$ & $\gamma^\sAGL$ & $\gamma^\sAGL$  \\
            &  & & & &  ($\theta_\acoustic^\sAGL$) & ($\theta_\acoustic^\sAGL$) & \\
            & & ($\sigma = 0.25$)\cite{curtarolo:art96} & & & ($\sigma = 0.25$) \cite{curtarolo:art96} & & & & ($\sigma = 0.25$)\cite{curtarolo:art96} &   \\
      \hline
      C &  3000 \cite{Morelli_Slack_2006} & 169.1 & 419.9  & 1450 \cite{slack, Morelli_Slack_2006} & 1536 & 2094 & 2222 & 0.75 \cite{Morelli_Slack_2006} & 1.74 & 1.77 \\
            & & & & & (1219) & (1662) & & 0.9 \cite{slack} & & \\
      SiC & 360 \cite{Ioffe_Inst_DB} & 67.19 & 113.0 & 740 \cite{slack} & 928 & 1106 & 1143 & 0.76 \cite{slack} & 1.84 & 1.85	\\
            & & & & & (737) & (878) & & & & \\
      Si & 166 \cite{Morelli_Slack_2006} & 20.58 & 26.19 & 395 \cite{slack, Morelli_Slack_2006} & 568 & 610 & 624 & 1.06 \cite{Morelli_Slack_2006} & 2.09 & 2.06	 \\
            & & & & & (451) & (484) &  & 0.56 \cite{slack} &  & \\
      Ge &  65 \cite{Morelli_Slack_2006} &  6.44 & 8.74 & 235 \cite{slack, Morelli_Slack_2006} & 296 & 329 & 342 & 1.06 \cite{Morelli_Slack_2006} & 2.3 & 2.31 	 \\
            & & & & & (235) & (261) &  & 0.76 \cite{slack} & &   \\
      BN & 760 \cite{Morelli_Slack_2006} & 138.4 & 281.6 & 1200 \cite{Morelli_Slack_2006} & 1409 & 1793 & 1887 & 0.7 \cite{Morelli_Slack_2006} & 1.73 & 1.75	\\
            & & & & & (1118) & (1423) & & & & \\
      BP & 350 \cite{Morelli_Slack_2006} & 52.56 & 105.0 & 670 \cite{slack, Morelli_Slack_2006} & 811 & 1025 & 1062 & 0.75 \cite{Morelli_Slack_2006} & 1.78 & 1.79	\\
            & & & & & (644) & (814) & & & & \\
      AlP & 90 \cite{Landolt-Bornstein, Spitzer_JPCS_1970} & 21.16 & 19.34 & 381 \cite{Morelli_Slack_2006} & 542 & 525 & 531 & 0.75 \cite{Morelli_Slack_2006} & 1.96 & 1.96	 \\
            & & & & & (430) & (417) & & & & \\
      AlAs & 98 \cite{Morelli_Slack_2006} &  12.03 & 11.64 & 270 \cite{slack, Morelli_Slack_2006} & 378 & 373 & 377 & 0.66 \cite{slack, Morelli_Slack_2006} & 2.04 & 2.04	 \\
            & & & & &  (300) & (296) & & & & \\
      AlSb & 56 \cite{Morelli_Slack_2006} & 7.22 & 6.83 & 210 \cite{slack, Morelli_Slack_2006} & 281 & 276 & 277 & 0.6 \cite{slack, Morelli_Slack_2006} & 2.12 & 2.13 	 \\
            & & & & & (223) & (219) & & & & \\
      GaP & 100  \cite{Morelli_Slack_2006} & 11.76 & 13.34 & 275 \cite{slack, Morelli_Slack_2006} & 396 & 412 & 423 & 0.75 \cite{Morelli_Slack_2006} & 2.15 & 2.15 	\\
            & & & &  & (314) & (327) & &  0.76 \cite{slack} &  &  \\
      GaAs & 45 \cite{Morelli_Slack_2006} & 7.2 & 8.0 & 220 \cite{slack, Morelli_Slack_2006} & 302 & 313	& 322 & 0.75 \cite{slack, Morelli_Slack_2006} & 2.23 & 2.24 \\
            & & & & & (240) & (248) & & & & \\
      GaSb & 40 \cite{Morelli_Slack_2006} & 4.62 & 4.96 & 165 \cite{slack, Morelli_Slack_2006} & 234 & 240 & 248 & 0.75 \cite{slack, Morelli_Slack_2006} & 2.27 & 2.28 	 \\
            & & & & & (186) & (190) & & & & \\
      InP & 93 \cite{Morelli_Slack_2006} & 7.78 & 6.53 & 220 \cite{slack, Morelli_Slack_2006} & 304 & 286 & 287 & 0.6 \cite{slack, Morelli_Slack_2006} & 2.22 & 2.21 	 \\
            & & & & & (241) & (227) & & & & \\
      InAs & 30 \cite{Morelli_Slack_2006} & 5.36 & 4.33 & 165 \cite{slack, Morelli_Slack_2006} &  246 & 229 & 231 & 0.57 \cite{slack, Morelli_Slack_2006} & 2.26 & 2.26	 \\
            & & & & & (195) & (182) & & & & \\
      InSb & 20 \cite{Morelli_Slack_2006} & 3.64 & 3.02 & 135 \cite{slack, Morelli_Slack_2006} & 199 & 187 & 190 & 0.56 \cite{slack, Morelli_Slack_2006} & 2.3 & 2.3 	 \\
            & 16.5 \cite{Snyder_jmatchem_2011} & & & & (158) & (148) &  & & &  \\
      ZnS & 27 \cite{Morelli_Slack_2006} & 11.33 & 8.38 & 230 \cite{slack, Morelli_Slack_2006} & 379 & 341 & 346 & 0.75 \cite{slack, Morelli_Slack_2006} & 2.01 & 2.00 	 \\
            & & & & & (301) & (271) & & & & \\
      ZnSe & 19 \cite{Morelli_Slack_2006} & 7.46 & 5.44 & 190 \cite{slack, Morelli_Slack_2006} & 290 & 260	& 263 & 0.75 \cite{slack, Morelli_Slack_2006} & 2.07 & 2.06 	\\
            & 33  \cite{Snyder_jmatchem_2011} & & &  & (230) & (206) & & & &  \\
      ZnTe & 18 \cite{Morelli_Slack_2006} &  4.87 & 3.83 & 155 \cite{slack, Morelli_Slack_2006} & 228 & 210 & 212 & 0.97 \cite{slack, Morelli_Slack_2006} & 2.14 & 2.13  \\
            & & & & & (181) & (167) & & & & \\
      CdSe & 4.4 \cite{Snyder_jmatchem_2011} & 4.99 & 2.04 & 130 \cite{Morelli_Slack_2006} & 234 & 173 & 174 & 0.6 \cite{Morelli_Slack_2006} & 2.19 & 2.18 \\
            & & & & & (186) & (137) & & & & \\
      CdTe & 7.5 \cite{Morelli_Slack_2006} & 3.49 & 1.71 & 120 \cite{slack, Morelli_Slack_2006} & 191 & 150 & 152 & 0.52 \cite{slack, Morelli_Slack_2006} & 2.23 & 2.22	 \\
            & & & & & (152) & (119) & & & & \\
      HgSe & 3 \cite{Whitsett_PRB_1973} & 3.22 & 1.32 & 110 \cite{slack} & 190 & 140	& 140 & 0.17 \cite{slack} & 2.4 & 2.38	 \\
            & & & & &  (151) & (111) & & & & \\
      HgTe & 2.5 \cite{Snyder_jmatchem_2011}  & 2.36 & 1.21 & 141 \cite{Snyder_jmatchem_2011}  & 162 & 129 & 130 & 1.9 \cite{Snyder_jmatchem_2011}  & 2.46 & 2.45 \\
            & & & & (100) \cite{slack} & (129) & (102) & & & & \\

      \hline
    \end{tabular}
  }
\end{table*}

The experimental values $\gamma^\EXP$ of the Gr{\"u}neisen parameter are plotted {against 
$\gamma^\sAGL$, $\gamma^\sBM$, $\gamma^\sVIN$ and $\gamma^\sBCN$} in Fig. \ref{fig:zincblende_thermal_elastic}(f), and the values
are listed in Table \ref{tab:zincblende_thermal} and in the supplementary information.
The very high \RMSrD\ values (see Table \ref{tab:zincblende_correlation}) show that \AGL\ has problems accurately predicting
the Gr{\"u}neisen parameter for this set of materials, as the calculated value is often 2 to 3 times larger than the experimental one.
Note also that there are quite large differences between the values obtained for different equations of state, with $\gamma^\sBCN$ generally
having the lowest values while $\gamma^\sVIN$ has the highest values.
On the other hand, in contrast to the case of $\theta_\sDebye^\sAGL$, the value of $\sigma$ used makes little difference to the value
of $\gamma^\sAGL$. The {correlations} between $\gamma^\EXP$ and the \AGL\ values, as shown in Table \ref{tab:zincblende_correlation},
are also quite poor, with no value higher than 0.2 for the Pearson correlations, and {negative Spearman} correlations.

The experimental thermal conductivity $\kappa^\EXP$ is compared in Fig. \ref{fig:zincblende_thermal_elastic}(d) to the
thermal conductivities calculated with \AGL\ using the Leibfried-Schl{\"o}mann equation (Eq.\ (\ref{thermal_conductivity})): $\kappa^\sAGL$, $\kappa^\sBM$, $\kappa^\sVIN$ and $\kappa^\sBCN$,
while the values are listed in  Table \ref{tab:zincblende_thermal} and in the supplementary information.
The absolute agreement between the \AGL\ values and $\kappa^\EXP$ is quite poor, with \RMSrD\ values of the order of 0.8 and discrepancies of tens, or even hundreds, of percent
quite common. Considerable disagreements also exist between different experimental reports of these properties, in
almost all cases where they exist. Unfortunately, the scarcity of experimental data from different sources on the thermal properties of these materials
prevents reaching definite conclusions regarding the true values of these properties. The available data can thus only be considered as a rough indication
of their order of magnitude.

{The Pearson} correlations between the \AGL\ calculated thermal conductivity values and the experimental
values are high, ranging from $0.871$ to $0.932$, while the Spearman correlations are even higher, ranging from $0.905$
to $0.954$, as shown in Table \ref{tab:zincblende_correlation}. In particular, note that using the $\sigma^\sAEL$ in the \AGL\ calculations
improves the correlations by about 5\%, from $0.878$ to $0.927$ and from $0.905$ to $0.954$. For the different equations of state,
$\kappa^\sAGL$ and $\kappa^\sBCN$ appear to correlate better with $\kappa^\EXP$ than $\kappa^\sBM$ and $\kappa^\sVIN$ for this set of
materials.

\begin{table}[t!]
  \caption{\small Correlations and deviations between experimental values and \AEL\ and \AGL\ results for
    elastic and thermal properties for zincblende and diamond structure semiconductors.
  }
  \label{tab:zincblende_correlation}
  {\footnotesize
    \begin{tabular}{l r r r}
      \hline
      Property  & Pearson & Spearman & \RMSrD\ \\
                & (Linear) & (Rank Order) \\
      \hline
      $\kappa^\EXP$ vs. $\kappa^\sAGL$  ($\sigma = 0.25$) \cite{curtarolo:art96} & 0.878 & 0.905 & 0.776 \\
      $\kappa^\EXP$ vs. $\kappa^\sAGL$ & 0.927 & 0.95 & 0.796 \\
      $\kappa^\EXP$ vs. $\kappa^\sBM$ & 0.871 & 0.954 & 0.787  \\
      $\kappa^\EXP$ vs. $\kappa^\sVIN$ & 0.908 &  0.954 & 0.815 \\
      $\kappa^\EXP$ vs. $\kappa^\sBCN$ & 0.932 & 0.954 & 0.771 \\
      $\theta_\acoustic^\EXP$ vs. $\theta_\acoustic^\sAGL$  ($\sigma = 0.25$) \cite{curtarolo:art96} & 0.995 & 0.984 & 0.200 \\
      $\theta_\acoustic^\EXP$ vs. $\theta_\acoustic^\sAGL$ & 0.999 & 0.998 & 0.132 \\
      $\theta_\acoustic^\EXP$ vs. $\theta_\acoustic^\sBM$ & 0.999 & 0.998  & 0.132 \\
      $\theta_\acoustic^\EXP$ vs. $\theta_\acoustic^\sVIN$ & 0.999 &  0.998 & 0.127 \\
      $\theta_\acoustic^\EXP$ vs. $\theta_\acoustic^\sBCN$ & 0.999 & 0.998 & 0.136 \\
      $\gamma^\EXP$ vs. $\gamma^\sAGL$  ($\sigma = 0.25$) \cite{curtarolo:art96} & 0.137 & -0.187 & 3.51 \\
      $\gamma^\EXP$ vs. $\gamma^\sAGL$ & 0.145 & -0.165 & 3.49 \\
      $\gamma^\EXP$ vs. $\gamma^\sBM$ & 0.169 & -0.178 & 3.41  \\
      $\gamma^\EXP$ vs. $\gamma^\sVIN$ & 0.171 &  -0.234  & 3.63 \\
      $\gamma^\EXP$ vs. $\gamma^\sBCN$ & 0.144 & -0.207 & 3.32 \\
      $B^\EXP$ vs. $B_\sVRH^\sAEL$ & 0.999 & 0.992  & 0.130 \\
      $B^\EXP$ vs. $B_\sStatic^\sAGL$ & 0.999 & 0.986 & 0.201 \\
      $B^\EXP$ vs. $B_\sStatic^\sBM$ & 0.999 & 0.986 & 0.189 \\
      $B^\EXP$ vs. $B_\sStatic^\sVIN$ & 0.999 & 0.986 & 0.205 \\
      $B^\EXP$ vs. $B_\sStatic^\sBCN$ & 0.999 & 0.986 & 0.185 \\
      $G^\EXP$ vs. $G_\sVRH^\sAEL$ & 0.998 & 0.980 & 0.111  \\
      $G^\EXP$ vs. $G_\sVoigt^\sAEL$ & 0.998 & 0.980 & 0.093 \\
      $G^\EXP$ vs. $G_\sReuss^\sAEL$ & 0.998 & 0.980 & 0.152 \\
      $\sigma^\EXP$ vs. $\sigma^\sAEL$ & 0.977 & 0.982 & 0.095 \\
      \hline
    \end{tabular}
  }
\end{table}

As we noted in our previous work on \AGL\ \cite{curtarolo:art96}, some of the inaccuracy in the thermal conductivity results may be due to the inability of the Leibfried-Schl{\"o}mann equation to fully
describe effects such as the suppression of phonon-phonon scattering due to large gaps between the branches of
the phonon dispersion \cite{Lindsay_PRL_2013}. This can be seen from the thermal conductivity values shown in Table 2.2 of Ref. \onlinecite{Morelli_Slack_2006}
calculated using the experimental values of $\theta_\acoustic$ and $\gamma$ in the Leibfried-Schl{\"o}mann equation. There are large discrepancies in certain cases such as diamond,
while the Pearson and Spearman correlations of $0.932$ and $0.941$ respectively are very similar to the correlations we calculated using the \AGL\ evaluations of
$\theta_\acoustic$ and $\gamma$.

Thus, the unsatisfactory quantitative reproduction of these quantities by the Debye quasi-harmonic model
has little impact on its effectiveness as a screening tool for identifying high or
low thermal conductivity materials. The model can be used when these
experimental values are unavailable to help determine the relative values of these quantities and for
ranking {materials conductivity}.

\subsection{Rocksalt structure materials}

The mechanical and thermal properties were calculated for a set of materials with the rocksalt structure
(spacegroup: Fm$\overline{3}$m,\ $\#$225; Pearson symbol: cF8; 
\AFLOW\ Prototype: {\sf AB\_cF8\_225\_a\_b} \cite{curtarolo:art121}). This {is the same set of materials
as} in Table II of Ref. \onlinecite{curtarolo:art96}, which in turn {are from} the
sets in Table III of Ref. \onlinecite{slack} and Table 2.1 of Ref. \onlinecite{Morelli_Slack_2006}.

\begin{figure*}[t!]
  \includegraphics[width=0.98\textwidth]{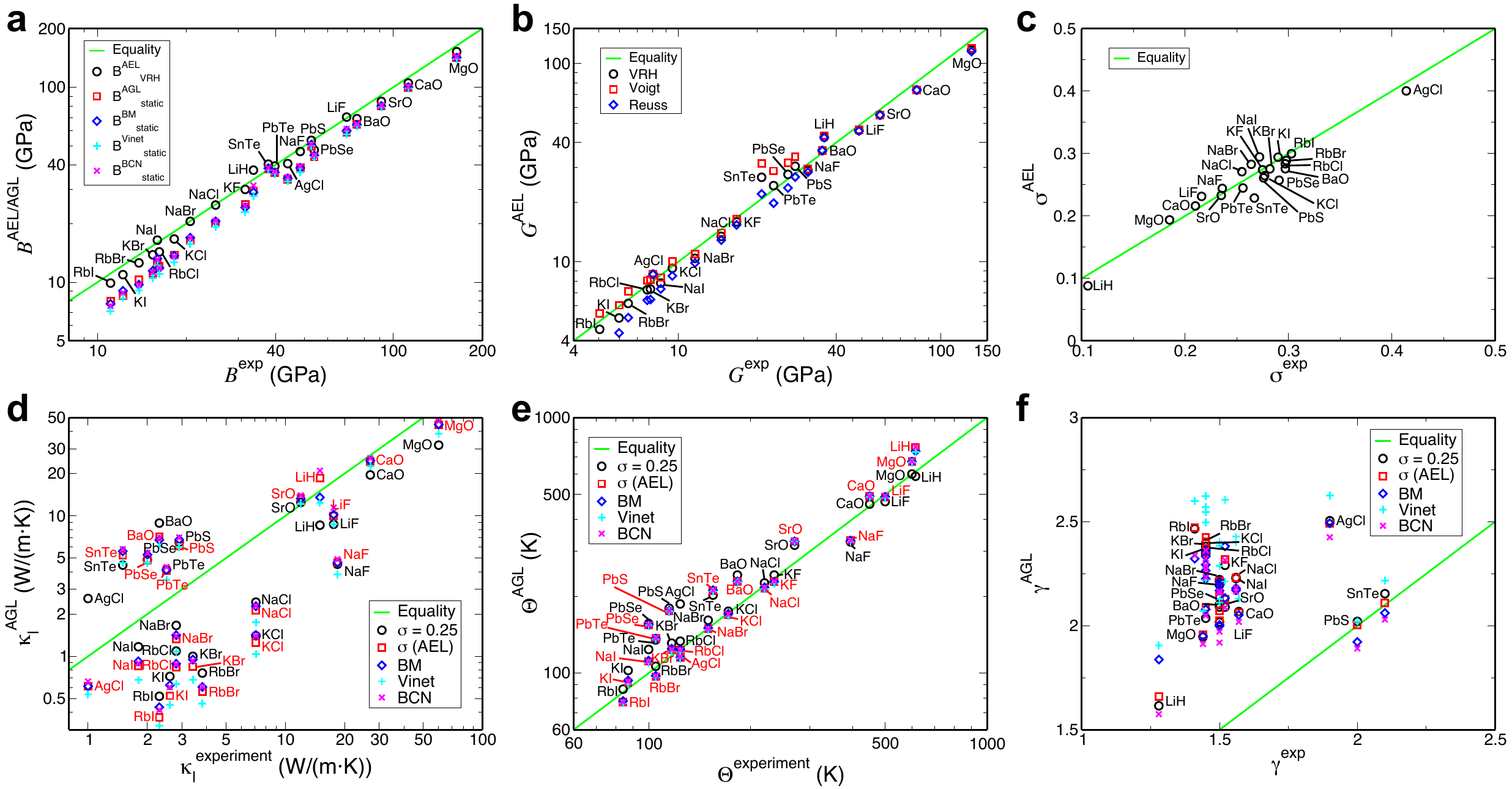}
  \vspace{-4mm}
  \caption{\small 
    {\bf (a)} Bulk modulus,
    {\bf (b)} shear modulus,
    {\bf (c)} Poisson ratio,
    {\bf (d)} lattice thermal conductivity at 300K,
    {\bf (e)} Debye temperature and
    {\bf (f)} Gr{\"u}neisen parameter of rocksalt structure (\AFLOW\ Prototype: {\sf AB\_cF8\_225\_a\_b} \cite{curtarolo:art121})
    semiconductors.
    The Debye temperatures plotted in {\bf (b)} are
    $\theta_\acoustic$, except for SnTe where $\theta_{\mathrm D}$ is
    quoted in Ref. \onlinecite{Snyder_jmatchem_2011}. }
  \label{fig:rocksalt_thermal_elastic}
\end{figure*}

The elastic properties of bulk modulus, shear modulus and Poisson ratio, as calculated using \AEL\ and \AGL\ are shown
in Table \ref{tab:rocksalt_elastic} and Fig. \ref{fig:rocksalt_thermal_elastic}, together
with experimental values from the literature where available. As can be seen
from the results in Table \ref{tab:rocksalt_elastic} and Fig. \ref{fig:rocksalt_thermal_elastic}(a), for this set of materials the
$B_\sVRH^\sAEL$ values are closest to experiment, with an \RMSrD\ of 0.078. The \AGL\ values from both the numerical
fit and the empirical equations of state are generally very similar to each other, while being slightly less than the $B_\sVRH^\sAEL$
values.

\begin{table*}[t!]
  \caption{\small Mechanical properties bulk modulus, shear modulus
    and Poisson ratio of 
    rocksalt (\AFLOW\ Prototype: {\sf AB\_cF8\_225\_a\_b} \cite{curtarolo:art121})
    structure semiconductors.
    ``N/A'' = Not available for that source.
    Units: $B$ and $G$ in \GPa.
  }
  \label{tab:rocksalt_elastic}
  \begin{tabular}{c c c c c c c c c c c c c c c c}
    \hline
    Comp. & $B^\EXP$  & $B_\sVRH^\sAEL$ & $B_\sVRH^\sMP$ & $B_\sStatic^\sAGL$ & $B_\sStatic^\sBM$ & $B_\sStatic^\sVIN$ &  $B_\sStatic^\sBCN$  & $G^\EXP$ & $G_\sVoigt^\sAEL$ & $G_\sReuss^\sAEL$ &  $G_\sVRH^\sAEL$ & $G_\sVRH^\sMP$ & $\sigma^\EXP$ & $\sigma^\sAEL$ & $\sigma^\sMP$    \\
    \hline
    LiH & 33.7 \cite{Laplaze_ElasticLiH_SSC_1976} & 37.7 & 36.1 & 29.5 & 29.0 & 27.7 & 31.4 & 36.0 \cite{Laplaze_ElasticLiH_SSC_1976} & 43.4 & 42.3 & 42.8 & 42.9 & 0.106 \cite{Laplaze_ElasticLiH_SSC_1976} & 0.088 & 0.07 \\
    LiF & 69.6 \cite{Haussuhl_ElasticRocksalt_ZP_1960} & 70.4 & 69.9 & 58.6 & 59.9 & 57.5 & 61.2 & 48.8 \cite{Haussuhl_ElasticRocksalt_ZP_1960} & 46.4 & 45.8 & 46.1 & 50.9 & 0.216 \cite{Haussuhl_ElasticRocksalt_ZP_1960} & 0.231 & 0.21 \\
    NaF & 48.5 \cite{Haussuhl_ElasticRocksalt_ZP_1960} & 46.9 & 47.6 & 38.7 & 38.6 & 36.8 & 39.3 & 31.2 \cite{Haussuhl_ElasticRocksalt_ZP_1960} & 29.5 & 28.4 & 28.9 & 30.0 & 0.236 \cite{Haussuhl_ElasticRocksalt_ZP_1960} & 0.244 & 0.24 \\
    NaCl & 25.1 \cite{Haussuhl_ElasticRocksalt_ZP_1960} & 24.9 & 22.6 & 20.0 & 20.5 & 19.2 & 20.7 & 14.6 \cite{Haussuhl_ElasticRocksalt_ZP_1960} & 14.0 & 12.9 & 13.5 & 14.3 & 0.255 \cite{Haussuhl_ElasticRocksalt_ZP_1960} & 0.271 & 0.24 \\
    NaBr & 20.6 \cite{Haussuhl_ElasticRocksalt_ZP_1960} & 20.5 & 27.1 & 16.3 & 16.9 & 15.7 & 16.9 & 11.6 \cite{Haussuhl_ElasticRocksalt_ZP_1960} & 11.0 & 9.9 & 10.4 & 11.6 & 0.264 \cite{Haussuhl_ElasticRocksalt_ZP_1960} & 0.283 & 0.31 \\
    NaI & 15.95 \cite{Haussuhl_ElasticRocksalt_ZP_1960} & 16.4 & 15.8 & 12.6 & 13.2 & 12.2 & 13.1 & 8.59 \cite{Haussuhl_ElasticRocksalt_ZP_1960} & 8.35 & 7.31 & 7.83 & 8.47 & 0.272 \cite{Haussuhl_ElasticRocksalt_ZP_1960} & 0.295 & 0.27 \\
    KF & 31.6 \cite{Haussuhl_ElasticRocksalt_ZP_1960} & 29.9 & 28.9 & 25.1 & 24.2 & 22.9  & 24.7 & 16.7 \cite{Haussuhl_ElasticRocksalt_ZP_1960} & 16.5 & 15.4 & 15.9 & 16.5 & 0.275 \cite{Haussuhl_ElasticRocksalt_ZP_1960} & 0.274 & 0.26 \\
    KCl & 18.2 \cite{Haussuhl_ElasticRocksalt_ZP_1960} & 16.7 & 15.8 & 13.8 & 13.7 & 12.7  & 13.6 & 9.51 \cite{Haussuhl_ElasticRocksalt_ZP_1960} & 10.1 & 8.51 & 9.30 & 9.24 & 0.277 \cite{Haussuhl_ElasticRocksalt_ZP_1960} & 0.265 & 0.26 \\
    KBr & 15.4 \cite{Haussuhl_ElasticRocksalt_ZP_1960} & 13.8 & 21.6 & 11.1 & 11.4 & 10.5 & 11.2 & 7.85 \cite{Haussuhl_ElasticRocksalt_ZP_1960} & 8.14 & 6.46 & 7.30 & 7.33 & 0.282 \cite{Haussuhl_ElasticRocksalt_ZP_1960} & 0.276 & 0.35 \\
    KI & 12.2 \cite{Haussuhl_ElasticRocksalt_ZP_1960} & 10.9 & 9.52 & 8.54 & 9.03 &  8.28 & 8.84 & 5.96 \cite{Haussuhl_ElasticRocksalt_ZP_1960} & 6.05 & 4.39 & 5.22 & 5.55 & 0.290 \cite{Haussuhl_ElasticRocksalt_ZP_1960} & 0.294 & 0.26 \\
    RbCl & 16.2 \cite{Haussuhl_ElasticRocksalt_ZP_1960} & 14.3 & 14.6 & 12.1 & 11.8 & 11.0 & 11.8 & 7.63 \cite{Haussuhl_ElasticRocksalt_ZP_1960} & 8.06 & 6.41 & 7.24 & 7.67 & 0.297 \cite{Haussuhl_ElasticRocksalt_ZP_1960} & 0.284 & 0.28 \\
    RbBr & 13.8 \cite{Haussuhl_ElasticRocksalt_ZP_1960} & 12.6 & 13.8 & 10.3 & 9.72 & 9.06 & 9.67 & 6.46 \cite{Haussuhl_ElasticRocksalt_ZP_1960} & 7.12 & 5.24 & 6.18 & 6.46 & 0.298 \cite{Haussuhl_ElasticRocksalt_ZP_1960} & 0.289 & 0.3 \\
    RbI & 11.1 \cite{Haussuhl_ElasticRocksalt_ZP_1960} & 9.90 & 9.66 & 8.01 & 7.74 & 7.12 & 7.54 & 5.03 \cite{Haussuhl_ElasticRocksalt_ZP_1960} & 5.50 & 3.65 & 4.57 & 4.63 & 0.303 \cite{Haussuhl_ElasticRocksalt_ZP_1960} & 0.300 & 0.29 \\
    AgCl & 44.0 \cite{Hughes_ElasticAgCl_PRB_1996} & 40.6 & N/A & 33.7 & 34.1 & 33.0 & 34.7 & 8.03 \cite{Hughes_ElasticAgCl_PRB_1996} & 8.68 & 8.66 & 8.67 & N/A & 0.414 \cite{Hughes_ElasticAgCl_PRB_1996} & 0.400 & N/A \\
    MgO  & 164 \cite{Sumino_ElasticMgO_JPE_1976} & 152 & 152 & 142 & 142 & 140 & 144 & 131 \cite{Sumino_ElasticMgO_JPE_1976}  & 119 & 115 & 117 & 119 & 0.185 \cite{Sumino_ElasticMgO_JPE_1976} & 0.194 & 0.19 \\
    CaO & 113 \cite{Chang_ElasticCaSrBaO_JPCS_1977} & 105 & 105 & 99.6 & 100 & 98.7 & 101 & 81.0 \cite{Chang_ElasticCaSrBaO_JPCS_1977} & 73.7 & 73.7 & 73.7 & 74.2 & 0.210 \cite{Chang_ElasticCaSrBaO_JPCS_1977} & 0.216 & 0.21 \\
    SrO & 91.2 \cite{Chang_ElasticCaSrBaO_JPCS_1977} & 84.7 & 87.4 &80.0 & 80.2 & 79.1 & 80.8 & 58.7 \cite{Chang_ElasticCaSrBaO_JPCS_1977} & 55.1 & 55.0 & 55.1 & 56.0 &  0.235 \cite{Chang_ElasticCaSrBaO_JPCS_1977} & 0.233 & 0.24 \\
    BaO & 75.4 \cite{Chang_ElasticCaSrBaO_JPCS_1977} & 69.1 & 68.4 & 64.6 & 64.3 & 63.0 & 64.6 & 35.4 \cite{Chang_ElasticCaSrBaO_JPCS_1977} & 36.4 & 36.4 & 36.4 & 37.8 & 0.297 \cite{Chang_ElasticCaSrBaO_JPCS_1977} & 0.276 & 0.27 \\
    PbS & 52.9 \cite{Semiconductors_BasicData_Springer, Peresada_ElasticPbS_PSSa_1976} & 53.5 & N/A & 49.9 & 50.8 & 50.0 & 51.0 & 27.9 \cite{Semiconductors_BasicData_Springer, Peresada_ElasticPbS_PSSa_1976} & 34.0 & 26.8 & 30.4 & N/A & 0.276 \cite{Semiconductors_BasicData_Springer, Peresada_ElasticPbS_PSSa_1976} & 0.261 & N/A \\
    PbSe & 54.1 \cite{Semiconductors_BasicData_Springer, Lippmann_ElasticPbSe_PSSa_1971} & 47.7 & N/A & 43.9 & 44.8 & 43.9 & 44.9 & 26.2 \cite{Semiconductors_BasicData_Springer, Lippmann_ElasticPbSe_PSSa_1971} & 31.7 & 23.6 & 27.6 &  N/A & 0.291 \cite{Semiconductors_BasicData_Springer, Lippmann_ElasticPbSe_PSSa_1971} & 0.257 & N/A \\
    PbTe & 39.8 \cite{Semiconductors_BasicData_Springer, Miller_ElasticPbTe_JPCSS_1981} & 39.5 & N/A & 36.4 & 36.6 & 35.8 & 36.8 & 23.1 \cite{Semiconductors_BasicData_Springer, Miller_ElasticPbTe_JPCSS_1981} & 28.7 & 19.8 & 24.3 & N/A & 0.256 \cite{Semiconductors_BasicData_Springer, Miller_ElasticPbTe_JPCSS_1981} & 0.245 & N/A \\
    SnTe & 37.8 \cite{Semiconductors_BasicData_Springer, Seddon_ElasticSnTe_SSC_1976} & 40.4 & 39.6 & 38.1 & 38.4 & 37.6 & 38.6 & 20.8 \cite{Semiconductors_BasicData_Springer, Seddon_ElasticSnTe_SSC_1976}  & 31.4 & 22.0 & 26.7 & 27.6 & 0.267 \cite{Semiconductors_BasicData_Springer, Seddon_ElasticSnTe_SSC_1976}  & 0.229 & 0.22 \\
    \hline
  \end{tabular}
\end{table*}

For the shear modulus, the experimental values $G^\EXP$  are compared to the \AEL\ values $G_\sVoigt^\sAEL$,
$G_\sReuss^\sAEL$ and $G_\sVRH^\sAEL$. As can be seen from the values in
Table \ref{tab:rocksalt_elastic} and Fig. \ref{fig:rocksalt_thermal_elastic}(b), the agreement with the experimental values is generally
good with an \RMSrD\ of 0.105 for $G_\sVRH^\sAEL$, with the Voigt approximation tending to overestimate and the Reuss approximation tending to underestimate, as would be
expected. The experimental values of the Poisson ratio $\sigma^\EXP$ and the \AEL\ values $\sigma^\sAEL$ (Eq.\ (\ref{Poissonratio})) are
also shown in Table \ref{tab:rocksalt_elastic} and Fig. \ref{fig:rocksalt_thermal_elastic}(c), and the values are generally in good
agreement. The Pearson (i.e. linear, Eq.\ (\ref{Pearson})) and Spearman (i.e. rank order, Eq.\ (\ref{Spearman})) correlations between all of
the the \AEL-\AGL\ elastic property values and experiment are shown in Table \ref{tab:rocksalt_correlation}, and are generally
very high for all of these properties, ranging from 0.959 and 0.827 respectively for $\sigma^\EXP$ vs. $\sigma^\sAEL$, up to 0.998
and 0.995 for $B^\EXP$ vs. $B_\sVRH^\sAEL$. These very high correlation values demonstrate the validity of using the \AEL-\AGL\
methodology to predict the elastic and mechanical properties of materials.

{The values} of $B_\sVRH^\sMP$, $G_\sVRH^\sMP$ and $\sigma^\sMP$ for rocksalt structure materials are also shown in 
Table \ref{tab:rocksalt_elastic}, where available. The Pearson {correlations for} the experimental results with the available values of 
$B_\sVRH^\sMP$, $G_\sVRH^\sMP$ and $\sigma^\sMP$ were calculated to be 0.997, 0.994 and 0.890, respectively, while the respective Spearman correlations
were 0.979, 0.998 and 0.817, and the \RMSrD\ values were 0.153, 0.105 and 0.126. For comparison, the corresponding Pearson correlations for the same 
subset of materials for $B_\sVRH^\sAEL$, $G_\sVRH^\sAEL$ and $\sigma^\sAEL$ are 0.998, 0.995, and 0.951 respectively,  while the respective Spearman correlations
were 0.996, 1.0 and 0.843, and the \RMSrD\ values were 0.079, 0.111 and 0.071. These correlation values are very similar, and the general close agreement
for the results for the values of $B_\sVRH^\sAEL$, $G_\sVRH^\sAEL$ and $\sigma^\sAEL$ with those of $B_\sVRH^\sMP$, $G_\sVRH^\sMP$ and $\sigma^\sMP$ 
demonstrate that the small differences in the parameters used for the DFT calculations make little difference to the results, 
indicating that the parameter set used here is robust for high-throughput calculations.

The thermal properties of Debye temperature, Gr{\"u}neisen parameter and thermal conductivity calculated using \AGL\ are
compared to the experimental values  taken from the literature in Table \ref{tab:rocksalt_thermal} and are also plotted in Fig. \ref{fig:rocksalt_thermal_elastic}.
For the Debye temperature, the experimental values $\theta^\EXP$ are compared {to
$\theta_\sDebye^\sAGL$}, $\theta_\sDebye^\sBM$, $\theta_\sDebye^\sVIN$ and $\theta_\sDebye^\sBCN$ in Fig. \ref{fig:rocksalt_thermal_elastic}(e), while the actual values for
the empirical equations of state are provided in the supplementary information.
Note that the $\theta^\EXP$ values taken from Ref. \onlinecite{slack} and
Ref. \onlinecite{Morelli_Slack_2006} are for $\theta_\acoustic$, and generally are in good agreement with the $\theta_\acoustic^\sAGL$ values. The
values obtained using the numerical $E(V)$ fit and the three different equations of state are also in good agreement with each other, whereas
the values of $\theta_\sDebye^\sAGL$ calculated using different $\sigma$ values differ significantly, indicating that, as in the case of the zincblende
and diamond structures, the value of $\sigma$ used is far more important  for this property than the equation of state used.  The correlation
between $\theta^\EXP$ and the various \AGL\ values is also quite high, of the order of 0.98 for the Pearson correlation and 0.92 for the Spearman
correlation.

\begin{table*}[t!]
  \caption{\small
    Thermal properties lattice thermal conductivity at 300K, Debye temperature and Gr{\"u}neisen parameter of rocksalt
    structure (\AFLOW\ Prototype: {\sf AB\_cF8\_225\_a\_b} \cite{curtarolo:art121}) semiconductors, comparing the effect of using the
    calculated value of the Poisson ratio to previous approximation of
    $\sigma = 0.25$.
    The values listed for $\theta^{\mathrm{exp}}$ are
    $\theta_\acoustic$, except 155K for SnTe which is $\theta_{\mathrm D}$ \cite{Snyder_jmatchem_2011}.
    ``N/A'' = Not available for that source.
    Units: $\kappa$ in \WmK, $\theta$ in \K.
  }
  \label{tab:rocksalt_thermal}
  \begin{tabular}{c c c c c c c c c c c}
    \hline
    Comp. & $\kappa^\EXP$  & $\kappa^\sAGL $ & $\kappa^\sAGL$  & $\theta^\EXP$  & $\theta_\sDebye^\sAGL$ & $\theta_\sDebye^\sAGL$ & $\theta_\sDebye^\sAEL$ & $\gamma^\EXP$ & $\gamma^\sAGL$ & $\gamma^\sAGL$ \\
          & & & & & ($\theta_\acoustic^\sAGL$) & ($\theta_\acoustic^\sAGL$) & & & &    \\
          & & ($\sigma = 0.25$) \cite{curtarolo:art96}  & & & ($\sigma = 0.25$) \cite{curtarolo:art96} & & & & ($\sigma = 0.25$) \cite{curtarolo:art96} & \\
    \hline
    LiH & 15 \cite{Morelli_Slack_2006} & 8.58 & 18.6 & 615 \cite{slack, Morelli_Slack_2006} & 743 & 962 & 1175 & 1.28 \cite{slack, Morelli_Slack_2006} & 1.62 & 1.66 \\
          & & & & & (590) & (764) & & & &  \\
    LiF & 17.6 \cite{Morelli_Slack_2006} & 8.71 & 9.96 & 500 \cite{slack, Morelli_Slack_2006} & 591 & 617 & 681 & 1.5 \cite{slack, Morelli_Slack_2006} & 2.02 & 2.03 	 \\
          & & & & & (469) &  (490) & & &  & \\
    NaF &  18.4 \cite{Morelli_Slack_2006} &  4.52 & 4.67 & 395 \cite{slack, Morelli_Slack_2006} & 411 & 416 & 455 & 1.5 \cite{slack, Morelli_Slack_2006} & 2.2 &  2.21 	  \\
          & & & & & (326) & (330) & & &  & \\
    NaCl & 7.1 \cite{Morelli_Slack_2006} & 2.43 & 2.12 & 220 \cite{slack, Morelli_Slack_2006} & 284 & 271 & 289 & 1.56 \cite{slack, Morelli_Slack_2006} & 2.23 & 2.23 	 \\
          & & & & & (225) & (215)  & & & & \\
    NaBr & 2.8 \cite{Morelli_Slack_2006} & 1.66 & 1.33 & 150 \cite{slack, Morelli_Slack_2006} & 203 & 188 & 198 & 1.5 \cite{slack, Morelli_Slack_2006} & 2.22 & 2.22 	 \\
          & & & & & (161) & (149) & & & &\\
    NaI & 1.8 \cite{Morelli_Slack_2006} & 1.17 & 0.851 & 100 \cite{slack, Morelli_Slack_2006} & 156 & 140 & 147 & 1.56 \cite{slack, Morelli_Slack_2006} & 2.23 & 2.23 	 \\
          & & & & & (124) & (111) & & & &\\
    KF & N/A & 2.68 & 2.21 & 235 \cite{slack, Morelli_Slack_2006} & 305 & 288 & 309	& 1.52 \cite{slack, Morelli_Slack_2006} & 2.29 & 2.32 	\\
          & & & & & (242) & (229) & & & &\\
    KCl & 7.1 \cite{Morelli_Slack_2006} & 1.4 & 1.25 & 172 \cite{slack, Morelli_Slack_2006} & 220 & 213 & 226 & 1.45 \cite{slack, Morelli_Slack_2006} & 2.38 & 2.40 	 \\
          & & & & & (175) & (169) & & & &\\
    KBr & 3.4 \cite{Morelli_Slack_2006} &  1.0 & 0.842 & 117 \cite{slack, Morelli_Slack_2006} & 165 & 156 & 162 & 1.45 \cite{slack, Morelli_Slack_2006} & 2.37 & 2.37 \\
          & & & & & (131) & (124) & & & &\\
    KI & 2.6 \cite{Morelli_Slack_2006} & 0.72 & 0.525 & 87 \cite{slack, Morelli_Slack_2006} & 129 & 116 & 120 & 1.45 \cite{slack, Morelli_Slack_2006} & 2.35 & 2.35 	 \\
          & & & & & (102) & (92) & & & &\\
    RbCl & 2.8 \cite{Morelli_Slack_2006} & 1.09 & 0.837 & 124 \cite{slack, Morelli_Slack_2006} & 168 & 155 & 160 & 1.45 \cite{slack, Morelli_Slack_2006} & 2.34 & 2.37 	 \\
          & & & & & (133) & (123) & & & &\\
    RbBr & 3.8 \cite{Morelli_Slack_2006} & 0.76 & 0.558 & 105 \cite{slack, Morelli_Slack_2006} & 134 & 122 & 129 & 1.45 \cite{slack, Morelli_Slack_2006} & 2.40 & 2.43 \\
          & & & & & (106) & (97) & & & &\\
    RbI & 2.3 \cite{Morelli_Slack_2006} & 0.52 & 0.368 & 84 \cite{slack, Morelli_Slack_2006} & 109 & 97 & 102 & 1.41 \cite{slack, Morelli_Slack_2006} & 2.47 & 2.47 	 \\
          & & & & & (87) & (77) & & & &\\
    AgCl & 1.0 \cite{Landolt-Bornstein, Maqsood_IJT_2003}  & 2.58 & 0.613 & 124 \cite{slack} & 235 & 145 & 148 & 1.9 \cite{slack} & 2.5 & 2.49 	 \\
          & & & & & (187) & (115) & & & &\\
    MgO  & 60 \cite{Morelli_Slack_2006} & 31.9 & 44.5 & 600 \cite{slack, Morelli_Slack_2006} & 758 & 849 & 890	& 1.44 \cite{slack, Morelli_Slack_2006} & 1.95 & 1.96 \\
          & & & & & (602) & (674) & & & &\\
    CaO & 27 \cite{Morelli_Slack_2006} & 19.5 & 24.3 & 450 \cite{slack, Morelli_Slack_2006} & 578 & 620 & 638 & 1.57 \cite{slack, Morelli_Slack_2006} & 2.07 & 2.06 	 \\
          & & & & & (459) & (492) & & & &\\
    SrO & 12 \cite{Morelli_Slack_2006} & 12.5 & 13.4 & 270 \cite{slack, Morelli_Slack_2006} & 399 & 413 & 421 & 1.52 \cite{slack, Morelli_Slack_2006} & 2.09 & 2.13 	 \\
          & & & & & (317) & (328) &  & & &\\
    BaO & 2.3 \cite{Morelli_Slack_2006} & 8.88 & 7.10 & 183 \cite{slack, Morelli_Slack_2006} & 305 & 288 & 292 & 1.5 \cite{slack, Morelli_Slack_2006} & 2.09 & 2.14 \\
          & & & & & (242) & (229) & & & &\\
    PbS & 2.9 \cite{Morelli_Slack_2006} & 6.48 & 6.11 & 115 \cite{slack, Morelli_Slack_2006} & 226 & 220 & 221 & 2.0 \cite{slack, Morelli_Slack_2006} & 2.02 & 2.00 	\\
          & & & & & (179) & (175) & & & &\\
    PbSe & 2.0 \cite{Morelli_Slack_2006} & 4.88 & 4.81 & 100 \cite{Morelli_Slack_2006} & 197 & 194 & 196 & 1.5 \cite{Morelli_Slack_2006} & 2.1 & 2.07 	 \\
          & & & & & (156) & (154) & & & &\\
    PbTe & 2.5 \cite{Morelli_Slack_2006} & 4.15 & 4.07 & 105 \cite{slack, Morelli_Slack_2006} & 170 & 172 & 175 & 1.45 \cite{slack, Morelli_Slack_2006} & 2.04 & 2.09 	 \\
          & & & & & (135) & (137) & & & &\\
    SnTe & 1.5 \cite{Snyder_jmatchem_2011} & 4.46 & 5.24 & 155 \cite{Snyder_jmatchem_2011} & 202 & 210 & 212 & 2.1 \cite{Snyder_jmatchem_2011} & 2.15 & 2.11 	 \\
          & & & & & (160) & (167) & & & &\\
    \hline
  \end{tabular}
\end{table*}

The experimental values $\gamma^\EXP$ of the Gr{\"u}neisen parameter are plotted {against
$\gamma^\sAGL$}, $\gamma^\sBM$, $\gamma^\sVIN$ and $\gamma^\sBCN$ in Fig. \ref{fig:rocksalt_thermal_elastic}(f), and the values
are listed in Table \ref{tab:rocksalt_thermal} and in the supplementary information.
These results show that \AGL\ has problems accurately predicting the Gr{\"u}neisen parameter for this set of materials as well, as the calculated values
are often 30\% to 50\% larger than the experimental ones and the \RMSrD\ values are of the order of 0.5. Note also that there are quite large differences between the values
obtained for different equations of state, with $\gamma^\sBCN$ generally having the lowest values while
$\gamma^\sVIN$ has the highest values, a similar pattern to that seen above for the zincblende and diamond structure materials. On the other hand, in contrast to the case of $\theta_\sDebye^\sAGL$,
the value of $\sigma$ used makes little difference to the value of $\gamma^\sAGL$. The correlation values between $\gamma^\EXP$
and the \AGL\ values, as shown in Table \ref{tab:rocksalt_correlation}, are also quite poor, with values ranging from -0.098 to
0.118 for the Pearson correlations, and negative values for the Spearman correlations.

The experimental thermal conductivity $\kappa^\EXP$ is compared in Fig. \ref{fig:rocksalt_thermal_elastic}(d) to the
thermal conductivities calculated with \AGL\ using the Leibfried-Schl{\"o}mann equation (Eq.\ (\ref{thermal_conductivity})): $\kappa^\sAGL$, $\kappa^\sBM$, $\kappa^\sVIN$ and $\kappa^\sBCN$,
while the values are listed in  Table \ref{tab:rocksalt_thermal} and in the supplementary information.
The linear correlation between the \AGL\ values and $\kappa^\EXP$ is somewhat better than for the zincblende materials set, with a Pearson
correlation as high as $0.94$, although the Spearman correlations are somewhat lower, ranging from $0.445$
to $0.556$. In particular, note that using the $\sigma^\sAEL$ in the \AGL\ calculations improves the correlations by about
2\% to 8\%, from $0.910$ to $0.932$ and from $0.445$ to $0.528$. For the different equations of state, the results for $\kappa^\sBM$
appear to correlate best with $\kappa^\EXP$ for this set of materials.

\begin{table}[t!]
  \caption{\small Correlations between experimental values and \AEL\ and \AGL\ results for
    elastic and thermal properties for rocksalt structure semiconductors.
  }
  \label{tab:rocksalt_correlation}
  {\footnotesize
    \begin{tabular}{l r r r}
      \hline
      Property  & Pearson & Spearman & \RMSrD\ \\
                & (Linear) & (Rank Order) \\
      \hline
      $\kappa^\EXP$ vs. $\kappa^\sAGL$  ($\sigma = 0.25$) \cite{curtarolo:art96} & 0.910 & 0.445 & 1.093  \\
      $\kappa^\EXP$ vs. $\kappa^\sAGL$ & 0.932 & 0.528 & 1.002  \\
      $\kappa^\EXP$ vs. $\kappa^\sBM$ & 0.940 & 0.556 & 1.038   \\
      $\kappa^\EXP$ vs. $\kappa^\sVIN$ & 0.933 &  0.540 & 0.920  \\
      $\kappa^\EXP$ vs. $\kappa^\sBCN$ & 0.930 & 0.554 & 1.082  \\
      $\theta_\acoustic^\EXP$ vs. $\theta_\acoustic^\sAGL$  ($\sigma = 0.25$) \cite{curtarolo:art96} & 0.985 & 0.948 & 0.253  \\
      $\theta_\acoustic^\EXP$ vs. $\theta_\acoustic^\sAGL$ & 0.978 & 0.928 & 0.222 \\
      $\theta_\acoustic^\EXP$ vs. $\theta_\acoustic^\sBM$ & 0.980 & 0.926 & 0.222  \\
      $\theta_\acoustic^\EXP$ vs. $\theta_\acoustic^\sVIN$ & 0.979 &  0.925 & 0.218 \\
      $\theta_\acoustic^\EXP$ vs. $\theta_\acoustic^\sBCN$ & 0.978 & 0.929 & 0.225 \\
      $\gamma^\EXP$ vs. $\gamma^\sAGL$  ($\sigma = 0.25$) \cite{curtarolo:art96} & 0.118 & -0.064 & 0.477 \\
      $\gamma^\EXP$ vs. $\gamma^\sAGL$ & 0.036 & -0.110 & 0.486 \\
      $\gamma^\EXP$ vs. $\gamma^\sBM$ & -0.019 & -0.088 &  0.462 \\
      $\gamma^\EXP$ vs. $\gamma^\sVIN$ & -0.098 &  -0.086 & 0.591 \\
      $\gamma^\EXP$ vs. $\gamma^\sBCN$ & 0.023 & -0.110 & 0.443 \\
      $B^\EXP$ vs. $B_\sVRH^\sAEL$ & 0.998 & 0.995 & 0.078 \\
      $B^\EXP$ vs. $B_\sStatic^\sAGL$ & 0.998 & 0.993 & 0.201 \\
      $B^\EXP$ vs. $B_\sStatic^\sBM$ & 0.997 & 0.993 & 0.199  \\
      $B^\EXP$ vs. $B_\sStatic^\sVIN$ & 0.997 & 0.990 & 0.239  \\
      $B^\EXP$ vs. $B_\sStatic^\sBCN$ & 0.998 & 0.993 & 0.197 \\
      $G^\EXP$ vs. $G_\sVRH^\sAEL$ & 0.994 & 0.997 & 0.105 \\
      $G^\EXP$ vs. $G_\sVoigt^\sAEL$ & 0.991 & 0.990 & 0.157 \\
      $G^\EXP$ vs. $G_\sReuss^\sAEL$ & 0.995 & 0.995 & 0.142 \\
      $\sigma^\EXP$ vs. $\sigma^\sAEL$ & 0.959 & 0.827 & 0.070 \\
      \hline
    \end{tabular}
  }
\end{table}

As in the case of the diamond and zincblende structure materials discussed in the previous Section,
Ref. \onlinecite{Morelli_Slack_2006} includes values of the thermal conductivity at 300K for rocksalt structure materials,
calculated using the experimental values of $\theta_\acoustic$ and $\gamma$ in the Leibfried-Schl{\"o}mann equation, in Table 2.1.
The correlation values of $0.986$ and $0.761$ with experiment are
better than those obtained for the \AGL\ results by a larger margin than for the zincblende materials.
Nevertheless, the Pearson correlation between the calculated and
experimental conductivities is high in both calculations, indicating that the \AGL\
approach may be used as a screening tool for high or low conductivity
compounds in cases where gaps exist in the experimental data for these
materials.

\subsection{Hexagonal structure materials}

The experimental data for this set of materials appears in Table III of Ref. \onlinecite{curtarolo:art96}, taken from Table 2.3 of 
Ref. \onlinecite{Morelli_Slack_2006}. Most of these materials have the wurtzite structure (P$6_3$mc,\ $\#$186;
Pearson symbol: hP4; \AFLOW\ prototype: {\sf AB\_hP4\_186\_b\_b} \cite{curtarolo:art121}) except InSe which is P$6_3$mmc,\ $\#$194,
Pearson symbol: hP8.

\begin{figure*}[t!]
  \includegraphics[width=0.98\textwidth]{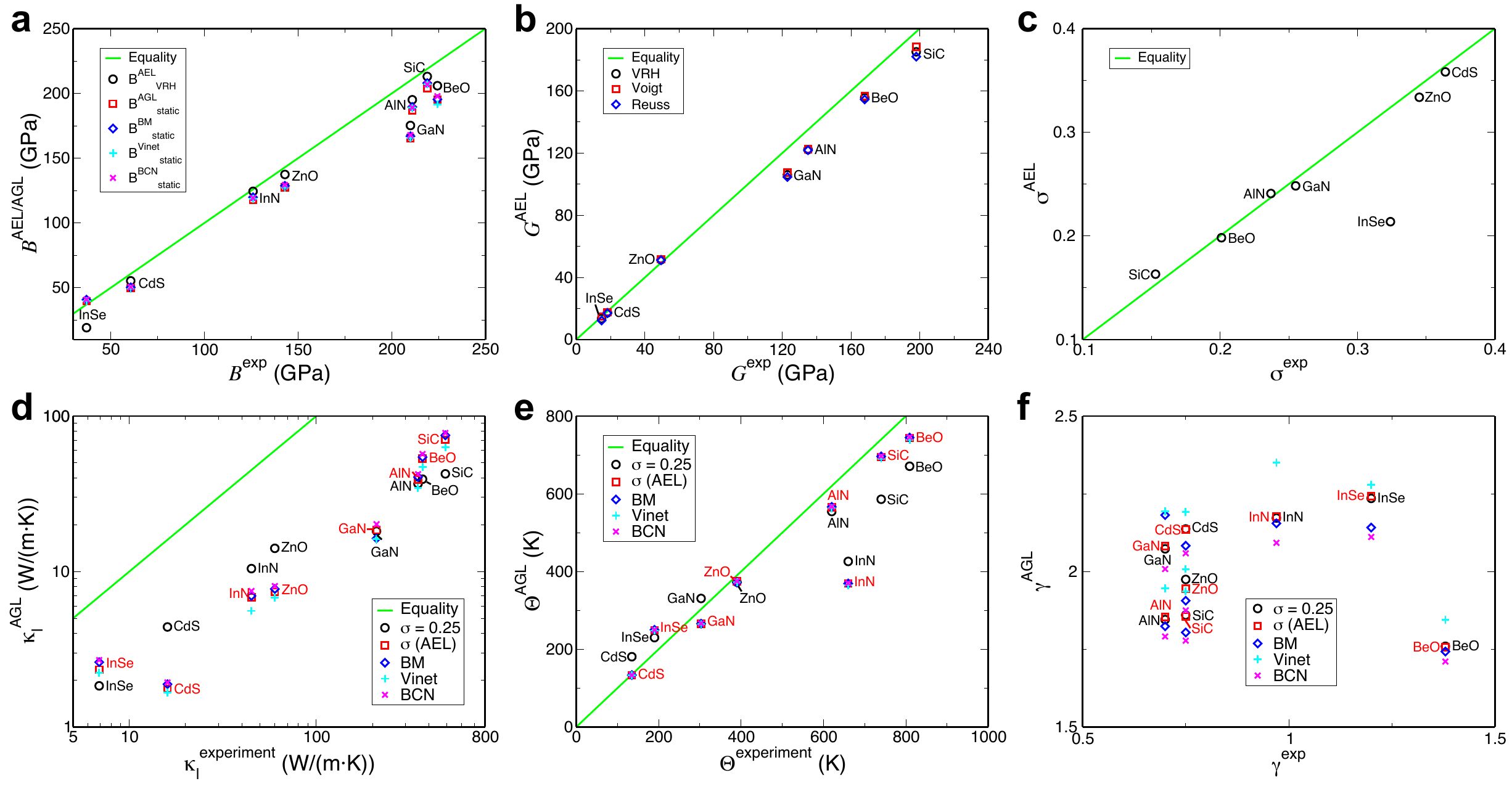}
  \vspace{-4mm}
  \caption{\small 
    {\bf (a)} Bulk modulus,
    {\bf (b)} shear modulus,
    {\bf (c)} Poisson ratio,
    {\bf (d)} lattice thermal conductivity,
    {\bf (e)} Debye temperature and
    {\bf (f)} Gr{\"u}neisen parameter of hexagonal structure
    semiconductors.
    The Debye temperatures plotted in {\bf (e)} are
    $\theta_\acoustic$, except for InSe and InN where $\theta_{\mathrm D}$
    values are quoted in Refs.
    \onlinecite{Snyder_jmatchem_2011, Ioffe_Inst_DB, Krukowski_jphyschemsolids_1998}. }
  \label{fig:wurzite_thermal_elastic}
\end{figure*}

The calculated elastic properties are shown in Table \ref{tab:wurzite_elastic} and Fig. \ref{fig:wurzite_thermal_elastic}. The bulk moduli 
values obtained from a direct calculation of the elastic tensor, $B_\sVRH^\sAEL$, are usually slightly higher than those obtained from the 
$E(V)$ curve and are also closer to experiment (Table \ref{tab:wurzite_elastic} and Fig. \ref{fig:wurzite_thermal_elastic}(a)), with the exception of 
InSe where it is noticeably lower.

\begin{table*}[t!]
  \caption{\small Bulk modulus, shear modulus and Poisson ratio of hexagonal structure semiconductors.
    ``N/A''= Not available for that source.
    Units: $B$ and $G$ in \GPa.
  }
  \label{tab:wurzite_elastic}
  \begin{tabular}{c c c c c c c c c c c c c c c c}
    \hline
    Comp. & $B^\EXP$  & $B_\sVRH^\sAEL$ & $B_\sVRH^\sMP$ & $B_\sStatic^\sAGL$ & $B_\sStatic^\sBM$ & $B_\sStatic^\sVIN$ &  $B_\sStatic^\sBCN$ & $G^\EXP$ & $G_\sVoigt^\sAEL$ & $G_\sReuss^\sAEL$ &  $G_\sVRH^\sAEL$ & $G_\sVRH^\sMP$ & $\sigma^\EXP$ & $\sigma^\sAEL$  & $\sigma^\sMP$     \\
    \hline
    SiC & 219 \cite{Arlt_ELasticSiC_JAAcS_1965} & 213 & 213 & 204 & 208 & 207 & 207 & 198 \cite{Arlt_ELasticSiC_JAAcS_1965} & 188 & 182 & 185 & 187 & 0.153 \cite{Arlt_ELasticSiC_JAAcS_1965} & 0.163 & 0.16 \\
    AlN & 211 \cite{Landolt-Bornstein, McNeil_ElasticAlN_JACerS_1993} & 195 & 194 & 187 & 190 & 189 & 189 & 135 \cite{Landolt-Bornstein, McNeil_ElasticAlN_JACerS_1993} & 123 & 122 & 122 & 122 & 0.237 \cite{Landolt-Bornstein, McNeil_ElasticAlN_JACerS_1993} & 0.241 & 0.24 \\
          & 200 \cite{Dodd_BulkmodAlN_JMS_2001} & & & & & & & 130 \cite{Dodd_BulkmodAlN_JMS_2001} & & & & & 0.234 \cite{Dodd_BulkmodAlN_JMS_2001} &\\
    GaN & 195 \cite{Semiconductors_BasicData_Springer, Savastenko_ElasticGaN_PSSa_1978} & 175 & 172 & 166 & 167 & 166 & 168 & 51.6 \cite{Semiconductors_BasicData_Springer, Savastenko_ElasticGaN_PSSa_1978}  & 107 & 105 & 106 & 105 & 0.378 \cite{Semiconductors_BasicData_Springer, Savastenko_ElasticGaN_PSSa_1978}  & 0.248 & 0.25 \\
          & 210 \cite{Polian_ElasticGaN_JAP_1996} & & & & & & & 123 \cite{Polian_ElasticGaN_JAP_1996} & & & & & 0.255 \cite{Polian_ElasticGaN_JAP_1996} & \\
    ZnO & 143 \cite{Semiconductors_BasicData_Springer, Kobiakov_ElasticZnOCdS_SSC_1980} & 137 & 130 & 128 & 129 & 127 & 129 & 49.4 \cite{Semiconductors_BasicData_Springer, Kobiakov_ElasticZnOCdS_SSC_1980} & 51.7 & 51.0 & 51.4 & 41.2 & 0.345 \cite{Semiconductors_BasicData_Springer, Kobiakov_ElasticZnOCdS_SSC_1980} & 0.334 & 0.36 \\
    BeO & 224.4 \cite{Cline_JAP_1967} & 206 & 208 & 195 & 195 & 192 & 198 & 168 \cite{Cline_JAP_1967} & 157 & 154 & 156 & 156 & 0.201 \cite{Cline_JAP_1967} & 0.198 & 0.2 \\
    CdS & 60.7 \cite{Semiconductors_BasicData_Springer, Kobiakov_ElasticZnOCdS_SSC_1980} & 55.4 & 53.3 & 49.7 & 50.3 & 49.4 & 50.6 & 18.2 \cite{Semiconductors_BasicData_Springer, Kobiakov_ElasticZnOCdS_SSC_1980}  & 17.6 & 17.0 & 17.3 & 17.6 & 0.364 \cite{Semiconductors_BasicData_Springer, Kobiakov_ElasticZnOCdS_SSC_1980} & 0.358 & 0.35 \\
    InSe & 37.1 \cite{Gatulle_ElasticInSe_PSSb_1983} & 19.2 & N/A & 39.8 & 40.8 & 39.7 & 41.0 & 14.8 \cite{Gatulle_ElasticInSe_PSSb_1983} & 14.9 & 12.3 & 13.6 & N/A & 0.324 \cite{Gatulle_ElasticInSe_PSSb_1983} & 0.214 & N/A \\
    InN & 126 \cite{Ueno_BulkmodInN_PRB_1994} & 124 & N/A & 118 & 120 & 119 & 119 & N/A & 55.4 & 54.4 & 54.9 & N/A & N/A & 0.308 & N/A \\
    \hline
  \end{tabular}
\end{table*}

For the shear modulus, the experimental values $G^\EXP$  are compared to the \AEL\ values $G_\sVoigt^\sAEL$,
$G_\sReuss^\sAEL$ and $G_\sVRH^\sAEL$. As can be seen in
Table \ref{tab:wurzite_elastic} and Fig. \ref{fig:wurzite_thermal_elastic}(b), the agreement with the experimental values is very
good. Similarly good agreement is obtained for the Poisson ratio of most materials (Table \ref{tab:wurzite_elastic}
and Fig. \ref{fig:wurzite_thermal_elastic}(c)), with
a single exception for InSe where the calculation deviates significantly from the experiment.
The Pearson (i.e. linear, Eq.\ (\ref{Pearson})) and Spearman (i.e. rank order, Eq.\ (\ref{Spearman})) correlations between the calculated
elastic properties and their experimental values are generally
quite high (Table \ref{tab:wurzite_correlation}), ranging from 0.851 and 0.893 respectively for $\sigma^\EXP$ vs. $\sigma^\sAEL$, up to 0.998
and 1.0 for $G^\EXP$ vs. $G_\sVRH^\sAEL$.

The Materials Project values of $B_\sVRH^\sMP$, $G_\sVRH^\sMP$ and $\sigma^\sMP$ for hexagonal structure materials are also shown in 
Table \ref{tab:wurzite_elastic}, where available. The Pearson correlations values for the experimental results with the available values of 
$B_\sVRH^\sMP$, $G_\sVRH^\sMP$ and $\sigma^\sMP$ were calculated to be 0.984, 0.998 and 0.993, respectively, while the respective Spearman correlations
were 0.943, 1.0 and 0.943, and the \RMSrD\ values were 0.117, 0.116 and 0.034. For comparison, the corresponding Pearson correlations for the same 
subset of materials for $B_\sVRH^\sAEL$, $G_\sVRH^\sAEL$ and $\sigma^\sAEL$ are 0.986, 0.998, and 0.998 respectively,  while the respective Spearman correlations
were 0.943, 1.0 and 1.0, and the \RMSrD\ values were 0.100, 0.091 and 0.036. These correlation values are very similar, and the general close agreement
for the results for the values of $B_\sVRH^\sAEL$, $G_\sVRH^\sAEL$ and $\sigma^\sAEL$ with those of $B_\sVRH^\sMP$, $G_\sVRH^\sMP$ and $\sigma^\sMP$ 
demonstrate that the small differences in the parameters used for the DFT calculations make little difference to the results, 
indicating that the parameter set used here is robust for high-throughput calculations.

The thermal properties calculated using \AGL\ are
compared to the experimental values in Table \ref{tab:wurzite_thermal} and are also plotted in Fig. \ref{fig:wurzite_thermal_elastic}.
For the Debye temperature, the $\theta^\EXP$ values taken from Ref. \onlinecite{Morelli_Slack_2006} are for $\theta_\acoustic$,
and are mostly in good agreement with the calculated $\theta_\acoustic^\sAGL$ values.  As in the case of the other materials sets, 
the values obtained using the numerical $E(V)$ fit and the three different
equations of state are very similar to each other, whereas $\theta_\sDebye^\sAGL$ calculated using $\sigma=0.25$ differs significantly. 
In fact, the values of $\theta_\sDebye^\sAGL$ calculated with $\sigma^\sAEL$ have a lower the correlation with $\theta^\EXP$ than the values calculated with 
$\sigma = 0.25$ do, although the \RMSrD\ values are lower when $\sigma^\sAEL$ is used. However, most of this discrepancy appears to be due to the clear 
outlier value for the material InN. When the values for this material are removed from the data set, the Pearson correlation values become very similar
when both the $\sigma = 0.25$ and $\sigma = \sigma^\sAEL$ values are used,  increasing to 0.995 and 0.994 respectively.

\begin{table*}[t!]
  \caption{\small
    Lattice thermal conductivity, Debye temperature and Gr{\"u}neisen parameter of hexagonal structure semiconductors, comparing the effect of using the
    calculated value of the Poisson ratio to previous approximation of $\sigma = 0.25$.
    The values listed for $\theta^{\mathrm{exp}}$ are $\theta_\acoustic$, except 190K for InSe \cite{Snyder_jmatchem_2011} and 660K for InN \cite{Ioffe_Inst_DB,
      Krukowski_jphyschemsolids_1998} which are $\theta_{\mathrm D}$.
    ``N/A'' = Not available for that source.
    Units: $\kappa$ in \WmK, $\theta$ in \K.
  }
  \label{tab:wurzite_thermal}
  \begin{tabular}{c c c c c c c c c c c}
    \hline
    Comp. & $\kappa^\EXP$  & $\kappa^\sAGL $ & $\kappa^\sAGL$  & $\theta^\EXP$  & $\theta_\sDebye^\sAGL$ & $\theta_\sDebye^\sAGL$ & $\theta_\sDebye^\sAEL$ & $\gamma^\EXP$ & $\gamma^\sAGL$ & $\gamma^\sAGL$ \\
          & & & & & ($\theta_\acoustic^\sAGL$) & ($\theta_\acoustic^\sAGL$) & & & &    \\
          & & ($\sigma = 0.25$) \cite{curtarolo:art96}  & & & ($\sigma = 0.25$) \cite{curtarolo:art96} & & & & ($\sigma = 0.25$) \cite{curtarolo:art96} & \\
    \hline
    SiC & 490 \cite{Morelli_Slack_2006} & 42.49 & 70.36 & 740 \cite{Morelli_Slack_2006} & 930 & 1103 & 1138 & 0.75 \cite{Morelli_Slack_2006} & 1.86 & 1.86 \\
          & & & & & (586) & (695) & & & &  \\
    AlN & 350 \cite{Morelli_Slack_2006} & 36.73 & 39.0 & 620 \cite{Morelli_Slack_2006} & 880 & 898 & 917 & 0.7 \cite{Morelli_Slack_2006} & 1.85 & 1.85 	 \\
          & & & & & (554) & (566) & & & & \\
    GaN &  210 \cite{Morelli_Slack_2006} &  18.17 & 18.54 & 390 \cite{Morelli_Slack_2006} & 592 & 595 & 606 & 0.7 \cite{Morelli_Slack_2006} & 2.07 & 2.08 	 \\
          & & & & & (373) & (375) &  & & & \\
    ZnO & 60 \cite{Morelli_Slack_2006} & 14.10 & 7.39 & 303 \cite{Morelli_Slack_2006} & 525 & 422 & 427 & 0.75 \cite{Morelli_Slack_2006} & 1.97 & 1.94 	 \\
          & & & & & (331) & (266) &  & & & \\
    BeO & 370 \cite{Morelli_Slack_2006} & 39.26 & 53.36 & 809 \cite{Morelli_Slack_2006} & 1065 & 1181 & 1235 & 1.38 \cite{Slack_JAP_1975, Cline_JAP_1967, Morelli_Slack_2006} & 1.76 & 1.76 	 \\
          & & & & & (671) & (744) & & & & \\
    CdS & 16 \cite{Morelli_Slack_2006} & 4.40 & 1.76 & 135 \cite{Morelli_Slack_2006} & 287 & 211 & 213 & 0.75 \cite{Morelli_Slack_2006} & 2.14 & 2.14 	 \\
          & & & & & (181) & (133) & & & & \\
    InSe & 6.9 \cite{Snyder_jmatchem_2011} &  1.84 &  2.34 & 190 \cite{Snyder_jmatchem_2011} & 230 & 249 & 168 & 1.2 \cite{Snyder_jmatchem_2011} & 2.24 & 2.24 	 \\
          & & & & & (115) & (125) &  & & & \\
    InN & 45 \cite{Ioffe_Inst_DB, Krukowski_jphyschemsolids_1998} & 10.44 & 6.82 & 660 \cite{Ioffe_Inst_DB, Krukowski_jphyschemsolids_1998} & 426 & 369 & 370 & 0.97 \cite{Krukowski_jphyschemsolids_1998} & 2.17 & 2.18 	 \\
          & & & & & (268) & (232)   & & & & \\
    \hline
  \end{tabular}
\end{table*}

The experimental and calculated values of the Gr{\"u}neisen parameter are listed in Table \ref{tab:wurzite_thermal}
and in the supplementary information, and are plotted in Fig. \ref{fig:wurzite_thermal_elastic}(f).
Again, the Debye model does not reproduce the experimental data, as the calculated values
are often 2 to 3 times too large and the \RMSrD\ is larger than 1.5.
The corresponding correlation, shown in Table \ref{tab:wurzite_correlation}, are also quite poor, with no value higher than 0.160 for
the Spearman correlations, and negative values for the Pearson correlations.

The comparison between the experimental thermal conductivity $\kappa^\EXP$ and the calculated values is also quite poor
(Fig. \ref{fig:wurzite_thermal_elastic}(d) and Table \ref{tab:wurzite_thermal}), with \RMSrD\ values of the order of 0.9.
Considerable disagreements also exist between different experimental reports for most materials.
Nevertheless, the Pearson correlations between the \AGL\ calculated thermal conductivity values and the experimental
values are high, ranging from $0.974$ to $0.980$, while the Spearman correlations are even higher, ranging from $0.976$
to $1.0$.

\begin{table}[t!]
  \caption{\small Correlations between experimental values and \AEL\ and \AGL\ results for
    elastic and thermal properties for hexagonal structure semiconductors.
  }
  \label{tab:wurzite_correlation}
  {\footnotesize
    \begin{tabular}{l r r r}
      \hline
      Property  & Pearson & Spearman & \RMSrD\ \\
                & (Linear) & (Rank Order) & \\
      \hline
      $\kappa^\EXP$ vs. $\kappa^\sAGL$  ($\sigma = 0.25$) \cite{curtarolo:art96} & 0.977 & 1.0 & 0.887  \\
      $\kappa^\EXP$ vs. $\kappa^\sAGL$ & 0.980 & 0.976 & 0.911 \\
      $\kappa^\EXP$ vs. $\kappa^\sBM$ & 0.974 & 0.976 & 0.904  \\
      $\kappa^\EXP$ vs. $\kappa^\sVIN$ & 0.980 &  0.976 & 0.926  \\
      $\kappa^\EXP$ vs. $\kappa^\sBCN$ & 0.980 & 0.976 & 0.895  \\
      $\theta_\acoustic^\EXP$ vs. $\theta_\acoustic^\sAGL$  ($\sigma = 0.25$) \cite{curtarolo:art96} & 0.960 & 0.976  & 0.233 \\
      $\theta_\acoustic^\EXP$ vs. $\theta_\acoustic^\sAGL$ & 0.921 & 0.929 & 0.216 \\
      $\theta_\acoustic^\EXP$ vs. $\theta_\acoustic^\sBM$ & 0.921 & 0.929 & 0.217  \\
      $\theta_\acoustic\EXP$ vs. $\theta_\acoustic^\sVIN$ & 0.920 &  0.929 & 0.218 \\
      $\theta_\acoustic^\EXP$ vs. $\theta_\acoustic^\sBCN$ & 0.921 & 0.929 & 0.216 \\
      $\gamma^\EXP$ vs. $\gamma^\sAGL$  ($\sigma = 0.25$) \cite{curtarolo:art96} & -0.039 & 0.160 & 1.566 \\
      $\gamma^\EXP$ vs. $\gamma^\sAGL$ & -0.029 & 0.160 & 1.563 \\
      $\gamma^\EXP$ vs. $\gamma^\sBM$ & -0.124 & -0.233 & 1.547  \\
      $\gamma^\EXP$ vs. $\gamma^\sVIN$ & -0.043 &  0.012 & 1.677 \\
      $\gamma^\EXP$ vs. $\gamma^\sBCN$ & -0.054 & 0.098 & 1.467 \\
      $B^\EXP$ vs. $B_\sVRH^\sAEL$ & 0.990 & 0.976 & 0.201 \\
      $B^\EXP$ vs. $B_\sStatic^\sAGL$ & 0.990 & 0.976 & 0.138 \\
      $B^\EXP$ vs. $B_\sStatic^\sBM$ & 0.988 & 0.976 & 0.133  \\
      $B^\EXP$ vs. $B_\sStatic^\sVIN$ & 0.988 & 0.976 & 0.139 \\
      $B^\EXP$ vs. $B_\sStatic^\sBCN$ & 0.990 & 0.976 & 0.130 \\
      $G^\EXP$ vs. $G_\sVRH^\sAEL$ & 0.998 & 1.0 & 0.090 \\
      $G^\EXP$ vs. $G_\sVoigt^\sAEL$ & 0.998 & 1.0 & 0.076 \\
      $G^\EXP$ vs. $G_\sReuss^\sAEL$ & 0.998 & 1.0 & 0.115 \\
      $\sigma^\EXP$ vs. $\sigma^\sAEL$ & 0.851 & 0.893 & 0.143 \\
      \hline
    \end{tabular}
  }
\end{table}

As for the rocksalt and zincblende material sets, Ref. \onlinecite{Morelli_Slack_2006} (Table 2.3) includes
values of the thermal conductivity at 300K for wurzite structure materials, calculated using the
experimental values of the Debye temperature and Gr{\"u}neisen parameter in the Leibfried-Schl{\"o}mann equation.
The Pearson and Spearman correlations are $0.996$ and $1.0$ respectively, which are slightly higher than the correlations obtained using
the \AGL\ calculated quantities. The difference is insignificant since all of these
correlations are very high and
could reliably serve as a screening tool of the thermal conductivity.
However, as we noted in our previous work on \AGL\ \cite{curtarolo:art96}, the high correlations calculated with the
experimental $\theta_\acoustic$ and $\gamma$ were obtained using
$\gamma=0.75$ for BeO. Table 2.3 of
Ref. \onlinecite{Morelli_Slack_2006} also cites an alternative value
of $\gamma=1.38$ for BeO (Table \ref{tab:wurzite_thermal}). Using this outlier
value would severely degrade the results down to $0.7$, for the
Pearson correlation, and $0.829$, for the Spearman correlation.
These values are too low for a reliable screening tool. This
demonstrates the ability of the
\AEL-\AGL\ calculations to compensate for anomalies in the
experimental data when
they exist and still provide a reliable screening method for the
thermal conductivity.

\subsection{Rhombohedral materials}

The elastic properties of a few materials with rhombohedral structures
(spacegroups: R$\overline{3}$mR,\ $\#$166, R$\overline{3}$mH,\ $\#$166; Pearson symbol: hR5; \AFLOW\ prototype: {\sf A2B3\_hR5\_166\_c\_ac} \cite{curtarolo:art121};
and spacegroup: R$\overline{3}$cH,\ $\#$167; Pearson symbol: hR10; \AFLOW\ prototype: {\sf A2B3\_hR10\_167\_c\_e} \cite{curtarolo:art121})
are shown in Table \ref{tab:rhombo_elastic} (we have left out the material Fe$_2$O$_3$ which was included in
the data set in Table IV of Ref. \onlinecite{curtarolo:art96}, due to convergence issues with some of the
strained structures required for the calculation of the elastic tensor). 
The comparison between experiment and calculation is qualitatively reasonable, but the scarcity of experimental results
does not allow for a proper correlation analysis.

\begin{table*}[t!]
  \caption{\small Bulk modulus, shear modulus and Poisson ratio  of rhombohedral
    semiconductors.
    ``N/A'' = Not available for that source.
    Units: $B$ and $G$ in \GPa.
  }
  \label{tab:rhombo_elastic}
  \begin{tabular}{c c c c c c c c c c c c c c c c}
    \hline
    Comp. & $B^\EXP$  & $B_\sVRH^\sAEL$ & $B_\sVRH^\sMP$ & $B_\sStatic^\sAGL$ & $B_\sStatic^\sBM$ & $B_\sStatic^\sVIN$ &  $B_\sStatic^\sBCN$ & $G^\EXP$ & $G_\sVoigt^\sAEL$ & $G_\sReuss^\sAEL$ &  $G_\sVRH^\sAEL$ & $G_\sVRH^\sMP$ & $\sigma^\EXP$ & $\sigma^\sAEL$ & $\sigma^\sMP$      \\
    \hline
    Bi$_2$Te$_3$ & 37.0 \cite{Semiconductors_BasicData_Springer, Jenkins_ElasticBi2Te3_PRB_1972} & 28.8 & 15.0 & 43.7 & 44.4 & 43.3 & 44.5 & 22.4 \cite{Semiconductors_BasicData_Springer, Jenkins_ElasticBi2Te3_PRB_1972} & 23.5 & 16.3 & 19.9 & 10.9 & 0.248 \cite{Semiconductors_BasicData_Springer, Jenkins_ElasticBi2Te3_PRB_1972} & 0.219 & 0.21 \\
    Sb$_2$Te$_3$ & N/A &  22.9 & N/A & 45.3 & 46.0 & 45.2 & 46.0 & N/A & 20.6 & 14.5 & 17.6 & N/A & N/A & 0.195 & N/A \\
    Al$_2$O$_3$ & 254 \cite{Goto_ElasticAl2O3_JGPR_1989} & 231 & 232 & 222 & 225 & 224 & 224 & 163.1 \cite{Goto_ElasticAl2O3_JGPR_1989} & 149 & 144 & 147 & 147 & 0.235 \cite{Goto_ElasticAl2O3_JGPR_1989} & 0.238 & 0.24 \\
    Cr$_2$O$_3$ & 234 \cite{Alberts_ElasticCr2O3_JMMM_1976} & 203 & 203 & 198 & 202 & 201 & 201 & 129 \cite{Alberts_ElasticCr2O3_JMMM_1976} & 115 & 112 & 113 & 113 & 0.266 \cite{Alberts_ElasticCr2O3_JMMM_1976} & 0.265 & 0.27 \\
    Bi$_2$Se$_3$ & N/A & 93.9 & N/A & 57.0 & 57.5 & 56.4 & 57.9 & N/A & 53.7 & 28.0 & 40.9 & N/A & N/A & 0.310 & N/A \\
    \hline
  \end{tabular}
\end{table*}

\begin{figure}
  \includegraphics[width=0.98\columnwidth]{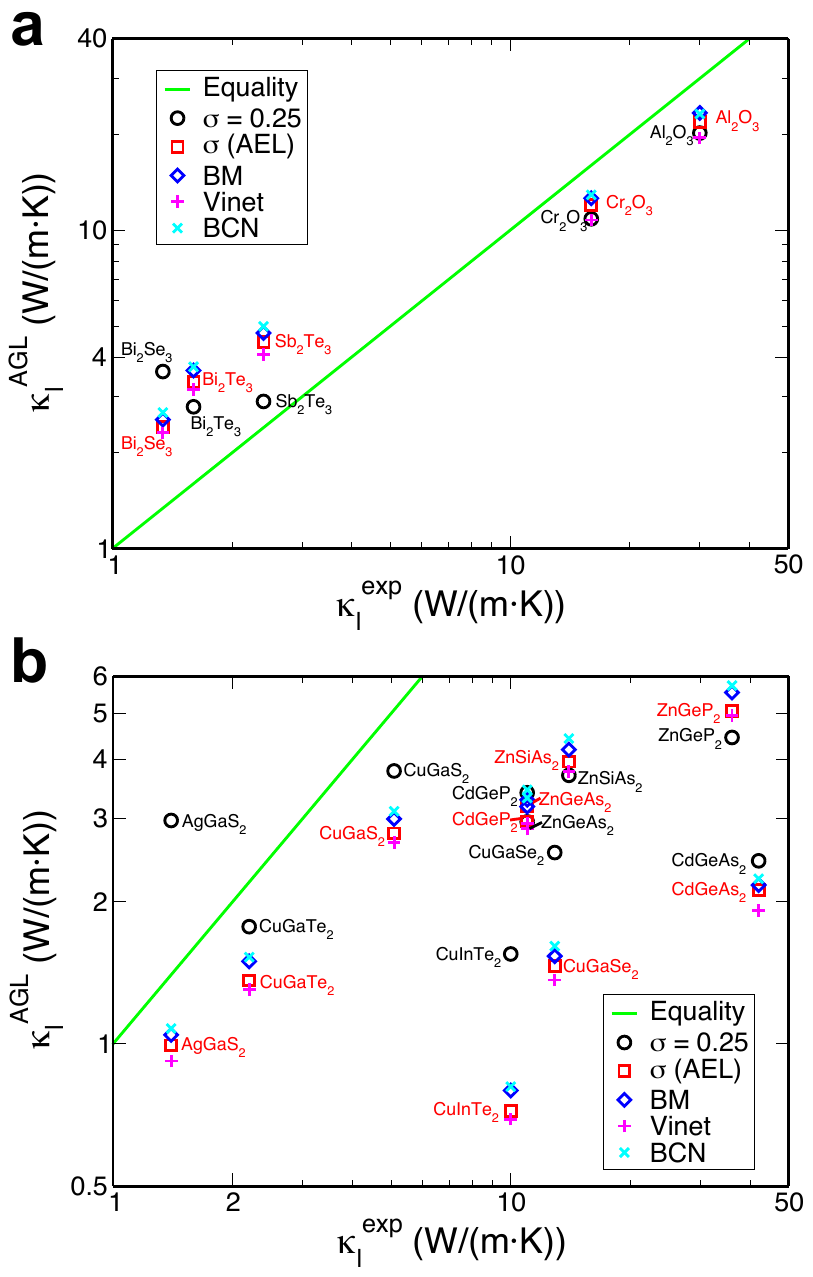}
  \vspace{-4mm}
  \caption{\small
    {\bf (a)}  Lattice thermal conductivity of rhombohedral semiconductors at 300K.
    {\bf (b)}  Lattice thermal conductivity of body-centred tetragonal semiconductors at 300K.
  }
  \label{fig:mixed_thermal}
\end{figure}

The thermal properties calculated using \AGL\ are
compared to the experimental values in Table \ref{tab:rhombo_thermal} and the thermal conductivity is also plotted in
Fig. \ref{fig:mixed_thermal}(a).
The experimental Debye temperatures are $\theta_\sDebye$ for Bi$_2$Te$_3$ and Sb$_2$Te$_3$, and
$\theta_\acoustic$ for Al$_2$O$_3$. The values obtained using the numerical $E(V)$ fit and the three different equations of state
(see supplementary material)
are very similar, but just roughly reproduce the experiments.

\begin{table*}[t!]
  \caption{\small Lattice thermal conductivity, Debye temperatures and Gr{\"u}neisen parameter of rhombohedral
    semiconductors, comparing the effect of using the
    calculated value of the Poisson ratio to previous approximation of
    $\sigma = 0.25$.
    The experimental Debye temperatures are $\theta_{\mathrm D}$ for
    Bi$_2$Te$_3$ and Sb$_2$Te$_3$, and  $\theta_\acoustic$ for Al$_2$O$_3$.
    ``N/A'' = Not available for that source.
    Units: $\kappa$ in \WmK, $\theta$ in \K.
  }
  \label{tab:rhombo_thermal}
  \begin{tabular}{c c c c c c c c c c c}
    \hline
    Comp. & $\kappa^\EXP$  & $\kappa^\sAGL $ & $\kappa^\sAGL$  & $\theta^\EXP$  & $\theta_\sDebye^\sAGL$ & $\theta_\sDebye^\sAGL$ & $\theta_\sDebye^\sAEL$ & $\gamma^\EXP$ & $\gamma^\sAGL$ & $\gamma^\sAGL$ \\
          & & & & & ($\theta_\acoustic^\sAGL$) & ($\theta_\acoustic^\sAGL$) & & & &    \\
          & & ($\sigma = 0.25$) \cite{curtarolo:art96}  & & & ($\sigma = 0.25$) \cite{curtarolo:art96} & & & & ($\sigma = 0.25$) \cite{curtarolo:art96} & \\
    \hline
    Bi$_2$Te$_3$ & 1.6 \cite{Snyder_jmatchem_2011} & 2.79 & 3.35 & 155 \cite{Snyder_jmatchem_2011} & 191 & 204 & 161 & 1.49 \cite{Snyder_jmatchem_2011} & 2.13 & 2.14 	 \\
          & & & & & (112) & (119) & & & & \\
    Sb$_2$Te$_3$ & 2.4 \cite{Snyder_jmatchem_2011} & 2.90 & 4.46 & 160 \cite{Snyder_jmatchem_2011} & 217 & 243 & 170 & 1.49 \cite{Snyder_jmatchem_2011} & 2.2 & 2.11 	\\
          & & & & & (127) & (142) & & & & \\
    Al$_2$O$_3$ & 30 \cite{Slack_PR_1962} &  20.21 & 21.92 & 390 \cite{slack} & 927 & 952 & 975 & 1.32 \cite{slack} & 1.91 & 1.91 	 \\
          & & & & & (430) & (442) & & & & \\
    Cr$_2$O$_3$  & 16 \cite{Landolt-Bornstein, Bruce_PRB_1977} & 10.87 & 12.03 & N/A & 733 &  717 & 720 & N/A & 2.26 & 2.10 	\\
          & & & & &  (340) & (333) & & & & \\
    Bi$_2$Se$_3$ & 1.34 \cite{Landolt-Bornstein} & 3.60 & 2.41 & N/A & 223 & 199 & 241 & N/A & 2.08 & 2.12 	\\
          & & & & & (130) & (116) & & & & \\
    \hline
  \end{tabular}
\end{table*}

The calculated Gr{\"u}neisen parameters are about 50\% larger than the experimental ones, and
the value of $\sigma$ used makes a little difference in the calculation.
The absolute agreement between the \AGL\ values and $\kappa^\EXP$ is also quite poor (Fig. \ref{fig:mixed_thermal}(a)).
However, despite all these discrepancies,
the Pearson correlations between the calculated thermal conductivities and the experimental
values are all high, of the order of $0.998$, while the Spearman correlations range from $0.7$ to $1.0$,
with all of the different equations of state having very similar correlations with experiment.
Using the calculated $\sigma^\sAEL$, vs. the rough Cauchy approximation, improves the Spearman correlation from $0.7$ to $1.0$.

\begin{table}[t!]
  \caption{\small Correlations between experimental values and \AEL\ and \AGL\ results for
    elastic and thermal properties for rhombohedral structure semiconductors.
  }
  \label{tab:rhombo_correlation}
  {\footnotesize
    \begin{tabular}{l r r r}
      \hline
      Property  & Pearson & Spearman & \RMSrD\ \\
                & (Linear) & (Rank Order) \\
      \hline
      $\kappa^\EXP$ vs. $\kappa^\sAGL$  ($\sigma = 0.25$) \cite{curtarolo:art96} & 0.997 & 0.7 & 0.955  \\
      $\kappa^\EXP$ vs. $\kappa^\sAGL$ & 0.998 & 1.0 & 0.821 \\
      $\kappa^\EXP$ vs. $\kappa^\sBM$ & 0.997 & 1.0 & 0.931  \\
      $\kappa^\EXP$ vs. $\kappa^\sVIN$ & 0.998 &  1.0 & 0.741 \\
      $\kappa^\EXP$ vs. $\kappa^\sBCN$ & 0.997 & 1.0 & 1.002 \\
      \hline
    \end{tabular}
  }
\end{table}


\subsection{Body-centred tetragonal materials}

The mechanical properties of the body-centred tetragonal materials (spacegroup:
I$\overline{4}2$d,\ $\#$122; Pearson symbol: tI16; \AFLOW\ prototype: {\sf ABC2\_tI16\_122\_a\_b\_d} \cite{curtarolo:art121})
of Table V of Ref. \onlinecite{curtarolo:art96} are reported in Table
\ref{tab:bct_elastic}. The calculated bulk moduli miss considerably the few available experimental results, while the shear moduli
are well reproduced. Reasonable estimates are also obtained for the Poisson ratio.

\begin{table*}
  \caption{\small Bulk modulus, shear modulus and Poisson ratio of body-centred tetragonal semiconductors. Note that there appears to be an error in Table 1 of Ref. \onlinecite{Fernandez_ElasticCuInTe_PSSa_1990}
    where the bulk modulus values are stated to be in units of $10^{12}$ Pa. This seems unlikely, as that would give a bulk modulus for CuInTe$_2$ an order of magnitude larger than
    that for diamond. Also, units of $10^{12}$ Pa would be inconsistent with the experimental results listed in Ref. \onlinecite{Neumann_ElasticCuInTe_PSSa_1986}, so therefore it seems that these values are in units of
    $10^{10}$ Pa, which are the values shown here.
    ``N/A'' = Not available for that source.
    Units: $B$ and $G$ {in} \GPa.
  }
  \label{tab:bct_elastic}
  \begin{tabular}{c c c c c c c c c c c c c c c c}
    \hline
    Comp. & $B^\EXP$  & $B_\sVRH^\sAEL$ & $B_\sVRH^\sMP$ & $B_\sStatic^\sAGL$ & $B_\sStatic^\sBM$ & $B_\sStatic^\sVIN$ &  $B_\sStatic^\sBCN$ & $G^\EXP$ & $G_\sVoigt^\sAEL$ & $G_\sReuss^\sAEL$ & $G_\sVRH^\sAEL$ & $G_\sVRH^\sMP$ & $\sigma^\EXP$ & $\sigma^\sAEL$  & $\sigma^\sMP$     \\
    \hline
    CuGaTe$_2$ & N/A & 47.0 & N/A & 42.5 & 43.2 & 42.0 & 43.5 & N/A & 25.1 & 22.1 & 23.6 & N/A & N/A & 0.285 & N/A \\
    ZnGeP$_2$ & N/A & 73.1 & 74.9 & 70.1 & 71.1 & 70.0 & 71.4 & N/A & 50.5 & 46.2 & 48.4 & 48.9 & N/A & 0.229 & 0.23 \\
    ZnSiAs$_2$ & N/A & 67.4 & 65.9 & 63.4 & 64.3 & 63.1 & 64.6 & N/A & 44.4 & 40.4 & 42.4 & 42.2 & N/A & 0.240 & 0.24 \\
    CuInTe$_2$ & 36.0 \cite{Neumann_ElasticCuInTe_PSSa_1986} & 53.9 & N/A & 38.6 & 39.2 & 38.2 & 39.4 & N/A & 20.4 & 17.2 & 18.8 & N/A & 0.313 \cite{Fernandez_ElasticCuInTe_PSSa_1990} & 0.344 & N/A \\
          & 45.4 \cite{Fernandez_ElasticCuInTe_PSSa_1990} & & & & & & & & & & &  \\
    AgGaS$_2$ & 67.0 \cite{Grimsditch_ElasticAgGaS2_PRB_1975} & 70.3 & N/A & 56.2 & 57.1 & 56.0 & 57.4 & 20.8 \cite{Grimsditch_ElasticAgGaS2_PRB_1975} & 20.7 & 17.4 & 19.1 & N/A & 0.359 \cite{Grimsditch_ElasticAgGaS2_PRB_1975} & 0.375 & N/A \\
    CdGeP$_2$ & N/A & 65.3 & 65.2 & 60.7 & 61.6 & 60.4 & 61.9  & N/A & 37.7 & 33.3 & 35.5 & 35.0 & N/A & 0.270 & 0.27 \\
    CdGeAs$_2$ & 69.9 \cite{Hailing_ElasticCdGeAs_JPCSS_1982} & 52.6 & N/A & 49.2 & 49.6 & 48.3 & 49.9 & 29.5 \cite{Hailing_ElasticCdGeAs_JPCSS_1982} & 30.9 & 26.2 & 28.6 & N/A & 0.315 \cite{Hailing_ElasticCdGeAs_JPCSS_1982} & 0.270 & N/A \\
    CuGaS$_2$ & 94.0 \cite{Bettini_ElasticCuGaS_SSC_1975} & 73.3 & N/A & 69.0 & 69.9 & 68.7 & 70.6 & N/A & 37.8 & 32.4 & 35.1 & N/A & N/A  & 0.293 & N/A  \\
    CuGaSe$_2$ & N/A & 69.9 & N/A & 54.9 & 55.6 & 54.4 & 56.0 & N/A & 30.3 & 26.0 & 28.1 & N/A & N/A & 0.322 & N/A \\
    ZnGeAs$_2$ & N/A & 59.0 & N/A & 56.2 & 56.7 & 55.5 & 57.1 & N/A & 39.0 & 35.6 & 37.3 & N/A & N/A & 0.239 & N/A \\
    \hline
  \end{tabular}
\end{table*}

The thermal properties are reported in Table \ref{tab:bct_thermal} and Fig. \ref{fig:mixed_thermal}(b).
The $\theta^\EXP$ values are all for $\theta_\sDebye$, and in most cases are in good agreement with the values obtained
with the \AEL\ calculated $\sigma$. The
values from the numerical $E(V)$ fit and the three different equations of state are again very similar, but differ significantly
from {$\theta_\sDebye^\sAGL$} calculated with $\sigma=0.25$.

\begin{table*}
  \caption{\small
    Lattice thermal conductivity at 300K, Debye temperatures and Gr{\"u}neisen parameter of  body-centred tetragonal
    semiconductors, comparing the effect of using the
    calculated value of the Poisson ratio to previous approximation of
    $\sigma = 0.25$.
    ``N/A'' = Not available for that source.
    Units: $\kappa$ in \WmK, $\theta$ in \K.
  }
  \label{tab:bct_thermal}
  \begin{tabular}{c c c c c c c c c c c}
    \hline
    Comp. & $\kappa^\EXP$  & $\kappa^\sAGL $ & $\kappa^\sAGL$  & $\theta^\EXP$  & $\theta_\sDebye^\sAGL$ & $\theta_\sDebye^\sAGL$ & $\theta_\sDebye^\sAEL$ & $\gamma^\EXP$ & $\gamma^\sAGL$ & $\gamma^\sAGL$ \\
          & & & & & ($\theta_\acoustic^\sAGL$) & ($\theta_\acoustic^\sAGL$) & & & &    \\
          & & ($\sigma = 0.25$) \cite{curtarolo:art96}  & & & ($\sigma = 0.25$) \cite{curtarolo:art96} & & & & ($\sigma = 0.25$) \cite{curtarolo:art96} & \\
    \hline
    CuGaTe$_2$ & 2.2  \cite{Snyder_jmatchem_2011} & 1.77 & 1.36 & 226 \cite{Snyder_jmatchem_2011} & 234 & 215 & 218 & 1.46 \cite{Snyder_jmatchem_2011} & 2.32 & 2.32 	 \\
          & & & & & (117) & (108) & & & & \\
    ZnGeP$_2$ & 35 \cite{Landolt-Bornstein, Beasley_AO_1994} & 4.45 & 5.07 & 500 \cite{Landolt-Bornstein} & 390 & 408 & 411 & N/A & 2.13 & 2.14 	   \\
          & 36 \cite{Landolt-Bornstein, Beasley_AO_1994} & & & 428 \cite{Abrahams_JCP_1975} & (195) & (204) & & & & \\
          & 18 \cite{Landolt-Bornstein, Shay_1975, Masumoto_JPCS_1966} & & & & & & & & & \\
    ZnSiAs$_2$ & 14\cite{Landolt-Bornstein, Shay_1975, Masumoto_JPCS_1966} & 3.70 & 3.96  & 347 \cite{Landolt-Bornstein, Bohnhammel_PSSa_1981} & 342 & 350 & 354 & N/A & 2.15 & 2.15  	 \\
          & & & & & (171) & (175) & & & & \\
    CuInTe$_2$ & 10\cite{Landolt-Bornstein, Rincon_PSSa_1995} & 1.55 & 0.722 & 185 \cite{Landolt-Bornstein, Rincon_PSSa_1995} & 215 & 166 & 185 & 0.93 \cite{Rincon_PSSa_1995} & 2.33 & 2.32 	 \\
          & & & & 195 \cite{Landolt-Bornstein, Bohnhammel_PSSa_1982}  & (108) & (83) & & & &\\
    AgGaS$_2$ & 1.4\cite{Landolt-Bornstein, Beasley_AO_1994} & 2.97 & 0.993 & 255 \cite{Landolt-Bornstein, Abrahams_JCP_1975} & 324 & 224 & 237 & N/A & 2.20 & 2.20 	\\
          & & & & & (162) & (112) & & & & \\
    CdGeP$_2$ & 11 \cite{Landolt-Bornstein, Shay_1975, Masumoto_JPCS_1966} & 3.40 & 2.96 & 340 \cite{Landolt-Bornstein, Abrahams_JCP_1975} & 335 & 320 & 324 & N/A & 2.20 & 2.21 \\
          & & & & & (168) & (160) & & & & \\
    CdGeAs$_2$ & 42 \cite{Landolt-Bornstein, Shay_1975} & 2.44 & 2.11 & 241 \cite{Bohnhammel_PSSa_1981} & 266 & 254 & 255 & N/A & 2.20 & 2.20 	\\
          & & & & & (133) & (127) & & & &\\
    CuGaS$_2$ & 5.09 \cite{Landolt-Bornstein} & 3.78 & 2.79 & 356 \cite{Landolt-Bornstein, Abrahams_JCP_1975} & 387 & 349 & 349 & N/A & 2.24 & 2.24 	 \\
          & & & & & (194) & (175) & & & &\\
    CuGaSe$_2$ & 12.9 \cite{Landolt-Bornstein, Rincon_PSSa_1995} & 2.54 & 1.46 & 262 \cite{Landolt-Bornstein, Bohnhammel_PSSa_1982} & 294 & 244 & 265 & N/A & 2.27 & 2.26 	 \\
          & & & & & (147) & (122) & & & &\\
    ZnGeAs$_2$ & 11\cite{Landolt-Bornstein, Shay_1975} & 2.95 & 3.18 & N/A & 299 & 307 & 308 & N/A & 2.16 & 2.17 	 \\
          & & & & & (150) & (154) & & & &\\
    \hline
  \end{tabular}
\end{table*}

The comparison of the experimental thermal conductivity $\kappa^\EXP$ to the calculated values, in Fig. \ref{fig:mixed_thermal}(b),
shows poor reproducibility. The available data can thus only be considered a rough indication of their order of magnitude.
The Pearson and Spearman correlations are also quite low for all types of calculation,
but somewhat better when the calculated $\sigma^\sAEL$ is used instead of the Cauchy approximation.

\begin{table}[t!]
  \caption{\small Correlations between experimental values and \AEL\ and \AGL\ results for
    elastic and thermal properties for body-centred tetragonal structure semiconductors.
  }
  \label{tab:bct_correlation}
  {\footnotesize
    \begin{tabular}{l r r r}
      \hline
      Property  & Pearson & Spearman & \RMSrD\ \\
                & (Linear) & (Rank Order) \\
      \hline
      $\kappa^\EXP$ vs. $\kappa^\sAGL$  ($\sigma = 0.25$) \cite{curtarolo:art96} & 0.265 & 0.201 & 0.812 \\
      $\kappa^\EXP$ vs. $\kappa^\sAGL$ & 0.472 & 0.608 & 0.766 \\
      $\kappa^\EXP$ vs. $\kappa^\sBM$ & 0.467 & 0.608 & 0.750 \\
      $\kappa^\EXP$ vs. $\kappa^\sVIN$ & 0.464 &  0.608 & 0.778 \\
      $\kappa^\EXP$ vs. $\kappa^\sBCN$ & 0.460 & 0.608 & 0.741 \\
      \hline
    \end{tabular}
  }
\end{table}

\subsection{Miscellaneous materials}

In this Section we consider materials with various other structures, as in Table VI of Ref. \onlinecite{curtarolo:art96}:
CoSb$_3$ and IrSb$_3$ (spacegroup: Im$\overline{3}$,\ $\#$204; Pearson
symbol: cI32; \AFLOW\ prototype: {\sf A3B\_cI32\_204\_g\_c} \cite{curtarolo:art121}), ZnSb (Pbca,\ $\#$61; oP16; \AFLOW\ prototype: {\sf AB\_oP16\_61\_c\_c} \cite{curtarolo:art121}),
Sb$_2$O$_3$ (Pccn,\ $\#$56; oP20), InTe (Pm$\overline{3}$m,\ $\#$221; cP2; \AFLOW\ prototype: {\sf AB\_cP2\_221\_b\_a} \cite{curtarolo:art121}, and I4/mcm,\ $\#$140; tI16), 
Bi$_2$O$_3$  ($P121/c1,\ \#14$; $mP20$); and 
SnO$_2$ ($P42/mnm,\ \#136$; $tP6$; {\sf A2B\_tP6\_136\_f\_a} \cite{curtarolo:art121}). 
Two different structures are listed for InTe. In Ref. \onlinecite{curtarolo:art96}, we
considered its simple cubic structure, but this is a high-pressure phase \cite{Chattopadhyay_BulkModInTe_JPCS_1985}, while the ambient
pressure phase is body-centred tetragonal. It appears that the thermal conductivity results should be for the body-centred tetragonal
phase \cite{Spitzer_JPCS_1970}, therefore both sets of results are reported here. The correlation values shown in the tables below
were calculated for the body-centred tetragonal structure.

The elastic properties are shown
in Table \ref{tab:misc_elastic}. Large discrepancies appear between the results of all calculations
and the few available experimental results.
\begin{table*}
  \caption{\small Bulk modulus, shear modulus and Poisson ratio of materials with various
    structures.
    ``N/A'' = Not available for that source.
    Units: $B$ and $G$ {in} \GPa.
  }
  \label{tab:misc_elastic}
  \begin{tabular}{c c c c c c c c c c c c c c c c c}
    \hline
    Comp. & Pearson  & $B^\EXP$  & $B_\sVRH^\sAEL$ & $B_\sVRH^\sMP$ & $B_\sStatic^\sAGL$ & $B_\sStatic^\sBM$ & $B_\sStatic^\sVIN$ &  $B_\sStatic^\sBCN$ & $G^\EXP$ & $G_\sVoigt^\sAEL$ & $G_\sReuss^\sAEL$ &  $G_\sVRH^\sAEL$ &  $G_\sVRH^\sMP$ & $\sigma^\EXP$ & $\sigma^\sAEL$ & $\sigma^\sMP$      \\
    \hline
    CoSb$_3$ & $cI32$ & N/A & 78.6 & 82.9 & 75.6 & 76.1 & 75.1 & 76.3 & N/A & 57.2 & 55.1 & 56.2 & 57.0 & N/A & 0.211 & 0.22 \\
    IrSb$_3$ & $cI32$ & N/A & 97.5 & 98.7 & 94.3 & 94.8 & 93.8 & 95.5 & N/A & 60.9 & 59.4 & 60.1 & 59.7& N/A & 0.244 & 0.25 \\
    ZnSb & $oP16$ & N/A & 47.7 & 47.8 & 46.7 & 47.0 & 46.0 & 47.7 & N/A & 29.2 & 27.0 & 28.1 & 28.2 & N/A & 0.253 & 0.25 \\
    Sb$_2$O$_3$ & $oP20$ & N/A & 16.5 & 19.1 & 97.8 & 98.7 & 97.8 & 98.7 & N/A & 22.8 & 16.4 & 19.6 & 20.4 & N/A & 0.0749 & 0.11 \\
    InTe & $cP2$ & 90.2 \cite{Chattopadhyay_BulkModInTe_JPCS_1985} & 41.7 & N/A & 34.9 & 34.4 & 33.6 & 34.7 & N/A & 8.41 & 8.31 & 8.36 & N/A& N/A & 0.406 & N/A \\
    InTe & $tI16$ & 46.5 \cite{Chattopadhyay_BulkModInTe_JPCS_1985} & 20.9 & N/A & 32.3 & 33.1 & 32.2 & 33.2 & N/A & 13.4 & 13.0 & 13.2 & N/A & N/A & 0.239 & N/A \\
    Bi$_2$O$_3$ & $mP20$ & N/A & 48.0 & 54.5 & 108 & 110 & 109 & 109 & N/A & 30.3 & 25.9 & 28.1 & 29.9 & N/A & 0.255 & 0.27 \\
    SnO$_2$ & $tP6$ & 212 \cite{Chang_ElasticSnO2_JGPR_1975} & 159 & N/A & 158 & 162 & 161 & 161 & 106 \cite{Chang_ElasticSnO2_JGPR_1975} & 86.7 & 65.7 & 76.2 & N/A & 0.285 \cite{Chang_ElasticSnO2_JGPR_1975} & 0.293 & N/A \\
    \hline
  \end{tabular}
\end{table*}

The thermal properties are
compared to the experimental values in Table \ref{tab:misc_thermal}.
The experimental Debye temperatures are for $\theta_\sDebye$, except ZnSb for which it is $\theta_\acoustic$. Good agreement
is found between calculation and the few available experimental values. Again, the numerical $E(V)$ fit and the three different
equations of state give similar results.
For the Gr{\"u}neisen parameter, experiment and calculations again differ considerably, while the changes due to the different
values of $\sigma$ used in the
calculations are negligible.

\begin{table*}
  \caption{\small Lattice thermal conductivity at 300K, Debye temperatures
    and Gr{\"u}neisen parameter of  materials with various structures, comparing the effect of using the
    calculated value of the Poisson ratio to previous approximation of
    $\sigma = 0.25$.
    The experimental Debye temperatures are $\theta_{\mathrm D}$,
    except ZnSb for which it is $\theta_\acoustic$.
    ``N/A'' = Not available for that source.
    Units: $\kappa$ in \WmK, $\theta$ in \K.
  }
  \label{tab:misc_thermal}
  \begin{tabular}{c c c c c c c c c c c c}
    \hline
    Comp. & Pearson  & $\kappa^\EXP$  & $\kappa^\sAGL $ & $\kappa^\sAGL$  & $\theta^\EXP$  & $\theta_\sDebye^\sAGL$ & $\theta_\sDebye^\sAGL$ & $\theta_\sDebye^\sAEL$ & $\gamma^\EXP$ & $\gamma^\sAGL$ & $\gamma^\sAGL$ \\
          & & & & & & ($\theta_\acoustic^\sAGL$) & ($\theta_\acoustic^\sAGL$) & & & &    \\
          & & & ($\sigma = 0.25$) \cite{curtarolo:art96}  & & & ($\sigma = 0.25$) \cite{curtarolo:art96} & & & & ($\sigma = 0.25$) \cite{curtarolo:art96} & \\
    \hline
    CoSb$_3$ & $cI32$ & 10 \cite{Snyder_jmatchem_2011} & 1.60 & 2.60 & 307 \cite{Snyder_jmatchem_2011} & 284 & 310 & 312 & 0.95 \cite{Snyder_jmatchem_2011} & 2.63 & 2.33 \\
          & & & & & & (113) & (123) & & & & \\
    IrSb$_3$ & $cI32$ & 16 \cite{Snyder_jmatchem_2011} & 2.64 & 2.73 & 308 \cite{Snyder_jmatchem_2011} & 283 & 286 & 286 & 1.42 \cite{Snyder_jmatchem_2011} & 2.34 & 2.34  \\
          & & & & & & (112) & (113) & & & & \\
    ZnSb & $oP16$ &  3.5 \cite{Madsen_PRB_2014, Bottger_JEM_2010} &  1.24 & 1.23 & 92 \cite{Madsen_PRB_2014} & 244 & 242 & 237 &  0.76 \cite{Madsen_PRB_2014, Bottger_JEM_2010} &  2.24 & 2.23 	 \\
          & & & & & & (97) & (96) & & & & \\
    Sb$_2$O$_3$ & $oP20$ & 0.4 \cite{Landolt-Bornstein} & 3.45 & 8.74 & N/A & 418 & 572 & 238 & N/A & 2.13 & 2.12 	\\
          & & & & & & (154) & (211) & & & & \\
    InTe & $cP2$ & N/A & 3.12 & 0.709 & N/A & 191 & 113 & 116 & N/A & 2.28 & 2.19 	\\
          & & & & & & (152) & (90) & & & & \\
    InTe & $tP16$ & 1.7 \cite{Snyder_jmatchem_2011, Spitzer_JPCS_1970} & 1.32 & 1.40 & 186 \cite{Snyder_jmatchem_2011} & 189 & 193 & 150 & 1.0 \cite{Snyder_jmatchem_2011} & 2.23 & 2.24 	\\
          & & & & & & (95) & (97) & & & & \\
    Bi$_2$O$_3$ & $mP20$ & 0.8 \cite{Landolt-Bornstein} & 3.04 & 2.98 & N/A & 345 & 342 & 223 & N/A & 2.10 & 2.10 	 \\
          & & & & & & (127) & (126) & & & & \\
    SnO$_2$ & $tP6$ & 98 \cite{Turkes_jpcss_1980} & 9.56 & 6.98 & N/A & 541 & 487 & 480 & N/A & 2.48 & 2.42 	 \\
          &  & 55 \cite{Turkes_jpcss_1980} & & & & (298) & (268) & & & & \\
    \hline
  \end{tabular}
\end{table*}

The experimental thermal conductivity $\kappa^\EXP$ is compared in Table \ref{tab:misc_thermal} to the thermal conductivity
calculated with \AGL\ using the
Leibfried-Schl{\"o}mann equation (Eq.\ (\ref{thermal_conductivity})) for $\kappa^\sAGL$, while the values obtained for $\kappa^\sBM$, $\kappa^\sVIN$
and $\kappa^\sBCN$ are listed in
the supplementary information. The absolute agreement between the \AGL\ values and $\kappa^\EXP$ is quite poor.
The scarcity of
experimental data from different sources
on the thermal properties of these materials prevents reaching definite conclusions regarding the true values of these
properties. The available data can thus
only be considered as a rough indication of their order of magnitude.

\begin{table}[t!]
  \caption{\small Correlations between experimental values and \AEL\ and \AGL\ results for
    elastic and thermal properties for materials with miscellaneous structures.
  }
  \label{tab:misc_correlation}
  {\footnotesize
    \begin{tabular}{l r r r}
      \hline
      Property  & Pearson & Spearman & \RMSrD\ \\
                & (Linear) & (Rank Order) \\
      \hline
      $\kappa^\EXP$ vs. $\kappa^\sAGL$  ($\sigma = 0.25$) \cite{curtarolo:art96} & 0.937 & 0.071 & 3.38 \\
      $\kappa^\EXP$ vs. $\kappa^\sAGL$ & 0.438 & -0.143 & 8.61 \\
      $\kappa^\EXP$ vs. $\kappa^\sBM$ & 0.498 & -0.143 &  8.81 \\
      $\kappa^\EXP$ vs. $\kappa^\sVIN$ & 0.445 &  0.0 & 8.01 \\
      $\kappa^\EXP$ vs. $\kappa^\sBCN$ & 0.525 & -0.143 & 9.08 \\
      \hline
    \end{tabular}
  }
\end{table}

For these materials, the Pearson  correlation between the calculated
and experimental values of the thermal conductivity ranges from $0.438$ to $0.937$, while the corresponding
Spearman correlations range from $-0.143$ to $0.071$. In this case, using $\sigma^\sAEL$ in the \AGL\
calculations does not improve the correlations, instead actually lowering the values somewhat.
However, it should be noted that the Pearson correlation is heavily influenced by the values for SnO$_2$. 
When this entry is removed from the list, the Pearson correlation values fall to $-0.471$ and $-0.466$
when the $\sigma = 0.25$ and $\sigma = \sigma^\sAEL$ values are used, respectively.
The low correlation values, particularly for the Spearman correlation, for this set of materials demonstrates the
importance of the information about the material structure when interpreting results obtained using the \AGL\ method
in order to identify candidate materials for specific thermal applications. This is partly due to the fact that the Gr{\"u}neisen
parameter values tend to be similar for materials with the same
structure. Therefore, the effect of the Gr{\"u}neisen parameter on the ordinal ranking of
the lattice thermal conductivity of materials with the same structure
is small.


\subsection{Thermomechanical properties from LDA}

{
The thermomechanical properties of a randomly-selected subset of the materials investigated in this work were calculated using LDA 
in order to check the impact of the choice of exchange-correlation functional on the results. For the LDA calculations, all structures were 
first re-relaxed using the LDA exchange-correlation functional with VASP using the appropriate parameters and potentials as
described in the \AFLOW\ standard \cite{curtarolo:art104}, and then the appropriate strained structures were calculated using LDA. 
These calculations were restricted to a subset of materials to limit the total number of additional first-principles calculations required, and the materials were 
selected randomly from each of the sets in the previous sections so as to cover as wide a range of different structure types as possible, given the available experimental data.
Results for elastic properties obtained using LDA, GGA and experimental measurements are shown in Table \ref{tab:LDA_elastic}, while the thermal properties are shown in 
Table \ref{tab:LDA_thermal}. All thermal properties listed in Table \ref{tab:LDA_thermal} were calculated using $\sigma^\sAEL$ in the expression 
for the Debye temperature.}

{
In general, the LDA values for elastic and thermal properties are slightly higher than the GGA values, as would be generally expected
due to their relative tendencies to overbind and underbind, respectively \cite{He_GGA_LDA_PRB_2014, Saadaoui_GGA_LDA_EPJB_2015}. 
The correlations and RMSrD of both the LDA and GGA results with experiment for this set of materials are listed in Table \ref{tab:LDA_correlation}.
The Pearson and Spearman correlation values for LDA and GGA are very close to each other for most of the listed properties. The RMSrD values show 
greater differences, although it isn't clear that one of the exchange-correlation functionals consistently gives better predictions than the other. 
Therefore, the choice of exchange-correlation functional will make little difference to the predictive capability of the workflow, so we choose to 
use GGA-PBE as it is the functional used for performing the structural relaxation for the entries in the \AFLOW\ data repository. 
}

\begin{table*}[t!]
  \caption{\small Bulk modulus, shear modulus and Poisson ratio of a subset of the materials investigated in this work, comparing the effect of using different exchange-correlation functionals.
    ``N/A''= Not available for that source.
    Units: $B$ and $G$ in \GPa.
  }
  \label{tab:LDA_elastic}
  {\footnotesize
    \begin{tabular}{c c c c c c c c c c c c c c c c}
      \hline
      Comp. & $B^\EXP$  & $B_\sVRH^\sGGA$ & $B_\sVRH^\sLDA$ & $B_\sStatic^\sGGA$ & $B_\sStatic^\sLDA$  & $G^\EXP$ & $G_\sVoigt^\sGGA$ & $G_\sVoigt^\sLDA$ & $G_\sReuss^\sGGA$ & $G_\sReuss^\sLDA$ &  $G_\sVRH^\sGGA$ & $G_\sVRH^\sLDA$ & $\sigma^\EXP$ & $\sigma^\sGGA$ & $\sigma^\sLDA$      \\
      \hline
      Si & 97.8 \cite{Semiconductors_BasicData_Springer, Hall_ElasticSi_PR_1967} & 89.1& 96.9 & 84.2 & 92.1 & 66.5 \cite{Semiconductors_BasicData_Springer, Hall_ElasticSi_PR_1967} & 64 & 65 & 61 & 61.9 & 62.5 & 63.4 & 0.223 \cite{Semiconductors_BasicData_Springer, Hall_ElasticSi_PR_1967} & 0.216 & 0.231  \\

      BN & 367.0 \cite{Lam_BulkMod_PRB_1987} & 372 & 402 & 353 & 382 & N/A & 387 & 411 & 374 & 395 & 380 & 403 & N/A & 0.119 & 0.124 \\

      GaSb & 57.0  \cite{Lam_BulkMod_PRB_1987} & 47.0 & 58.3 & 41.6 & 52.3 & 34.2  \cite{Boyle_ElasticGaPSb_PRB_1975} & 30.8 & 35.3 & 28.3 & 32.2 & 29.6 & 33.7  &  0.248  \cite{Boyle_ElasticGaPSb_PRB_1975} & 0.240 & 0.258 \\
 
      InAs & 60.0  \cite{Lam_BulkMod_PRB_1987} & 50.1 & 62.3 & 45.7 & 57.4 & 29.5 \cite{Semiconductors_BasicData_Springer, Gerlich_ElasticAlSb_JAP_1963} & 27.3 & 30.1 & 24.2 & 26.4 & 25.7 & 28.2 & 0.282 \cite{Semiconductors_BasicData_Springer, Gerlich_ElasticAlSb_JAP_1963} & 0.281 & 0.303 \\

      ZnS & 77.1  \cite{Lam_BulkMod_PRB_1987} & 71.2 & 88.4 & 65.8 & 83.3 & 30.9 \cite{Semiconductors_BasicData_Springer} & 36.5 & 42.1 & 31.4 & 35.7 & 33.9 & 38.9 & 0.318 \cite{Semiconductors_BasicData_Springer} & 0.294 & 0.308 \\

    NaCl & 25.1 \cite{Haussuhl_ElasticRocksalt_ZP_1960} & 24.9 & 33.3 & 20.0 & 27.6 & 14.6 \cite{Haussuhl_ElasticRocksalt_ZP_1960} & 14.0 & 19.8 & 12.9 & 16.6 & 13.5 & 18.2 & 0.255 \cite{Haussuhl_ElasticRocksalt_ZP_1960} & 0.271 & 0.269 \\

    KI & 12.2 \cite{Haussuhl_ElasticRocksalt_ZP_1960} & 10.9 & 16.3 & 8.54 & 13.3 & 5.96 \cite{Haussuhl_ElasticRocksalt_ZP_1960} & 6.05 & 9.39 & 4.39 & 5.3 & 5.22 & 7.35 & 0.290 \cite{Haussuhl_ElasticRocksalt_ZP_1960} & 0.294 & 0.305 \\

    RbI & 11.1 \cite{Haussuhl_ElasticRocksalt_ZP_1960} & 9.90 & 14.8 & 8.01 & 12.1 & 5.03 \cite{Haussuhl_ElasticRocksalt_ZP_1960} & 5.50 & 8.54 & 3.65 & 3.94 & 4.57 & 6.24 & 0.303 \cite{Haussuhl_ElasticRocksalt_ZP_1960} & 0.300 & 0.315 \\

    MgO  & 164 \cite{Sumino_ElasticMgO_JPE_1976} & 152 & 164 & 142 & 163 & 131 \cite{Sumino_ElasticMgO_JPE_1976}  & 119 & 138 & 115 & 136 & 117 & 137 & 0.185 \cite{Sumino_ElasticMgO_JPE_1976} & 0.194 & 0.173 \\

   CaO & 113 \cite{Chang_ElasticCaSrBaO_JPCS_1977} & 105 & 129 & 99.6 & 122 & 81.0 \cite{Chang_ElasticCaSrBaO_JPCS_1977} & 73.7 & 87.4 & 73.7 & 86.3 & 73.7 & 86.9 & 0.210 \cite{Chang_ElasticCaSrBaO_JPCS_1977} & 0.216 & 0.225 \\

    GaN & 195 \cite{Semiconductors_BasicData_Springer, Savastenko_ElasticGaN_PSSa_1978} & 175 & 202 & 166 & 196 & 51.6 \cite{Semiconductors_BasicData_Springer, Savastenko_ElasticGaN_PSSa_1978}  & 107 & 116 & 105 & 113 & 106 & 114 & 0.378 \cite{Semiconductors_BasicData_Springer, Savastenko_ElasticGaN_PSSa_1978}  & 0.248 & 0.262 \\
         & 210 \cite{Polian_ElasticGaN_JAP_1996} & & & & & 123 \cite{Polian_ElasticGaN_JAP_1996} & & & & & & & 0.255 \cite{Polian_ElasticGaN_JAP_1996} & & \\

    CdS & 60.7 \cite{Semiconductors_BasicData_Springer, Kobiakov_ElasticZnOCdS_SSC_1980} & 55.4 & 68.2 & 49.7 & 64.1 & 18.2 \cite{Semiconductors_BasicData_Springer, Kobiakov_ElasticZnOCdS_SSC_1980}  & 17.6 & 18.4 & 17.0 & 17.8 & 17.3 & 18.1 & 0.364 \cite{Semiconductors_BasicData_Springer, Kobiakov_ElasticZnOCdS_SSC_1980} & 0.358 & 0.378 \\


    Al$_2$O$_3$ & 254 \cite{Goto_ElasticAl2O3_JGPR_1989} & 231 & 259 & 222 & 250 & 163.1 \cite{Goto_ElasticAl2O3_JGPR_1989} & 149 & 166 & 144 & 163 & 147 & 165 & 0.235 \cite{Goto_ElasticAl2O3_JGPR_1989} & 0.238 & 0.238 \\

    CdGeP$_2$ & N/A & 65.3 & 78.4 & 60.7 & 74.5 & N/A & 37.7 & 42.1 & 33.3 & 36.8 & 35.5 & 39.4 & N/A & 0.270 & 0.285 \\

    CuGaSe$_2$ & N/A & 69.9 & 76.4 & 54.9 & 72.1 & N/A & 30.3 & 34.7 & 26.0 & 30.0 & 28.1 & 32.3  & N/A & 0.322 & 0.315 \\

    CoSb$_3$ & N/A & 78.6 & 99.6 & 75.6 & 96.1 & N/A & 57.2 & 67.1 & 55.1 & 64.2 & 56.2 & 65.7 & N/A & 0.211 & 0.23 \\

      \hline
    \end{tabular}
  }
\end{table*}

\begin{table*}[t!]
  \caption{\small Thermal properties lattice thermal conductivity at
    300K, Debye temperature and Gr{\"u}neisen parameter of 
    a subset of materials, comparing the effect of using different exchange-correlation functionals.
    The values listed for $\theta^{\mathrm{exp}}$ are $\theta_\acoustic$, except 340K for CdGeP$_2$ \cite{Landolt-Bornstein, Abrahams_JCP_1975}, 262K for CuGaSe$_2$ \cite{Landolt-Bornstein, Bohnhammel_PSSa_1982} and 307K for CoSb$_3$ \cite{Snyder_jmatchem_2011} which are $\theta_{\mathrm D}$.
    Units: $\kappa$ in \WmK, $\theta$ in \K.
  }
  \label{tab:LDA_thermal}
  {\footnotesize
    \begin{tabular}{c c c c c c c c c c c}
      \hline
      Comp. & $\kappa^\EXP$  & $\kappa^\sGGA $ & $\kappa^\sLDA$ & $\theta^\EXP$  & $\theta_\sDebye^\sGGA$ & $\theta_\sDebye^\sLDA$ & $\gamma^\EXP$ & $\gamma^\sGGA$ & $\gamma^\sLDA$  \\
            & & & & &  ($\theta_\acoustic^\sGGA$) & ($\theta_\acoustic^\sLDA$) & \\
     \hline
      Si & 166 \cite{Morelli_Slack_2006} & 26.19 & 27.23 & 395 \cite{slack, Morelli_Slack_2006} & 610 & 614 & 1.06 \cite{Morelli_Slack_2006} & 2.06 & 2.03	 \\
            & & & & & (484) & (487) & 0.56 \cite{slack} &  & \\
      BN & 760 \cite{Morelli_Slack_2006} & 281.6 & 312.9 & 1200 \cite{Morelli_Slack_2006} & 1793 & 1840 & 0.7 \cite{Morelli_Slack_2006} & 1.75 & 1.72	\\
            & & & & & (1423) & (1460) & & & \\
      GaSb & 40 \cite{Morelli_Slack_2006} & 4.96 & 5.89 & 165 \cite{slack, Morelli_Slack_2006} & 240 & 254 & 0.75 \cite{slack, Morelli_Slack_2006} & 2.28 & 2.25 	 \\
            & & & & & (190) & (202) & & &  \\
      InAs & 30 \cite{Morelli_Slack_2006} & 4.33 & 4.92 & 165 \cite{slack, Morelli_Slack_2006} & 229 & 238 & 0.57 \cite{slack, Morelli_Slack_2006} & 2.26 & 2.22	 \\
            & & & & & (182) & (189) & & & \\
      ZnS & 27 \cite{Morelli_Slack_2006} & 8.38 & 9.58 & 230 \cite{slack, Morelli_Slack_2006} & 341 & 363 & 0.75 \cite{slack, Morelli_Slack_2006} & 2.00 & 2.02 	 \\
            & & & & & (271) & (288) & & & \\
    NaCl & 7.1 \cite{Morelli_Slack_2006} & 2.12 & 2.92 & 220 \cite{slack, Morelli_Slack_2006} & 271 & 312 & 1.56 \cite{slack, Morelli_Slack_2006} & 2.23 & 2.29 	 \\
          & & & & & (215) & (248)  & & & \\
    KI & 2.6 \cite{Morelli_Slack_2006} & 0.525 & 0.811 & 87 \cite{slack, Morelli_Slack_2006} & 116 & 137 & 1.45 \cite{slack, Morelli_Slack_2006} & 2.35 & 2.37 	 \\
          & & & & & (92) & (109) & & & \\
    RbI & 2.3 \cite{Morelli_Slack_2006} & 0.368 & 0.593 & 84 \cite{slack, Morelli_Slack_2006} & 97 & 115 & 1.41 \cite{slack, Morelli_Slack_2006} & 2.47 & 2.45 	 \\
          & & & & & (77) & (91) & & & \\
    MgO  & 60 \cite{Morelli_Slack_2006} & 44.5 & 58.4 & 600 \cite{slack, Morelli_Slack_2006} & 849 & 935 & 1.44 \cite{slack, Morelli_Slack_2006} & 1.96 & 1.95 \\
          & & & & & (674) & (742) & & &\\
    CaO & 27 \cite{Morelli_Slack_2006} & 24.3 & 28.5 & 450 \cite{slack, Morelli_Slack_2006} & 620 & 665 & 1.57 \cite{slack, Morelli_Slack_2006} & 2.06 & 2.09 	 \\
          & & & & & (492) & (528) & & & \\
    GaN &  210 \cite{Morelli_Slack_2006} & 18.54 & 21.34 & 390 \cite{Morelli_Slack_2006} & 595 & 619 & 0.7 \cite{Morelli_Slack_2006} & 2.08 & 2.04 	 \\
          & & & & & (375) & (390) &  & & \\
    CdS & 16 \cite{Morelli_Slack_2006} & 1.76 & 1.84 & 135 \cite{Morelli_Slack_2006} & 211 & 217 & 0.75 \cite{Morelli_Slack_2006} & 2.14 & 2.14 	 \\
          & & & & & (133) & (137) & & & \\
    Al$_2$O$_3$ & 30 \cite{Slack_PR_1962} & 21.92 & 25.36 & 390 \cite{slack} & 952 & 1002 & 1.32 \cite{slack} & 1.91 & 1.91 	 \\
          & & & & & (442) & (465) & & & & \\
    CdGeP$_2$ & 11 \cite{Landolt-Bornstein, Shay_1975, Masumoto_JPCS_1966} & 2.96 & 3.47 & 340 \cite{Landolt-Bornstein, Abrahams_JCP_1975} & 320 & 337 & N/A & 2.21 & 2.18 \\
          & & & & & (160) & (169) & & & \\
    CuGaSe$_2$ & 12.9 \cite{Landolt-Bornstein, Rincon_PSSa_1995} & 1.46 & 2.23 & 262 \cite{Landolt-Bornstein, Bohnhammel_PSSa_1982} & 244 & 281 & N/A & 2.26 & 2.23 	 \\
          & & & & & (122) & (141) & & & &\\
    CoSb$_3$ & 10 \cite{Snyder_jmatchem_2011} & 2.60 & 3.25 & 307 \cite{Snyder_jmatchem_2011} & 310 & 332 & 0.95 \cite{Snyder_jmatchem_2011} & 2.33 & 2.28 \\
          & & & & & (123) & (132) & & & \\
      \hline
    \end{tabular}
  }
\end{table*}

\begin{table}[t!]
  \caption{\small Correlations between experimental values and \AEL\ and \AGL\ results for
    elastic and thermal properties comparing the LDA and GGA exchange-correlation functionals
    for this subset of materials.
  }
  \label{tab:LDA_correlation}
  {\footnotesize
    \begin{tabular}{l r r r}
      \hline
      Property  & Pearson & Spearman & \RMSrD\ \\
                & (Linear) & (Rank Order) & \\
      \hline
      $\kappa^\EXP$ vs. $\kappa^\sGGA$ & 0.963 & 0.867 & 0.755  \\
      $\kappa^\EXP$ vs. $\kappa^\sLDA$ & 0.959 & 0.848 & 0.706 \\
      $\theta^\EXP$ vs. $\theta^\sGGA$ & 0.996 & 0.996  & 0.119 \\
      $\theta^\EXP$ vs. $\theta^\sLDA$ & 0.996 & 0.996 & 0.174 \\
      $\gamma^\EXP$ vs. $\gamma^\sGGA$ & 0.172 & 0.130 & 1.514 \\
      $\gamma^\EXP$ vs. $\gamma^\sLDA$ & 0.265 & 0.296 & 1.490 \\
      $B^\EXP$ vs. $B_\sVRH^\sGGA$ & 0.995 & 1.0 & 0.111 \\
      $B^\EXP$ vs. $B_\sVRH^\sLDA$ & 0.996 & 1.0 & 0.185 \\
      $B^\EXP$ vs. $B_\sStatic^\sGGA$ & 0.996 & 1.0 & 0.205 \\
      $B^\EXP$ vs. $B_\sStatic^\sLDA$ & 0.998 & 1.0 & 0.072 \\
      $G^\EXP$ vs. $G_\sVRH^\sGGA$ & 0.999 & 0.993 & 0.108 \\
      $G^\EXP$ vs. $G_\sVRH^\sLDA$ & 0.997 & 0.986 & 0.153 \\
      $G^\EXP$ vs. $G_\sVoigt^\sGGA$ & 0.998 & 0.993 & 0.096 \\
      $G^\EXP$ vs. $G_\sVoigt^\sLDA$ & 0.996 & 0.986 & 0.315 \\
      $G^\EXP$ vs. $G_\sReuss^\sGGA$ & 0.999 & 0.993 & 0.163 \\
      $G^\EXP$ vs. $G_\sReuss^\sLDA$ & 0.997 & 0.993 & 0.111 \\
      $\sigma^\EXP$ vs. $\sigma^\sGGA$ & 0.982 & 0.986 & 0.037 \\
      $\sigma^\EXP$ vs. $\sigma^\sLDA$ & 0.983 & 0.993 & 0.052 \\
      \hline
    \end{tabular}
  }
\end{table}

\subsection{\AGL\ predictions for thermal conductivity}

The \AEL-\AGL\ methodology has been applied for
high-throughput screening of the elastic and thermal properties of
over 3000 materials included in the  \AFLOW\ database \cite{curtarolo:art92}.
Tables \ref{tab:highkappa} and \ref{tab:lowkappa} {list those} found
to have the highest and lowest thermal conductivities, respectively.
The high conductivity list is unsurprisingly dominated by various phases of elemental 
carbon{, boron nitride, boron carbide and boron carbon nitride,} while {all other} 
high-conductivity materials also contain at least one of the elements C, B or N.
\begin{table}[t!]
  \caption{\small
    Materials from \AFLOW\ database with the highest thermal conductivities as predicted using
    the \AEL-\AGL\ methodology.
    Units: $\kappa$ in \WmK.
  }
  \label{tab:highkappa}
  \begin{tabular}{l c c c}
    \hline
    Comp. & Pearson & Space Group \# & $\kappa^\sAGL $     \\

    \hline
    C & cF8 & 227 & 420 \\
    BN & cF8 & 216 & 282 \\
    C & hP4 & 194 & 272 \\
    C & tI8 & 139 & 206 \\
    BC$_2$N & oP4 & 25 & 188 \\
    BN & hP4 & 186 & 178 \\
    C & hP8 & 194 & 167 \\
    C & cI16 & 206 & 162 \\
    C & oS16 & 65 & 147 \\
    C & mS16 & 12 & 145 \\
    BC$_7$ & tP8 & 115 & 145 \\
    BC$_5$ & oI12 & 44 & 137 \\
    Be$_2$C & cF12 & 225 & 129 \\
    CN$_2$ & tI6 & 119 & 127 \\
    C & hP12 & 194 & 127 \\
    BC$_7$ & oP8 & 25 & 125 \\
    B$_2$C$_4$N$_2$ & oP8 & 17 & 120 \\
    MnB$_2$ & hP3 & 191 & 117 \\
    C & hP4 & 194 & 117 \\
    SiC & cF8 & 216 & 113 \\
    TiB$_2$ & hP3 & 191 & 110 \\
    AlN & cF8 & 225 & 107 \\
    BP & cF8 & 216 & 105 \\
    C & hP16 & 194 & 105 \\
    VN & hP2 & 187 & 101  \\
    \hline
  \end{tabular}
\end{table}
The low thermal conductivity list tends to contain materials
with large unit cells and heavier elements such as Hg, Tl, Pb and Au.

\begin{table}[t!]
  \caption{\small
    Materials from \AFLOW\ database with the lowest thermal conductivities as predicted using
    the \AEL-\AGL\ methodology.
    Units: $\kappa$ in \WmK.
  }
  \label{tab:lowkappa}
  \begin{tabular}{l c c c}
    \hline
    Comp. & Pearson & Space Group \# & $\kappa^\sAGL $     \\

    \hline
    Hg$_{33}$Rb$_3$ & cP36 & 221 & 0.0113 \\
    Hg$_{33}$K$_3$ & cP36 & 221 & 0.0116 \\
    Cs$_6$Hg$_{40}$ & cP46 & 223 & 0.0136 \\
    Ca$_{16}$Hg$_{36}$ & cP52 & 215 & 0.0751 \\
    CrTe & cF8 & 216 & 0.081 \\
    Hg$_4$K$_2$ & oI12 & 74 & 0.086 \\
    Sb$_6$Tl$_{21}$ & cI54 & 229 & 0.089 \\
    Se & cF24 & 227 & 0.093 \\
    Cs$_8$I$_{24}$Sn$_4$ & cF36 & 225 & 0.104 \\
    Ag$_2$Cr$_4$Te$_8$ & cF56 & 227 & 0.107 \\
    AsCdLi & cF12 & 216 &	0.116 \\
    Au$_{36}$In$_{16}$ & cP52 & 215 & 0.117 \\
    Cd$_3$In & cP4 & 221 & 0.128 \\
    AuLiSb & cF12 & 216 & 0.130 \\
    K$_5$Pb$_{24}$ & cI58 & 217 & 0.135 \\
    K$_8$Sn$_{46}$ & cP54 & 223 & 0.142 \\
    Au$_7$Cd$_{16}$Na$_6$ & cF116 & 225 & 0.145 \\
    Cs & cI2 & 229 & 0.148 \\
    Cs$_8$Pb$_4$Cl$_{24}$ & cF36 & 225 & 0.157 \\
    Au$_{4}$In$_8$Na$_{12}$ & cF96 & 227 & 0.158 \\
    SeTl & cP2 & 221 & 0.164 \\
    Cd$_{33}$Na$_6$ & cP39 & 200 & 0.166 \\
    Au$_{18}$In$_{15}$Na$_6$ & cP39 & 200 & 0.168 \\
    Cd$_{26}$Cs$_2$ & cF112 & 226 & 0.173 \\
    Ag$_2$I$_2$ & hP4 & 186 & 0.192 \\ 
    \hline
  \end{tabular}
\end{table}



By combining the AFLOW search for thermal conductivity values with other properties such as chemical, electronic or structural factors,
candidate materials for specific engineering applications can be rapidly identified for further in-depth analysis using more accurate
computational methods and for experimental examination. {The full set of thermomechanical properties calculated using 
\AEL-\AGL\ for over 3500 entries can be accessed online at \AFLOW.org \cite{aflowlib.org}, which incorporates search and sort functionality to 
generate customized lists of materials.}

\section{Conclusions}

We have implemented the ``Automatic Elasticity Library'' framework for {\it ab-initio}
elastic constant calculations, and integrated it with the ``Automatic \GIBBS\ Library'' implementation of the \GIBBS\ quasi-harmonic Debye model within
the  \AFLOW\ and Materials Project ecosystems.
We used it
to  automatically calculate the bulk modulus, shear modulus, Poisson ratio, thermal conductivity, Debye temperature and Gr{\"u}neisen parameter of materials with
various structures and compared them with available experimental results.

A major aim of high-throughput calculations is to identify useful
property descriptors for screening large datasets of structures \cite{curtarolo:art81}. 
Here, we have examined whether the {\it inexpensive} Debye model, despite its well known deficiencies, can be usefully leveraged for estimating thermal properties of materials by analyzing
correlations between calculated and corresponding experimental quantities.

It is found that the \AEL\ calculation of the elastic moduli
reproduces the experimental results quite well, within 5\% to 20\%,
particularly for materials with cubic and
hexagonal structures. The \AGL\ method, using an isotropic approximation
for the bulk modulus, tends to provide a slightly worse quantitative
agreement but still reproduces trends equally well. 
The correlations are very high, often above $~0.99$.
Using different values of the Poisson ratio mainly affects Debye temperatures, 
while having very little effect on Gr{\"u}neisen parameters.
Several different numerical and empirical equations of state have also been investigated. The differences
between the results obtained from them are
small, but in some cases they are found to introduce an additional
source of error compared to a direct evaluation of the bulk modulus
from the elastic tensor or from the $E(V)$ curve.
Using the different equations of state has very little effect on Debye temperatures,
but has more of an effect on Gr{\"u}neisen parameters. 
Currently, the values for \AGL\ properties available in the \AFLOW\ repository are those calculated by numerically fitting the $E_\sDFT(V)$
data and calculating the bulk modulus using Eq.\ (\ref{bulkmod}).
{The effect of using different exchange-correlation functionals was investigated for a subset of 16 materials. The results showed that 
LDA tended to overestimate thermomechanical properties such as bulk modulus or Debye temperature, compared to GGA's tendency
to underestimate. However, neither functional was consistently better than the other at predicting trends. We therefore use GGA-PBE for
the automated \AEL-\AGL\ calculations in order to maintain consistency with the rest of the \AFLOW\ data.} 

The \AEL-\AGL\ evaluation of the Debye temperature provides good
agreement with experiment for this set of materials, whereas the predictions of the Gr{\"u}neisen parameter
are quite poor. However, since the Gr{\"u}neisen parameter is slowly varying for materials sharing crystal structures, the \AEL-\AGL\
methodology provides a reliable screening tool for identifying materials with very high or very low thermal conductivity.
The correlations between the experimental values of the thermal conductivity and those calculated with  \AGL\ are summarized in
Table \ref{tab:kappa_correlation}. For the entire set of materials examined we find high values of the Pearson correlation
between $\kappa^\EXP$  and $\kappa^\sAGL$, ranging from $0.880$ to $0.933$. It is particularly high, above $0.9$, for materials
with high symmetry (cubic, hexagonal or rhombohedral) structures, but significantly lower for anisotropic materials.
In our previous work on \AGL\ \cite{curtarolo:art96}, we used an approximated the value of $\sigma = 0.25$ in Eq.\ (\ref{fpoisson}).
Using instead the Poisson ratio calculated in \AEL, $\sigma^\sAEL$, the overall correlations are improved
by about 5\%, from $0.880$ to $0.928$, in the agreement with previous
work on metals \cite{Liu_Debye_CMS_2015}. The correlations for
anisotropic materials, such as the body-centred tetragonal set
examined here, improved even more, demonstrating the significance of a
direct evaluation of the Poisson ratio.
This combined algorithm demonstrates the advantage of an integrated high-throughput materials design framework such as \AFLOW,
which enables the calculation of interdependent properties within a single automated workflow.

A direct \AEL\ evaluation of the Poisson ratio, instead of assuming a
simple approximation, e.g.\ a Cauchy solid with $\sigma = 0.25$,
consistently improves the correlations of the \AGL-Debye temperatures
with experiments.
However, it has very little effect on the values obtained for the Gr{\"u}neisen parameter.
Simple approximations lead to more numerically-robust and better system-size scaling calculations,
as they avoid the complications inherent in obtaining the elastic tensor.
{Therefore, \AGL\ could also be used on its own for initial rapid screening,
with \AEL\ being performed later for potentially interesting materials to increase the accuracy of the results.} 


\begin{table}[t!]
  \caption{\small Correlations between experimental values and \AEL\ and \AGL\ results for
    elastic and thermal properties for the entire set of materials.
  }
  \label{tab:kappa_correlation}
  {\footnotesize
    \begin{tabular}{l r r r}
      \hline
      Property  & Pearson & Spearman & \RMSrD\ \\
                & (Linear) & (Rank Order) \\
      \hline
      $\kappa^\EXP$ vs. $\kappa^\sAGL$  ($\sigma = 0.25$) \cite{curtarolo:art96} & 0.880 & 0.752 & 1.293 \\
      $\kappa^\EXP$ vs. $\kappa^\sAGL$ & 0.928 & 0.720 & 2.614 \\
      $\kappa^\EXP$ vs. $\kappa^\sBM$ & 0.879 & 0.735 & 2.673  \\
      $\kappa^\EXP$ vs. $\kappa^\sVIN$ & 0.912 &  0.737 & 2.443 \\
      $\kappa^\EXP$ vs. $\kappa^\sBCN$ & 0.933 & 0.733 & 2.751 \\
      \hline
    \end{tabular}
  }
\end{table}

With respect to rapid estimation of thermal conductivities, 
the approximations in the Leibfried-Schl{\"o}mann formalism
miss some of the details affecting the lattice thermal conductivity, such as the suppression of phonon-phonon scattering due to
large gaps between the branches of the phonon dispersion \cite{Lindsay_PRL_2013}.
Nevertheless, the high correlations between  $\kappa^\EXP$ and
$\kappa^\sAGL$ found for most of the structure families in this study demonstrate the utility of the \AEL-\AGL\ approach
as a screening method for large databases of materials where
experimental data is lacking or ambiguous.  
Despite its intrinsic limitations, the synergy presented by the \AEL-\AGL\ approach
provides the right balance between accuracy and complexity in identifying materials with
promising properties for further investigation.

\section{Acknowledgments}
We thank 
Drs. Kristin Persson, Gerbrand Ceder, Geoffory Hautier, Anubhav Jain, Shyue Ping Ong, 
Wei Chen, Patrick Huck, {Kiran Mathew,} Joseph Montoya and Donald Winston for various technical discussions.
We acknowledge support by the DOE (DE-AC02-05CH11231), specifically the Basic Energy Sciences program under Grant \# EDCBEE.
C.T.,  M.F., M.B.N. and S.C. acknowledge partial support by DOD-ONR (N00014-13-1-0635, N00014-11-1-0136, N00014-15-1-2863).
The consortium \AFLOW.org acknowledges Duke University -- Center for Materials Genomics --- and the CRAY corporation for computational support.

\appendix

\section{AFLOW AEL-AGL REST-API}
\label{restapi_keywords}

The \AEL-\AGL\ methodology described in this work is being used to calculate the elastic and thermal properties of materials in a high-throughput 
fashion by the \AFLOW\ consortium. The results are now available on the \AFLOW\ database \cite{aflowlib.org, curtarolo:art75}
via the \AFLOW\ \RESTAPI\ \cite{curtarolo:art92}. The following optional materials keywords have now been added to the  \AFLOW\ \RESTAPI\
to facilitate accessing this data.

\def\description{\item {{\it Description.}\ }}
\def\type{\item {{\it Type.}\ }}
\def\example{\item {{\it Example.}\ }}
\def\units{\item {{\it Units.}\ }}
\def\syntax{\item {{\it Request syntax.}\ }}


\begin{widetext}

\begin{itemize}

\item
\verb|ael_bulk_modulus_reuss| 
\begin{itemize}
\description Returns \AEL\ bulk modulus as calculated using the Reuss average.
\type \verb|number|.
\units GPa.
\example \verb|ael_bulk_modulus_reuss=105.315|.
\syntax \verb|$aurl/?ael_bulk_modulus_reuss|.
\end{itemize}

\item
\verb|ael_bulk_modulus_voigt| 
\begin{itemize}
\description Returns \AEL\ bulk modulus as calculated using the Voigt average.
\type \verb|number|.
\units GPa.
\example \verb|ael_bulk_modulus_voigt=105.315|.
\syntax \verb|$aurl/?ael_bulk_modulus_voigt|.
\end{itemize}

\item
\verb|ael_bulk_modulus_vrh| 
\begin{itemize}
\description Returns \AEL\ bulk modulus as calculated using the
Voigt-Reuss-Hill (\VRH) average.
\type \verb|number|.
\units GPa.
\example \verb|ael_bulk_modulus_vrh=105.315|.
\syntax \verb|$aurl/?ael_bulk_modulus_vrh|.
\end{itemize}

\item
\verb|ael_elastic_anistropy| 
\begin{itemize}
\description Returns \AEL\ elastic anisotropy.
\type \verb|number|.
\units dimensionless.
\example \verb|ael_elastic_anistropy=0.000816153|.
\syntax \verb|$aurl/?ael_elastic_anisotropy|.
\end{itemize}

\item
\verb|ael_poisson_ratio| 
\begin{itemize}
\description Returns \AEL\ Poisson ratio.
\type \verb|number|.
\units dimensionless.
\example \verb|ael_poisson_ratio=0.21599|.
\syntax \verb|$aurl/?ael_poisson_ratio|.
\end{itemize}

\item
\verb|ael_shear_modulus_reuss| 
\begin{itemize}
\description Returns \AEL\ shear modulus as calculated using the Reuss average.
\type \verb|number|.
\units GPa.
\example \verb|ael_shear_modulus_reuss=73.7868|.
\syntax \verb|$aurl/?ael_shear_modulus_reuss|.
\end{itemize}

\item
\verb|ael_shear_modulus_voigt| 
\begin{itemize}
\description Returns \AEL\ shear modulus as calculated using the Voigt average.
\type \verb|number|.
\units GPa.
\example \verb|ael_shear_modulus_voigt=73.7989|.
\syntax \verb|$aurl/?ael_shear_modulus_voigt|.
\end{itemize}

\item
\verb|ael_shear_modulus_vrh| 
\begin{itemize}
\description Returns \AEL\ shear modulus as calculated using the
Voigt-Reuss-Hill (\VRH) average.
\type \verb|number|.
\units GPa.
\example \verb|ael_shear_modulus_vrh=73.7929|.
\syntax \verb|$aurl/?ael_shear_modulus_vrh|.
\end{itemize}

\item
\verb|ael_speed_of_sound_average| 
\begin{itemize}
\description Returns \AEL\ average speed of sound calculated from the transverse and longitudinal speeds of sound.
\type \verb|number|.
\units m/s.
\example \verb|ael_speed_of_sound_average=500.0|.
\syntax \verb|$aurl/?ael_speed_of_sound_average|.
\end{itemize}

\item
\verb|ael_speed_of_sound_longitudinal| 
\begin{itemize}
\description Returns \AEL\ speed of sound in the longitudinal direction.
\type \verb|number|.
\units m/s.
\example \verb|ael_speed_of_sound_longitudinal=500.0|.
\syntax \verb|$aurl/?ael_speed_of_sound_longitudinal|.
\end{itemize}

\item
\verb|ael_speed_of_sound_transverse| 
\begin{itemize}
\description Returns \AEL\ speed of sound in the transverse direction.
\type \verb|number|.
\units m/s.
\example \verb|ael_speed_of_sound_transverse=500.0|.
\syntax \verb|$aurl/?ael_speed_of_sound_transverse|.
\end{itemize}

\end{itemize}


\begin{itemize}

\item
\verb|agl_acoustic_debye| 
\begin{itemize}
\description Returns \AGL\ acoustic Debye temperature.
\type \verb|number|.
\units K.
\example \verb|agl_acoustic_debye=492|.
\syntax \verb|$aurl/?agl_acoustic_debye|.
\end{itemize}

\item
\verb|agl_bulk_modulus_isothermal_300K| 
\begin{itemize}
\description Returns \AGL\ isothermal bulk modulus at 300K and zero pressure.
\type \verb|number|.
\units GPa.
\example \verb|agl_bulk_modulus_isothermal_300K=96.6|.
\syntax \verb|$aurl/?agl_bulk_modulus_isothermal_300K|.
\end{itemize}

\item
\verb|agl_bulk_modulus_static_300K| 
\begin{itemize}
\description Returns \AGL\ static bulk modulus at 300K and zero pressure.
\type \verb|number|.
\units GPa.
\example \verb|agl_bulk_modulus_static_300K=99.59|.
\syntax \verb|$aurl/?agl_bulk_modulus_static_300K|.
\end{itemize}

\item
\verb|agl_debye| 
\begin{itemize}
\description Returns \AGL\ Debye temperature.
\type \verb|number|.
\units K.
\example \verb|agl_debye=620|.
\syntax \verb|$aurl/?agl_debye|.
\end{itemize}

\item
\verb|agl_gruneisen| 
\begin{itemize}
\description Returns \AGL\ Gr{\"u}neisen parameter.
\type \verb|number|.
\units dimensionless.
\example \verb|agl_gruneisen=2.06|.
\syntax \verb|$aurl/?agl_gruneisen|.
\end{itemize}

\item
\verb|agl_heat_capacity_Cv_300K| 
\begin{itemize}
\description Returns \AGL\ heat capacity at constant volume (C$_V$) at 300K and zero pressure.
\type \verb|number|.
\units k$_\mathrm{B}$/cell.
\example \verb|agl_heat_capacity_Cv_300K=4.901|.
\syntax \verb|$aurl/?agl_heat_capacity_Cv_300K|.
\end{itemize}

\item
\verb|agl_heat_capacity_Cp_300K| 
\begin{itemize}
\description Returns \AGL\ heat capacity at constant pressure (C$_p$) at 300K and zero pressure.
\type \verb|number|.
\units k$_\mathrm{B}$/cell.
\example \verb|agl_heat_capacity_Cp_300K=5.502|.
\syntax \verb|$aurl/?agl_heat_capacity_Cp_300K|.
\end{itemize}

\item
\verb|agl_poisson_ratio_source| 
\begin{itemize}
\description Returns source of Poisson ratio used to calculate Debye temperature in \AGL. Possible sources include \verb|ael_poisson_ratio_<value>|, in
which case the Poisson ratio was calculated from first principles using \AEL; \verb|empirical_ratio_<value>|, in which case the value was taken 
from the literature; and \verb|Cauchy_ratio_0.25|, in which case the default value of 0.25 of the Poisson ratio of a Cauchy solid
was used.
\type \verb|string|.
\example \verb|agl_poisson_ratio_source=ael_poisson_ratio_0.193802|.
\syntax \verb|$aurl/?agl_poisson_ratio_source|.
\end{itemize}

\item
\verb|agl_thermal_conductivity_300K| 
\begin{itemize}
\description Returns \AGL\ thermal conductivity at 300K.
\type \verb|number|.
\units W/m*K.
\example \verb|agl_thermal_conductivity_300K=24.41|.
\syntax \verb|$aurl/?agl_thermal_conductivity_300K|.
\end{itemize}

\item
\verb|agl_thermal_expansion_300K| 
\begin{itemize}
\description Returns \AGL\ thermal expansion at 300K and zero pressure.
\type \verb|number|.
\units 1/K.
\example \verb|agl_thermal_expansion_300K=4.997e-05|.
\syntax \verb|$aurl/?agl_thermal_expansion_300K|.
\end{itemize}

\end{itemize}

\end{widetext}


\newcommand{\Ozolins}{Ozoli\c{n}\v{s}}

\end{document}